\documentclass{aa}  

\usepackage{txfonts}
\usepackage{graphicx,longtable,lscape,natbib,amssymb,multirow,sidecap}
\usepackage{epstopdf}
\usepackage{natbib}
\usepackage{color}
\usepackage[utf8]{inputenc}
\bibpunct{(}{)}{;}{a}{}{,} 
\newcommand{\ltsima} {$\; \buildrel < \over \sim \;$}  
\newcommand{\gtsima} {$\; \buildrel > \over \sim \;$}  
\newcommand{\lta} {\lower.5ex\hbox{\ltsima}}  
\newcommand{\gta} {\lower.5ex\hbox{\gtsima}}  
\newcommand{\Ha} {H$\alpha$\,}  
\newcommand{\Hb} {H$\beta$\,}

\newcommand{\ergscm}{$\>{\rm erg}\,{\rm s}^{-1}\,{\rm cm}^{-2}$}

\newcommand{\kms}{$\rm{\,km \,s}^{-1}$}

\newcommand{\loiii}{L$_{\rm{\tiny{ [O~III]}}}$}

\newcommand{\forb}[2]{\mbox{$[{\rm #1\, #2}]$}}
\newcommand{\oiii}{\forb{O}{III}\,}
\newcommand{\oii}{\forb{O}{II}\,}

\newcommand{\mc}{${\rm \mu}$m\,}
\newcommand{\sigoiii}{$\sigma_{\rm [O~III]}$\,}
\newcommand{\sigoii}{$\sigma_{\rm [O~II]}$\,}

\newcommand{\loii}{L$_{\rm{\tiny{ [O~II]}}}$}

\newcommand{\vblue}{$\nu_{\rm blue}$}
\newcommand{\voff}{$\Delta\nu_{\rm offset}$\,}
\newcommand{\dsig}{$\Delta\nu_{\rm \sigma[O~III]}$}

\usepackage[utf8]{inputenc}
 
\usepackage[normalem]{ulem}

\begin{document}

\title{Is there any evidence that ionized outflows quench star formation in type 1 quasars at z$<$ 1?}
\subtitle{} \titlerunning{Ionized gas outflow and star formation} \authorrunning{Balmaverde \& Marconi}

\author{B.~Balmaverde\inst{1,2}
\and A.~Marconi\inst{1,2}
\and M.~Brusa\inst{3,4}
\and S.~Carniani\inst{6}
\and G.~Cresci\inst{1}
\and E.~Lusso\inst{1}
\and R.~Maiolino\inst{5,6}
\and F.~Mannucci\inst{1}
\and T.~Nagao\inst{7}}

\institute {Dipartimento di Fisica e
  Astronomia, Università di Firenze, via G. Sansone 1, 50019 Sesto
  Fiorentino (Firenze), Italy
\and INAF - Osservatorio Astrofisico di Arcetri, Largo Enrico
  Fermi 5, I-50125 Firenze, Italy
\and  Dipartimento di Fisica e Astronomia, Università di Bologna, viale Berti Pichat 6/2, I-40127 Bologna, Italy
\and INAF - Osservatorio Astronomico di Bologna, via Ranzani 1, I-40127 Bologna, Italy
\and Cavendish Laboratory, University of Cambridge, 19 J. J. Thomson Ave, Cambridge CB3 0HE, UK
\and Kavli Institute for Cosmology, University of Cambridge, Madingley Road, Cambridge CB3 0HA, UK
\and  Research Center for Space and Cosmic Evolution,
Ehime University, Bunkyo-cho 2-5, Matsuyama 790-8577, Japan}

\offprints{balmaverde@oato.inaf.it}

\abstract{}{The aim of this paper is to test the basic model of negative AGN
  feedback. According to this model,  once the central black hole accretes at the Eddington limit and reaches
  a certain critical mass, AGN driven outflows blow out gas,
  suppressing star formation in the host galaxy and
  self-regulating black hole growth. 
  }  {We consider a sample of 224 quasars selected
  from the Sloan Digital Sky Survey (SDSS) at  z $<$ 1
  observed in the infrared band by the Herschel Space Observatory in
  point source photometry mode. 
  We evaluate the star
formation rate in 
relation to several outflow signatures traced by the \oiii$\lambda$4959,5007 and \oii$\lambda$3726,3729 emission lines in about half of the sample
with high quality spectra.
 }  {Most of the quasars show asymmetric and broad wings in [O~III], which we interpret as outflow signatures. 
   We separate the quasars in two
  groups,  ``weakly'' and ``strongly'' outflowing, using three
  different criteria. When we compare the mean star
  formation rate in five redshift  bins  in the two groups, we find that the SFRs
  are comparable or slightly larger in the strongly outflowing
  quasars.
  We estimate the stellar mass from SED fitting and the quasars 
 are distributed along the star formation main sequence, although with a large scatter. The scatter from this relation
 is uncorrelated with respect to the kinematic properties of the outflow.  
  Moreover, for quasars
  dominated in the infrared by starburst or by AGN emission, we do not find any correlation between
   the star formation rate and the velocity of
  the outflow, a trend previously reported in the literature for pure
  starburst galaxies.}{We conclude
  that the basic AGN negative feedback scenario seems not to agree
  with our results. 
  Although we use a large sample of quasars, we did not find
  any evidence that the star formation rate is suppressed in the presence 
  of AGN driven outflows on large scale. A possibility is that feedback is effective over much longer timescales than those of single 
  episodes of quasar activity.}

\maketitle

\section{Introduction}

An important discovery in the last 15 years is that
the black hole mass at the centre of the galaxies is a constant
fraction of the mass of the stellar bulge
(\citealt{ferrarese00}, \citealt{gebhardt00}).  From that moment it has
been clear that the evolution of galaxies is closely
connected with the growth of the central black hole and that the 
influence of the black hole on the host galaxy cannot be ignored.

In particular AGN feedback (i.e. the energy released by the AGN into the interstellar medium) is invoked to solve many astrophysical problems. For example,
 the very sharp cut-off seen at the bright end of the galaxies luminosity function cannot be reproduced
without invoking a feedback mechanism. In other words, super-winds may be responsible for the lack of bright galaxies that are not observed in reality 
(\citealt{benson03}) and for the red, dead elliptical galaxies we observe today. Another
crucial problem that requires AGN feedback is the so-called cooling flow problem at the centre of galaxy clusters, which requires a source of heating
able to balance or quench cooling (\citealt{fabian12} and references therein).

A wide variety of quenching mechanisms have been proposed to self-regulate the growth of the galaxy (i.e. the star formation rate) and
the black hole mass accretion rate (e.g. \citealt{dimatteo05}; \citet{hopkins06}; 
\citealt{croton06}; \citealt{martig09}).  Generally, the quenching is
identified with a triggering event, for example a major merger,
that channels a large amount of gas into the central region in a short
period of time, activating both star formation (SF) and quasar
activity.  The resulting AGN feedback rapidly exhausts or sweeps away
the gas from the galaxy.  Theoretical models (e.g.
\citealt{hopkins06}, \citealt{granato04}, \citealt{dimatteo05}) have proposed that feedback from AGNs is
responsible for the regulation of SF, transforming
a  starburst galaxy into a red elliptical galaxy.

In a considerable fraction
of AGNs there are clear signatures of outflows likely able to remove significant amounts of cold gas from the
galaxy.  AGN-driven outflows with velocities of $\sim$1000\kms\ have
been identified in ultra luminous infra red galaxies (ULIRGs), e.g. \citealt{cicone13}, \citealt{rupke13},
and fast outflows are seen in broad absorption line QSOs
displaying broad UV absorption lines with widths of several thousand
\kms (\citealt{turnshek88}). 
The highly ionized Fe absorption lines blueshifted by 0.05-0.3c observed in the X-ray bands 
(e.g.  \citealt{tombesi10}) have been found associated with the much slower ($\sim$1000 km/s) 
outflow components seen in the molecular gas (e.g. \citealt{feruglio15}, \citealt{tombesi15}). 
On the galactic
scale, integral field spectroscopic observations have revealed the
presence of ionized gas outflows with velocites up to 1000 \kms in low
and high redshift galaxies (\citealt{alexander10}, \citealt{harrison12}, \citealt{harrison14}, \citealt{cano12}).

However, studies exploring the role of  AGNs with respect to
star formation have  led to ambiguous and contrasting results, with
evidence for both negative and positive impact of AGN outflows on
star formation.  After many years of extensive research,  conclusive
proof that AGN feedback is able to halt galaxy-wide star
formation is still lacking. Moreover, many studies show that the
opposite may actually be true. The squeezing and compression of cold
gas induced by accretion related winds could cause local density
fluctuations leading to star formation \citep{ishibashi12}.
Theoretical models able to
explain the connection between the central engine and the host galaxy
properties with secular evolution without invoking AGN feedback  have been developed (e.g. \citealt{jahnke11}, \citealt{ciotti07}).
Finally, even if it is well known that the AGN driven outflows are
common in both  radio loud and radio quiet quasars, it is not obvious
if  there is an efficient coupling mechanism between the AGN and the ISM
(e.g. \citealt{bicknell00}, \citealt{kalfountzou12}).

In this paper we focus on ionized gas outflows traced by the optical
\oiii and \oii lines.  Since the forbidden line emission is isotropic and
self-absorption in narrow lines is negligible, the blue wings are
generally interpreted as the result of  outflowing gas.  The [OIII]
line is typical for AGNs and originates from the ionized narrow-line
region (NLR) gas surrounding the accreting super massive black hole in
the centre of the galaxy (see \citealt{osterbrock89}).  \citealt{mullaney13}
constructed average \oiii$\lambda$5007 profiles derived from stacking
SDSS spectra of type 1 AGNs and found profiles displaying prominent blue
wings shifted with respect to the narrow components.  Similar results are
found by \citet{zakamska14} who analysed the SDSS spectra of obscured
luminous quasars. The \oiii$\lambda$5007 emission line typically shows
blueshifts and blue excess, likely signatures that the NLR is undergoing an
outflow.

The question we address in this paper is whether SMBH outflows and
feedback have the ability to regulate the starburst on a galaxy-wide scale as predicted by the negative feedback scenario.
To this end, 
we compare the kinematic of the
ionized gas outflows derived from the \oiii line profile to the star
formation rate of the host galaxies estimated through the fit of
photo-metric points in the infrared. Since the two processes likely act on
different timescales (e.g. \citealt{stanley15} and references therein) we need a large sample of luminous powerful
quasars in order to recover the signature that AGN outflows affect the star
formation rate of the host galaxy.  
Direct {\rm Herschel} FIR measurements
of individual galaxies is a unique tool used to properly estimate the
ongoing star formation rate, especially in low luminosity IR galaxies,
that in large sky surveys would not be detected.  Therefore, we selected
a large sample of quasars at redshift $\lesssim$ 1 (to cover the \oiii
emission line wavelength in SDSS optical spectra) observed by the
infrared telescope {\rm Herschel} in photometric point source mode.

The paper is organized as follows. In Sect. 2 we define the sample of
quasars, and in Sect. 3 we present the analysis of the SDSS
spectra and of the Hershel data.  The main results and the comparison
between the outflow properties, the SFR and AGN properties, are presented in
Sects. 4 and 5. In Sect. 6 we provide a discussion, and in Sect. 7 our summary
and conclusions.

In this paper we assume the following cosmology parameters: $\Omega_m$=0.3, $\Lambda$ = 0.7 and
H$_0$ = 70 km s$^{-1}$ Mpc$^{-1}$.

\section{The sample}

\begin{figure}
\includegraphics[width=8cm,angle=0]{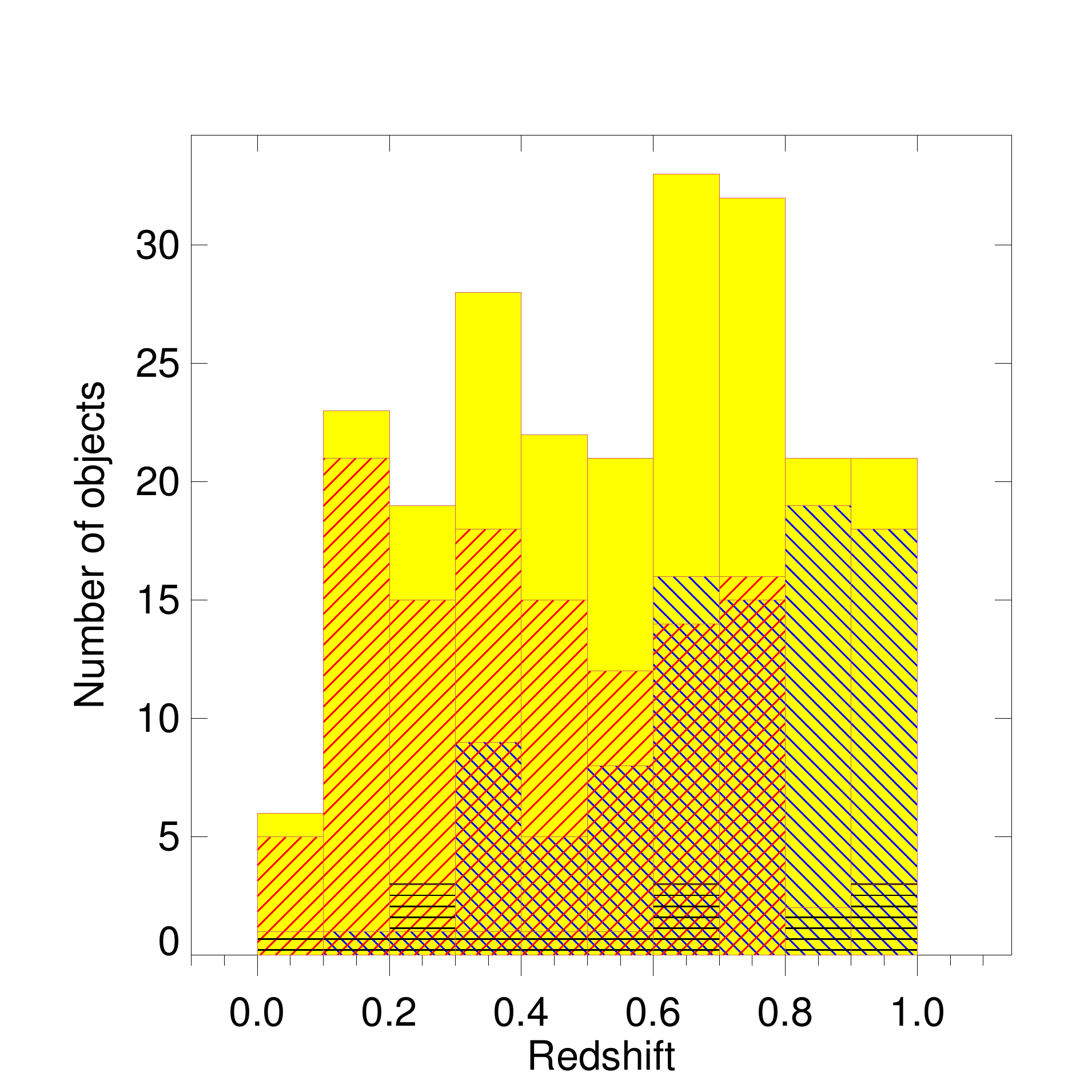}
\caption{Histogram of the redshift distribution for the quasars in our sample. The red,
 blue, and  black hatched areas  identify quasars from the DR7Q, DR9Q,  and SDSS archives, respectively. The yellow area 
denotes the sum of the three subsamples.}
\label{hist}
\end{figure}

\begin{figure}
\includegraphics[scale=0.32,angle=0]{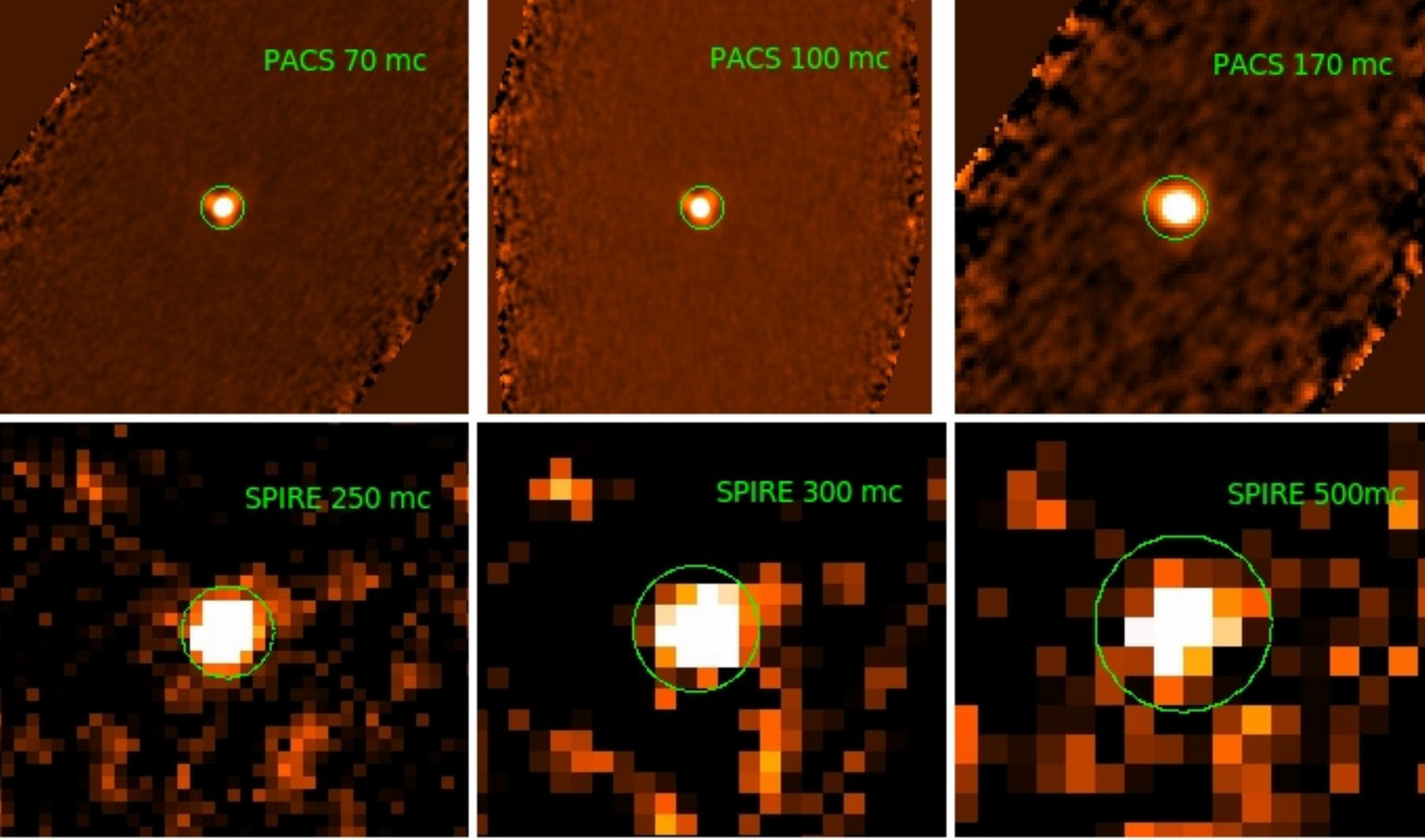}
\caption{Herschel PACS and SPIRE images in
different bands for a quasar from our sample. The green circle is centred
on the source and represents the extraction region.}
\label{herschel}
\end{figure}

The AGN parent sample includes bona fide quasars extracted from the
Sloan Digital Sky Survey (SDSS) archive or the SDSS quasar catalogue that have high resolution
photometric images from the infrared satellite {\rm Herschel}.  

First of all, we selected a list of reliable quasars candidates from

\begin{itemize}

\item  the
fifth edition of the spectroscopic quasar catalogue of
\citet{schneider10} based on the SDSS seventh data release (DR7Q). This catalogue contains 19552
quasars at redshifts of less than 0.8 (this redshift limit ensures that \oiii is in the observed frame).


\item the quasar catalogue of \citet{paris12} based on the SDSS ninth data
release (DR9Q). There are  23964 quasars at redshifts of less than 1.

\item the SDSS archive. We searched for   
objects classified as `QSO'  in the latest SDSS data
release (DR10) not present in the two quasar
catalogues with
redshifts of less than 1 and a median signal-to-noise ratio greater than 5
(35830 quasars).

\end{itemize}

These subsamples are selected according to several different criteria.
The DR7Q catalogue contains quasars mostly at z $<$ 2 (see
\citealt{shen11}, for their properties) selected with a complex
algorithm described in \citet{richards02}.  
The DR9Q quasar catalogue contains mostly
newly discovered quasars, but also some DR7 quasars 
(about 16420) 
that were re-observed during the Baryon Oscillation Spectroscopic Survey
(BOSS, \citealt{schlegel07}).  All quasars have
absolute i-band magnitudes brighter than -22, contain at least one
broad emission line (with FWHM $>$ 500 \kms) or -- if they do not -- have
interesting/complex absorption features, and have highly reliable
redshifts. The threshold in absolute magnitude was relaxed for DR9Q
(M$_i$ > -20.5) because BOSS were targeting mostly z $>$ 2.15
quasars.  

We then searched in the Herschel archive for PACS and/or SPIRE images covering a portion of a circle of five
arcmin of radius around the quasars position for all the $\sim$79000 quasars. We searched for
the nearest source within a circular region of 20\arcsec\ centred on SDSS quasar position.
We found 224 quasars
observed by {\rm Herschel} in point source photometry mode providing the
best instrumental sensitivity and resolution available (114 from DR7Q,
92 from DR9Q, and 18 from DR10 data archive). 
 In
Fig. \ref{hist} we show the distribution of the redshift values for the quasars in our
sample.

\section{Data analysis}

\subsection{Herschel photometry}

The {\rm Herschel Space Observatory} is equipped with a 3.5 m telescope that
has performed photometry and spectroscopy in approximately the
55-671 \mc\, range, mapping the ``warm'' dust heated by young 
stars. There are two photometric instruments, the Photodetecting Array Camera and Spectrometer (PACS), which  observes in 
three colours (blue 60-85 \mc, green 85-130 \mc, and red 130-210 \mc),
and  the Spectral and Photometric Imaging Receiver (SPIRE), which observes
 at 250, 350, and 500 \mc (\citealt{poglitsch10}, \citealt{griffin10}).

We perform aperture photometry on PACS and SPIRE pipeline-processed images using the
annularSkySperturePhotometry task in HIPE version 12.0.
We adopt a circular target aperture centred on the position of the
target of radius 10\arcsec\, and 15\arcsec\,  for blue/green and red PACS
images, respectively, and 22\arcsec, 30\arcsec, and 42\arcsec\  for the three SPIRE bands (see Fig. \ref{herschel}). 
The background annulus is
chosen between 20\arcsec\ and 40\arcsec\  for PACS blue/green bands, between 25\arcsec\ 
and 45\arcsec\, for PACS red band, and between 65\arcsec\, and 95\arcsec\, for
SPIRE images (see PACS and SPIRE data reduction guide\footnote{http://herschel.esac.esa.int/hcss-doc-14.0/}
for details on the PSF shape and the suggested photometry aperture). 
The default algorithm for the sky estimation is the same
used in the IDL daophot package, which eliminate the outlier pixels,
i.e.  values with a low probability given a Gaussian with specified
average and sigma.  
We define the luminosity errors $\sigma_{src}$ as the square root of the
quadratic sum of two terms,
$\sqrt{N_{src\_pix}}\sigma_{sky}$,
and the error on the mean of sky $\sqrt{N_{src\_pix}}\frac{\sigma_{sky}}{\sqrt{N_{sky}}}$
(however the second term is irrelevant because the sky is measured on a much larger area with respect to 
the extraction region); $\sigma_{src}$
is therefore given by the uncertainty of the background times the square root of the
number of pixels in the extraction region, $N_{src\_pix}$.  

If we do not
reach a signal-to-noise ratio (S/N, defined as the ratio of the flux density S over the
uncertainty error $\sigma_{src}$) of at least a factor 3, we estimate
an upper limit for the flux density at 2$\sigma$ confidence
level as 2$\times\sigma_{src}\times
\sqrt{N_{src\_pix}}$ (in this case $\sigma_{src}$ is the standard deviation of
the flux density mean values in the extraction region). Finally we
applied aperture correction using the task
photApertureCorrectionPointSource.

\begin{figure*}
\centering{
\includegraphics[scale=0.40,angle=0]{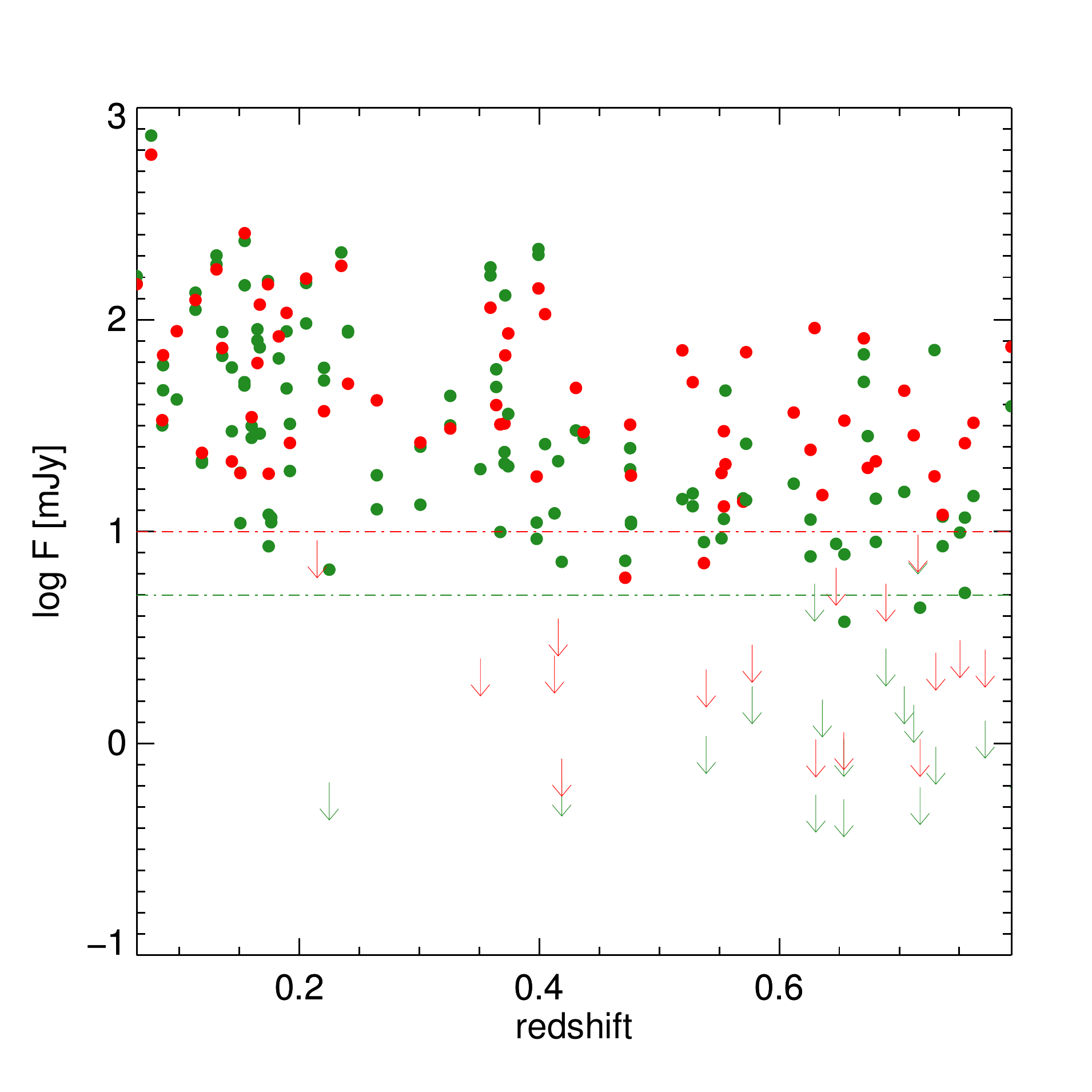}
\includegraphics[scale=0.40,angle=0]{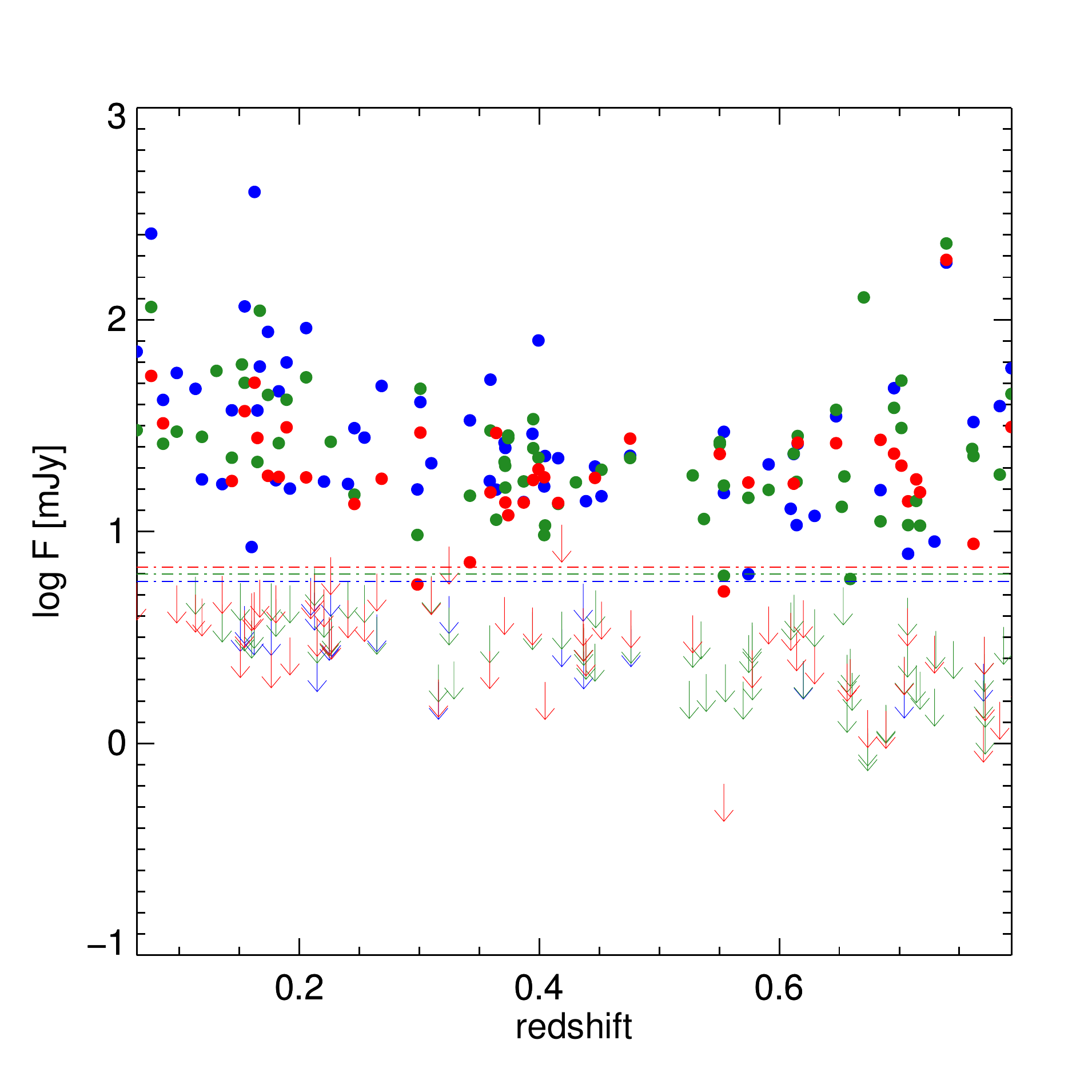}
\caption{Left panel: observed source fluxes in the PACS bands at 70/100 \mc and 160 \mc (green and red points, respectively),
compared with the  detector sensitivity threshold at 5$\sigma$ for one hour of observation (shown as dash-dotted green and red lines). 
Right panel: same figure for the SPIRE band at 250 \mc, 350 \mc, 500 \mc
(blue, green, and red points, respectively). The three dash-dotted lines are the  detector sensitivity thresholds in the three bands at 5$\sigma$ for one-hour observations.
}
\label{sensitivity}}
\end{figure*}

The sensitivity of the Herschel instruments depends on many factors, but
to a first order in all the observing modes it scales with the inverse of the square root of the on-source observation time. 
The typical sensitivity for a  5$\sigma$ detection in one hour of observation is of the order of 
 5 mJy for PACS in the 70/100 \mc band  and 10 mJy for SPIRE in the 160 \mc band\footnote{http://herschel.esac.esa.int/Docs/Herschel/html/ch03s02.html}.
In Fig. \ref{sensitivity}, we compare the PACS AND SPIRE fluxes in all the bands with this approximate sensitivity threshold. The median
exposure time for our data is comparable, and all our upper limits are located below
the threshold for detectability in one hour of observation and the detections (with very few exceptions) are located above these lines.

\begin{figure*}
\centering{
\includegraphics[scale=0.19,angle=0]{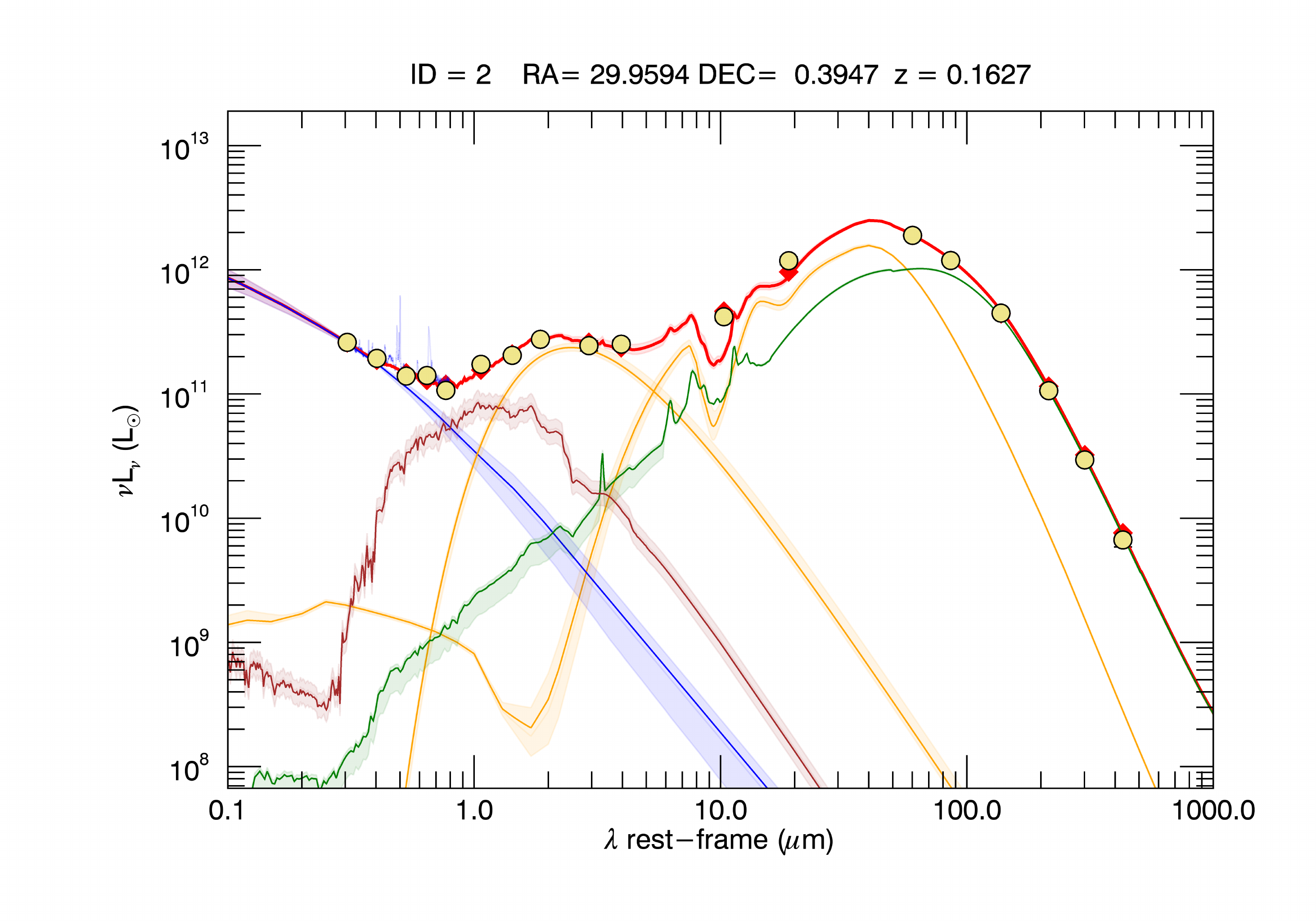}
\includegraphics[scale=0.19,angle=0]{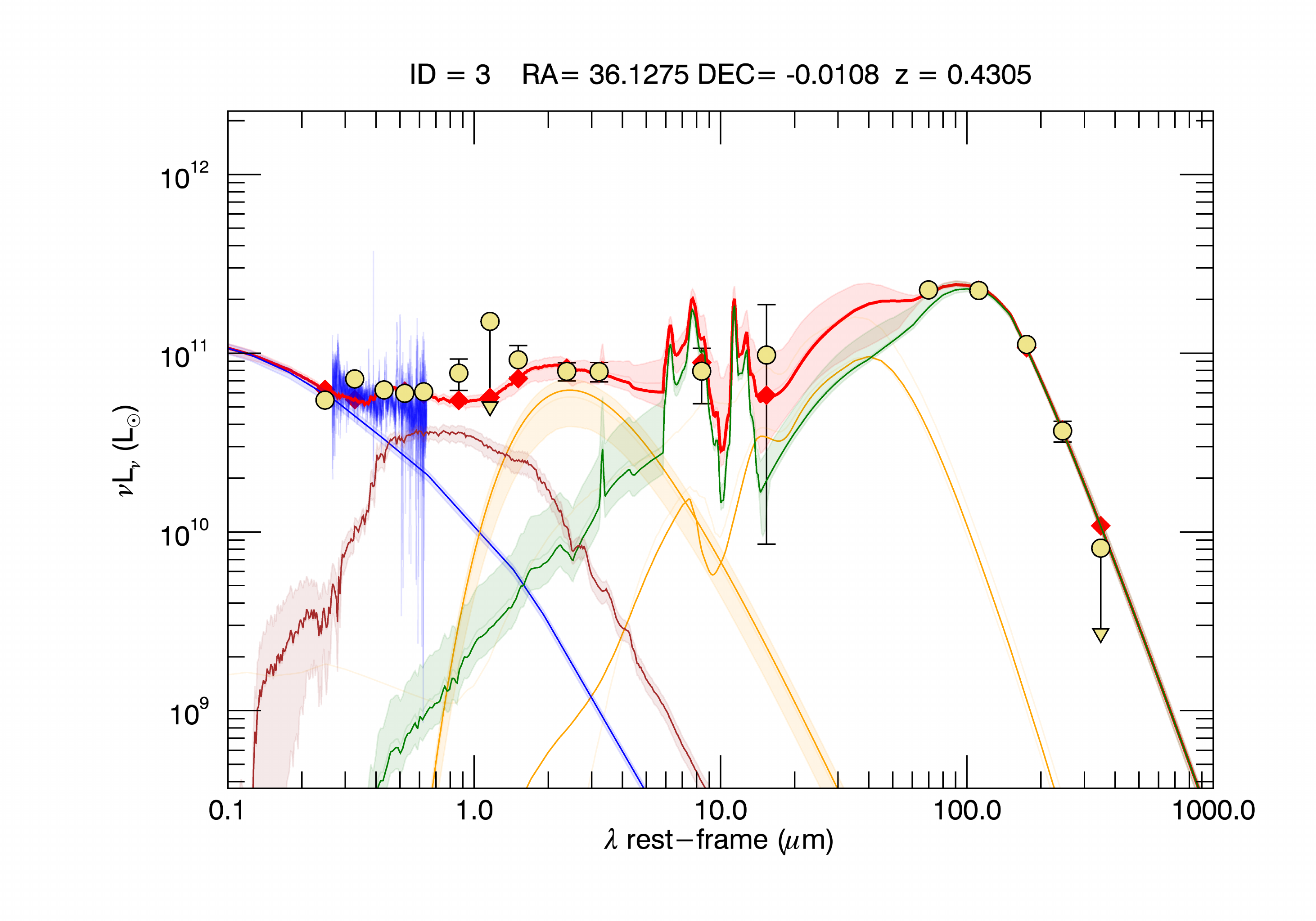}
\includegraphics[scale=0.19,angle=0]{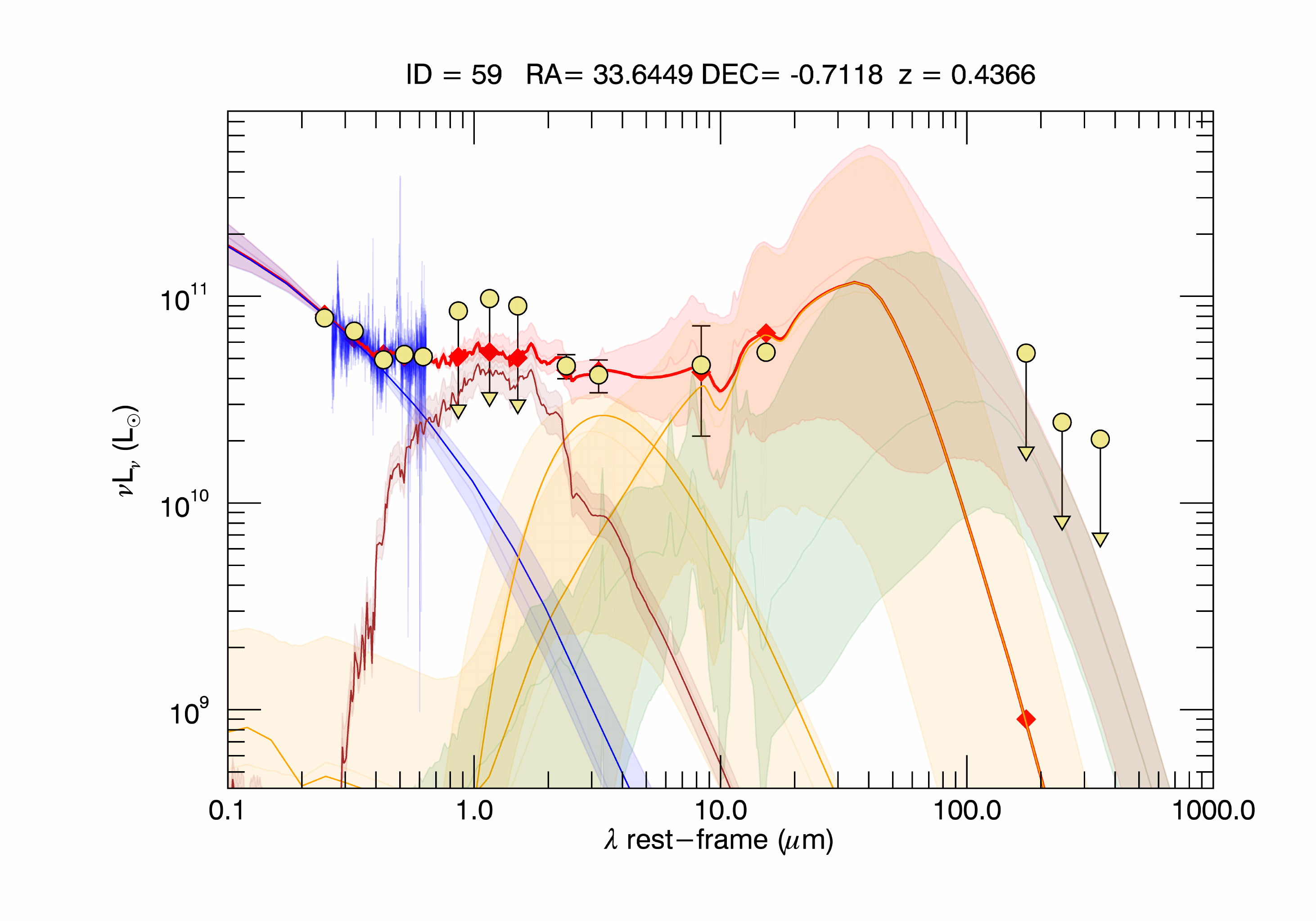}}
\centering{
\includegraphics[scale=0.19,angle=0]{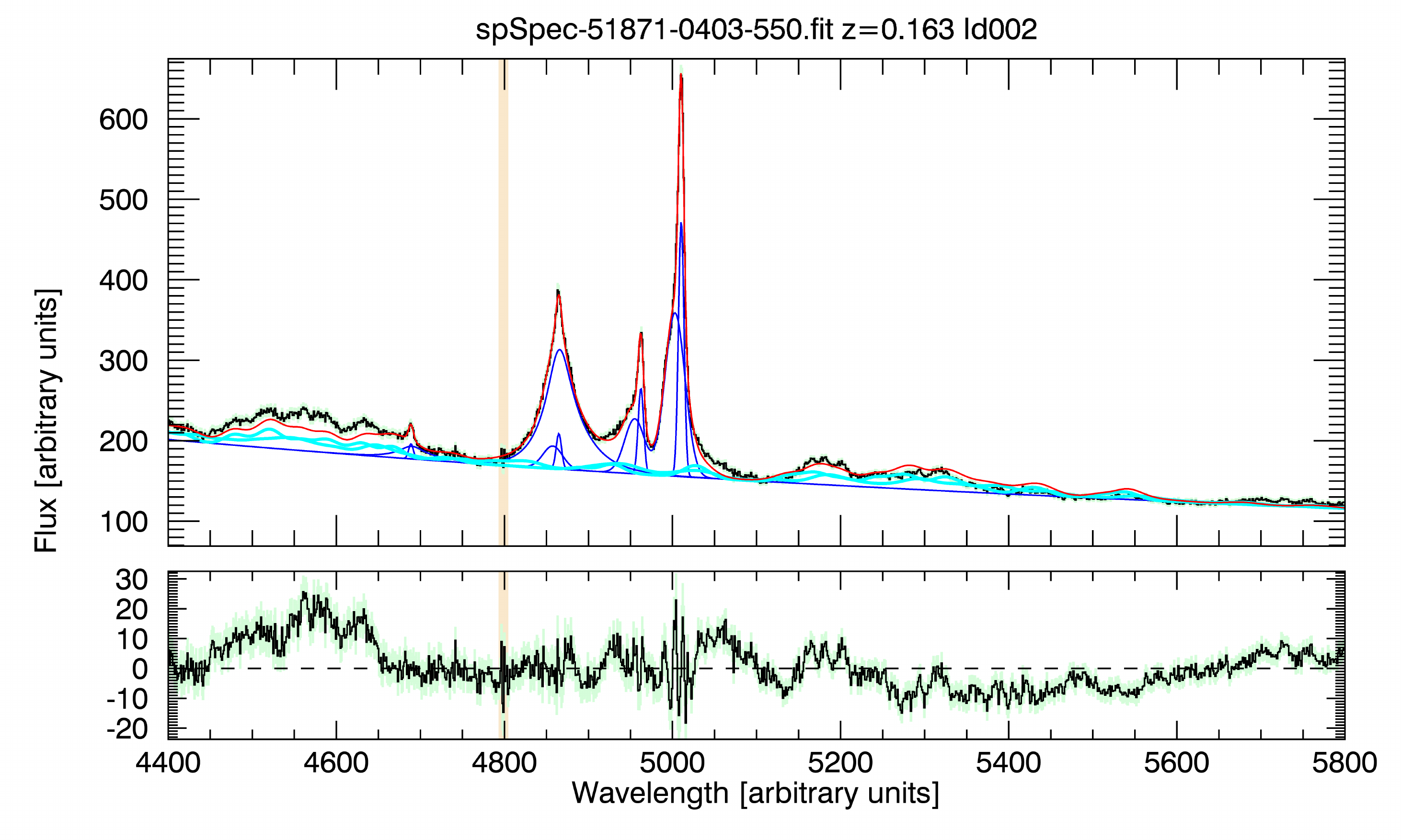}
\includegraphics[scale=0.19,angle=0]{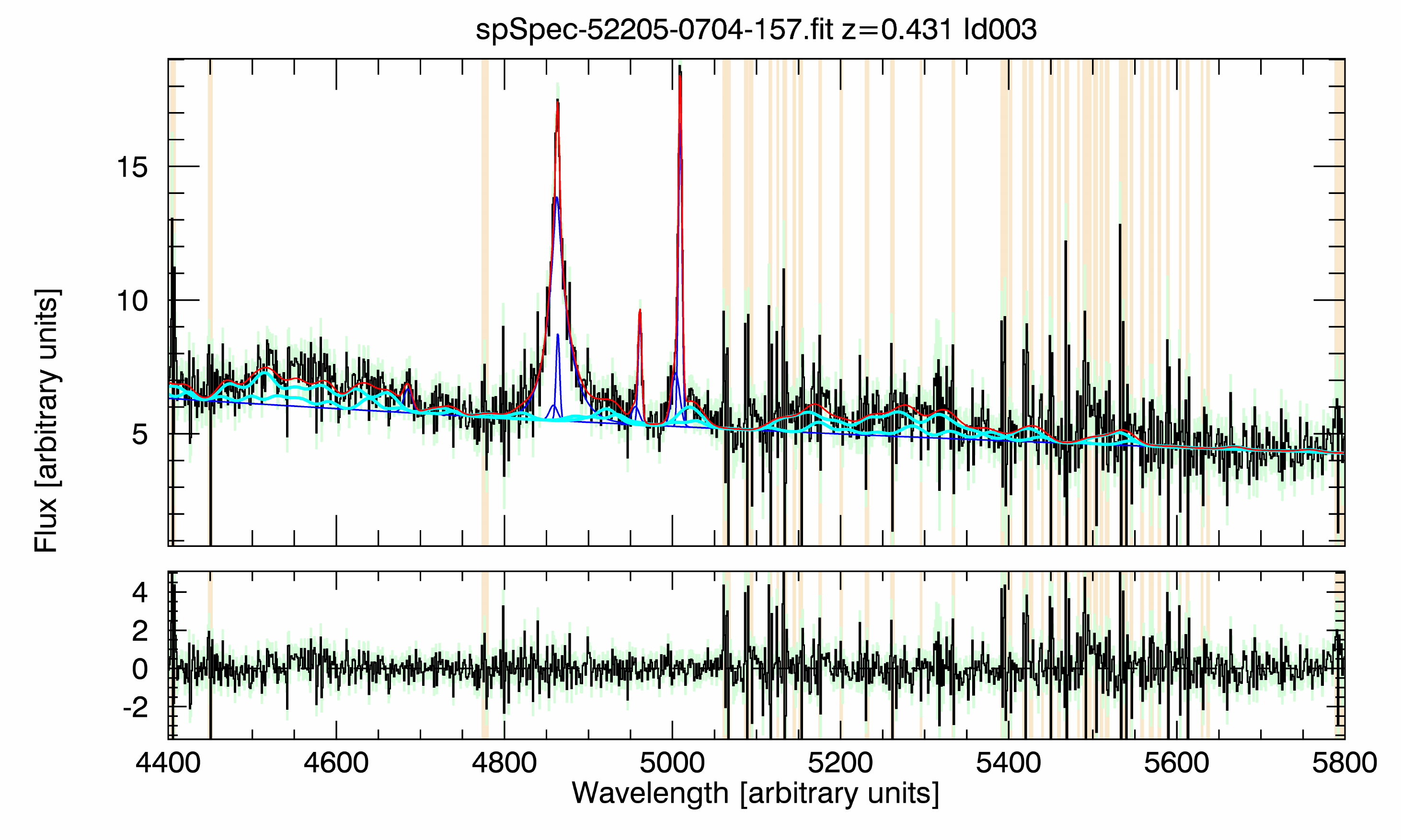}
\includegraphics[scale=0.19,angle=0]{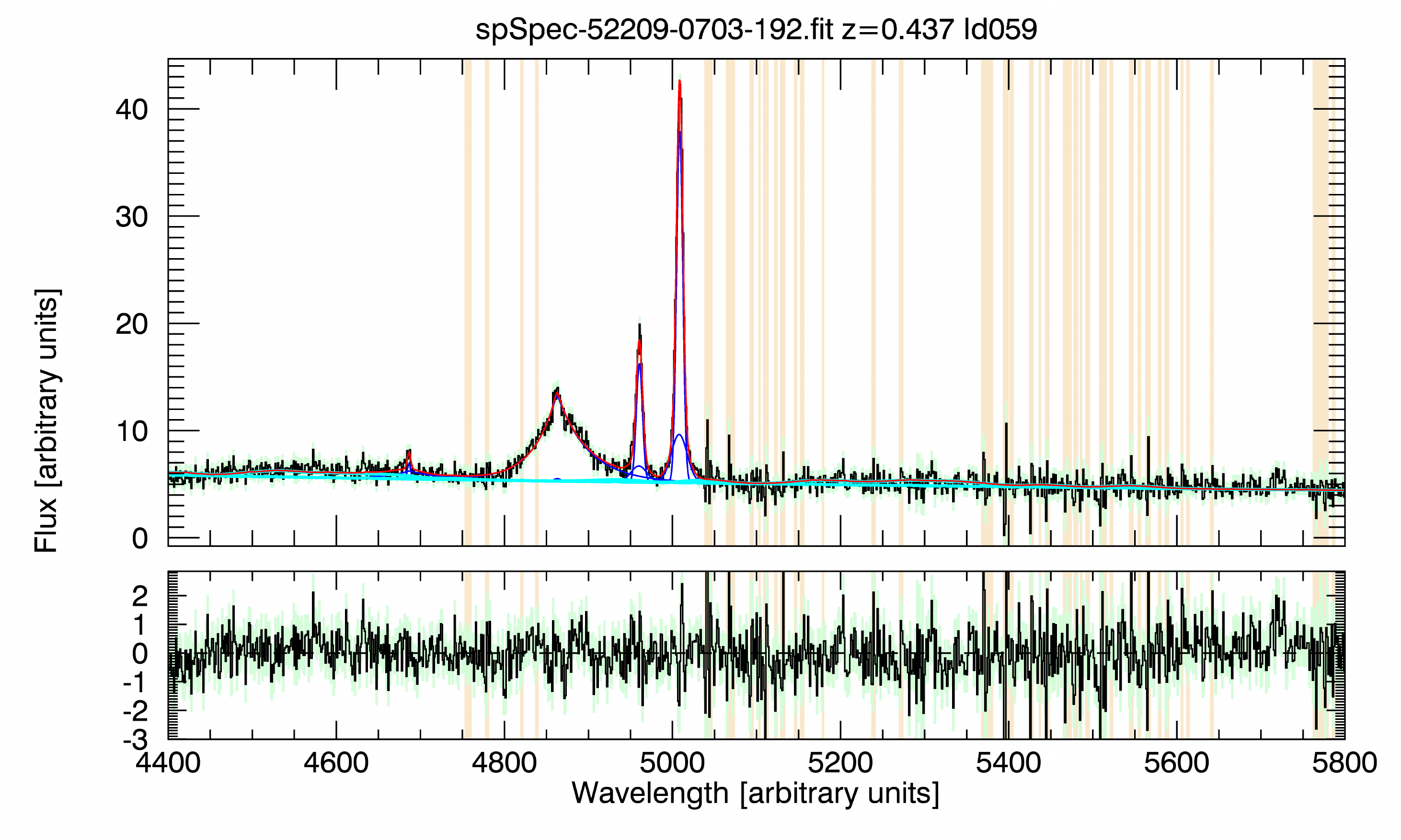}
\caption{ Top panel: Example of SED fitting for three quasars SDSS J015950.24+002340.8, SDSS J022430.60-000038.9, and  SDSS J021434.78-004242.6 of different
SED and spectra quality.
Points in yellow represent the
  luminosity of quasars (in solar luminosity units) from SDSS, 2MASS, WISE, PACS, and SPIRE 
  images plotted versus the rest-frame wavelength (from 0.1 to 1000
  \mc).  We plot in green the starburst template, in orange the torus and the hot black-body model, in brown the old star, and in blue
the accretion disk component. The results of the composition is the best
  fit, shown in red. The hatched area represents the uncertainties for the normalization of each model components obtained from the Monte Carlo analysis. 
  Bottom panel: example of the fit of the SDSS optical spectra for the same quasars presented above. 
In blue we plot the different components of the emission lines (narrow and broad) and in cyan  the Fe emission lines template.
In the bottom panel we show the residuals of the fit.}
\label{seds}}
\end{figure*}

\subsection{Ancillary data}

We build the spectral energy distribution (SED) of the 224 quasars by adding 
photometric points from $\sim$ 0.3 to 22 \mc band to the Herschel data, taking advantage of 
optical and infrared all-sky surveys (SDSS, 2MASS, WISE).

In the optical range, SDSS photometric system comprises five colour
bands (u, g , r, i, z) from about 0.3 to 0.9 \mc. 
We choose the quantity $psfMag$, 
for which the total flux is determined by fitting a PSF model to the object.

The Two Micron Sky Survey mapped the entire sky in the J (1.24 $\mu$m), H
(1.66 $\mu$m), and K (2.16 $\mu$m) near-infrared bands with a pixel size of
2.0\arcsec.  We considered sources in 2MASS from the All Sky Point Source Catalogue and the Reject Table within 2\arcsec\,
matching radius. If the QSO is not detected,
 we downloaded  the image of the field of
view around the quasars position from the 2MASS archive, and we used the IDL routine aper (adapted
from DAOPHOT) to estimate the limiting magnitude at the position of the quasar. In detail we derive the magnitude from a 4\arcsec\ radius region 
centred on the expected source position and 
we estimate the sky background in an annular region with inner radius of 14\arcsec and outer radius of 20\arcsec. 
To decide whether the source is detected  in the image, 
 we repeat the measurement in 1000 circular regions randomly selected around 100\arcsec\, from the source. 
 We then build the distribution of all 1000 flux values (mostly from empty regions of the sky) and if 
 the flux of the source is higher than the 95th percentile of this distribution, we consider this 
measurement  a detection, otherwise the 95th percentile as a 2$\sigma$ upper limit. 

The Wide-field Infrared Survey Explorer (WISE) is an infrared space
telescope that performed an all-sky astronomical survey in the W1
(3.4 $\mu$m), W2 (4.6 $\mu$m), W3 (12 $\mu$m), and W4 (22 $\mu$m) bands.  All  224
quasars in our sample are detected in at least one WISE band and are
present in the WISE All-Sky Source Catalog.  When the flux measurement has
a signal-to-noise ratio less than 2, the magnitudes are replaced with
the 2$\sigma$ brightness upper limit in magnitude units. 


To characterize the radio properties of our quasars, we follow the
same approach adopted by \citet{shen11} to include the radio
loudness parameter in the SDSS-DR7 catalogue quasars.  The standard
radio loudness parameter is defined as the ratio between the flux
density (f$_\nu$) at rest-frame 6 cm and 2500 \AA\, (R = {f$_{6
    cm}$/f$_{2500\AA\,}$, e.g.  \citealt{jiang07})\footnote{We match
  our sample with the VLA FIRST Survey at 20 cm with a searching
  radius of 30\arcsec\, and we derive the rest-frame 6 cm flux density
  from the FIRST integrated flux density at 20 cm assuming a power-law
  slope of $\alpha_\nu$ = $-$0.5. The rest-frame 2500 \AA\, flux
  density is determined from the fit to the continuum around 5100
  \AA\, fitted with a power-law shape (f$_\lambda$ =
  A$\lambda^\alpha$ ).}.  About 25\% of the quasars (i.e., 50/224) have been
  detected in the FIRST catalogue  and 34/224 have a
  radio loudness parameter greater than 10 ($\sim$ 15\% of our quasars
  are radio loud).


\subsection{SED fitting}

The infrared emission is dominated by the black-body emission of dust at
different temperatures, heated by the optical and UV emission of the
AGN and of young and old stars. A major difficulty in deriving
the star formation rate from infrared luminosity is to disentangle
contributions from different sources of heating. Typically the
dusty torus can reach dust sublimation temperatures of the order of
1700 K and the peak of the emission is at shorter wavelength (less
than 100 \mc) than that produced by warm dust around young
stars. In  star-forming galaxies there is a considerable overlap around 3-20 \mc where
the polycyclic aromatic hydrocarbon (PAH) emission may dominate the IR emission.  
However, this is not  problematic  because PAH features are almost absent in AGNs  since
molecules should be destroyed by the extreme optical-UV and soft X-ray radiation in AGN (\citealt{roche91}; \citealt{siebenmorgen04}). 
At shorter
wavelengths (around 3-4 \mc), the emission  is
dominated by older stars and luminosities are thus strongly correlated
with stellar mass \citep{meidt12}.  

In the literature there are many infrared templates for modelling
 galaxy infrared spectral energy distribution.  Some of these models
 are multi-dimensional, in the sense that there is a grid of
 parameters from which one can derive many physical characteristics of
 the system (e.g. \citealt{silva98}, \citealt{draine07}). Other are one-dimensional, i.e. they rely on a single parameter to characterize the
 shape of the model (e.g. \citealt{chary01} and \citealt{rieke09}). 
 
The measurements in low IR-luminosity quasars is of fundamental importance 
in this analysis since the sources that are not detected in {\rm Herschel} images 
are likely the best candidates to
reveal the quenching effect of powerful AGN winds on the star
formation.  Unfortunately, there are no public SED fitting codes that
can properly deal with upper limit photometric points. Therefore we developed our SED fitting code, in which
the best fit is the result of the sum of different components that describe the accretion disk emission, the unabsorbed stellar
population,  the emission of dust heated by the AGNs and by blue young stars.
Each component is the result of the combination of many templates that we specify below. The weight w$_j$ of each template
can be found by minimizing the  $\chi^2$ function
$$
\chi^2=\Sigma_i\,\Bigg(\frac{(F_i-\Sigma_jw_jF_{ij})}{\Delta F_i}\Bigg)^2,
$$
where F$_i$ and $\Delta$F$_i$ are the fluxes and error measured in each photometric band and F$_{ij}$ is the flux in band i for  template j.
 Since this is a linear bounded problem, we determine the weight of each template
using the idl routine BVLS.pro (bounded variable
least-squares) by Michele Cappellari\footnote{http://www-astro.physics.ox.ac.uk/~mxc/software/}. This routine 
solves the linear least-squares
problem
by finding the weights  of each template in order to minimize the $\chi^2$ function.
To estimate errors on fitting parameters, we adopt a Monte Carlo procedure. 
For each object we generate 1000 realizations of the observed SED with random extraction of photometric points 
normally distributed around the observed flux with sigma given by the errors. 
 The photometric upper limits are treated as null
detections with a positive error bar at 2$\sigma$ and we extract 
random values
uniformly distributed in this range.
For each realization of the observed SED we computed all the relevant physical quantities, for example the luminosity of each component (L$_{\rm AGN}$, L$_{\rm SB}$, etc.).
Finally, we estimate the errors 
as the percentile at 10\% and 90\% levels of the
distribution of all the luminosities. If the best model
does not require a starburst component, or the distribution is consistent with a null median value, we consider the 90th percentile of the distribution 
as a 2$\sigma$ upper limit.

For the SED fitting for each component we adopt the following templates:
\begin{itemize}
\item the infrared emission of the starburst component was modelled
  with a linear combination of the \citet{chary01} templates. These
  models provide synthesized SEDs of 105 average SED templates of
  local luminous and ultra-luminous purely star-forming infrared
  galaxies (LIRGs and ULIRGs) in a sequence of increasing infrared
  luminosity\footnote {We have verified that the choice of the
templates does not affect our results: adopting the starburst
templates by \citet{rieke09} or \citet{dale14}, the resultant SB and
AGN luminosity differs by less than a factor of 3.};
\item the AGN torus emission was modelled with a linear combination of
  DUSTY models by \citet{nenkova08}. To select a few likely independent
  models (from more than 10$^5$), we follow \citet{roseboom13} who
  demonstrate that it is possible to adequately fit their SED of luminous type
  1 quasars from the SDSS survey with only three models. These three models
  are for tori with inclinations i=0 and i=20 (see table 2 of their 
  paper for details);  to these we add the models at inclinations of  30, 35, 45, 50, 60, 70, and 75 degrees;
\item   the emission from three pure black bodies was modelled at temperatures 1100, 1300, and
  1500 K in order to account for the hot dust emission
  in the near-infrared not adequately reproduced by torus models (e.g. \citealt{leipski14});
\item the stellar emission from an unobscured population was modelled
  with stellar templates from \citet{bruzual03} computed with
  a \citet{chabrier03} IMF and metallicities $Z=0.008, 0.02, 0.05$; we
  considered models with ages $5.1\times 10^8, 2.0\times 10^9,
  5.0\times 10^9, 1.0\times 10^{10}, 1.3\times 10^{10}$ years;
\item the accretion disk emission was modelled with a linear combination of the four model
  spectra from figure 1 of \citet{slone12}
  which are representative of a physical conditions of quasar.
\item for the radio loud quasars it is possible that part of the
  infrared emission is dominated by synchrotron processes originating
  in the radio jets. However, the power-law slope of the
  sub-mm/far-infrared continuum is difficult to constrain in the infrared
  band. For this reason we prefer not to add a power-law component to the fitting templates and 
  highlight the radio loud sources in all the plots, remembering that in some of these 
quasars the star formation rate may be overestimated.

\end{itemize}
 In
Fig. \ref{seds}, top panel, we present examples of SED
fitting.

We estimate the star formation rate from the starburst model in the rest frame 8--1000 \mc\,
adopting a Chabrier initial mass function
(Chabrier 2003). In detail we use
the relation
$$
\left(\frac{\rm SFR}{\rm M_\odot yr^{-1}}\right)=0.94\times 3.88\times 10^{-44}\left(\frac{L_{\rm IR}}{\rm erg\, s^{-1}}\right)
$$
from \citet{murphy11}. The factor 0.94 accounts for a slight SFR underestimation
derived using the \cite{chabrier03} IMF with respect to the \citet{kroupa01} IMF 
(see \citealt{bolzonella10}; \citealt{pozzetti10};  \citealt{hainline11}). The stellar masses are derived 
from the normalization of the  \citet{bruzual03}  templates, since these models 
are normalized to a total mass of one $M_{\odot}$ in stars.
In the Appendix we test the accuracy of our SED fitting code, in particular with respect to the problem of model degeneracy.

\subsection{Fit of the optical spectra}

\begin{figure}
\includegraphics[width=9cm,angle=0]{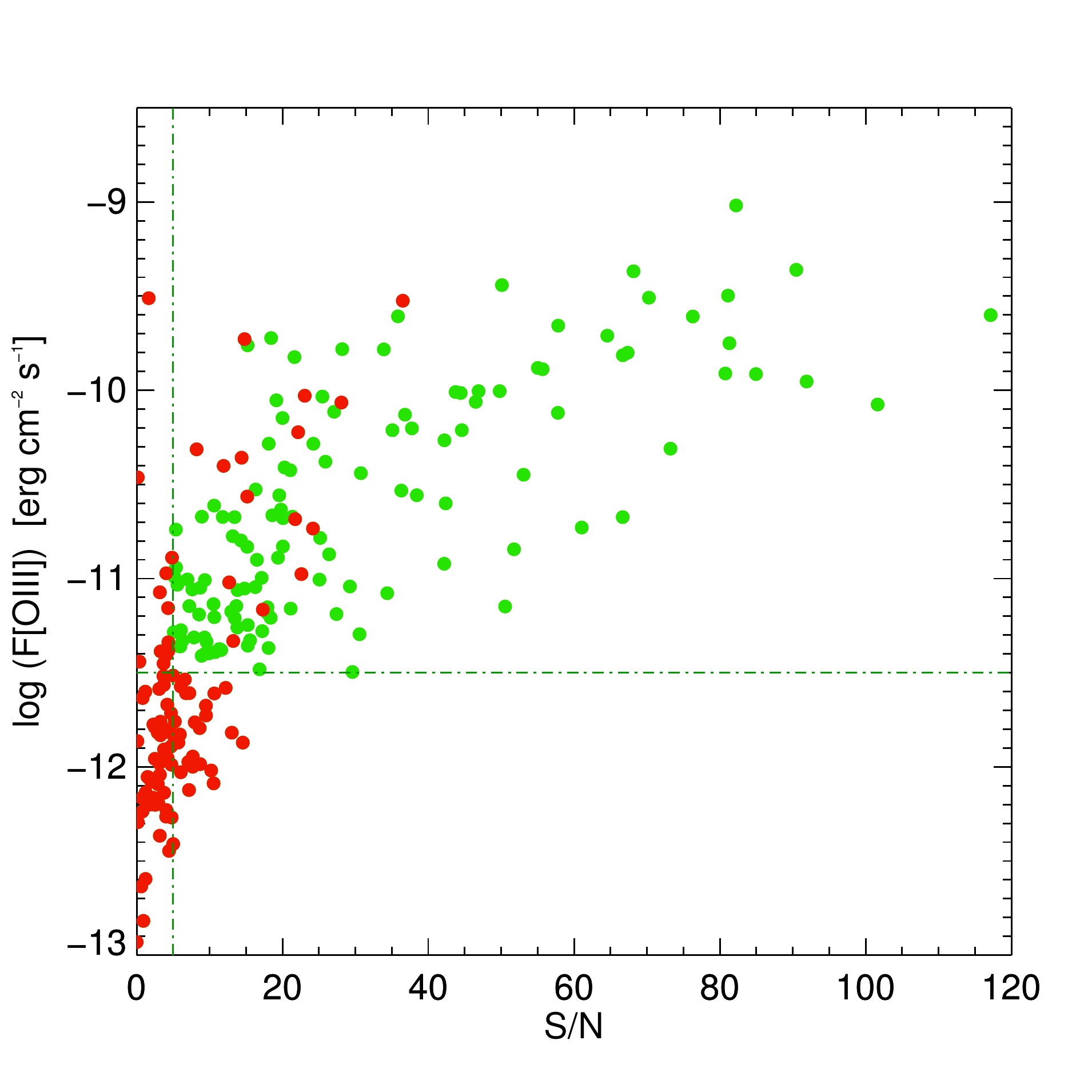}
\caption{Logarithmic flux (in \ergscm) versus the signal-to-noise ratio of the \oiii emission line. We adopt the dotted line as the threshold in flux and S/N
to select only spectra with a visible \oiii emission line and reliable fit (green points). Red points represent the quasars for which we consider the kinematic parameters based
on unreliable \oiii lines  and we do not consider these QSO in the following analysis.}
\label{selection}
\end{figure}

\begin{figure}
\centering{
\includegraphics[width=7.5cm,angle=0]{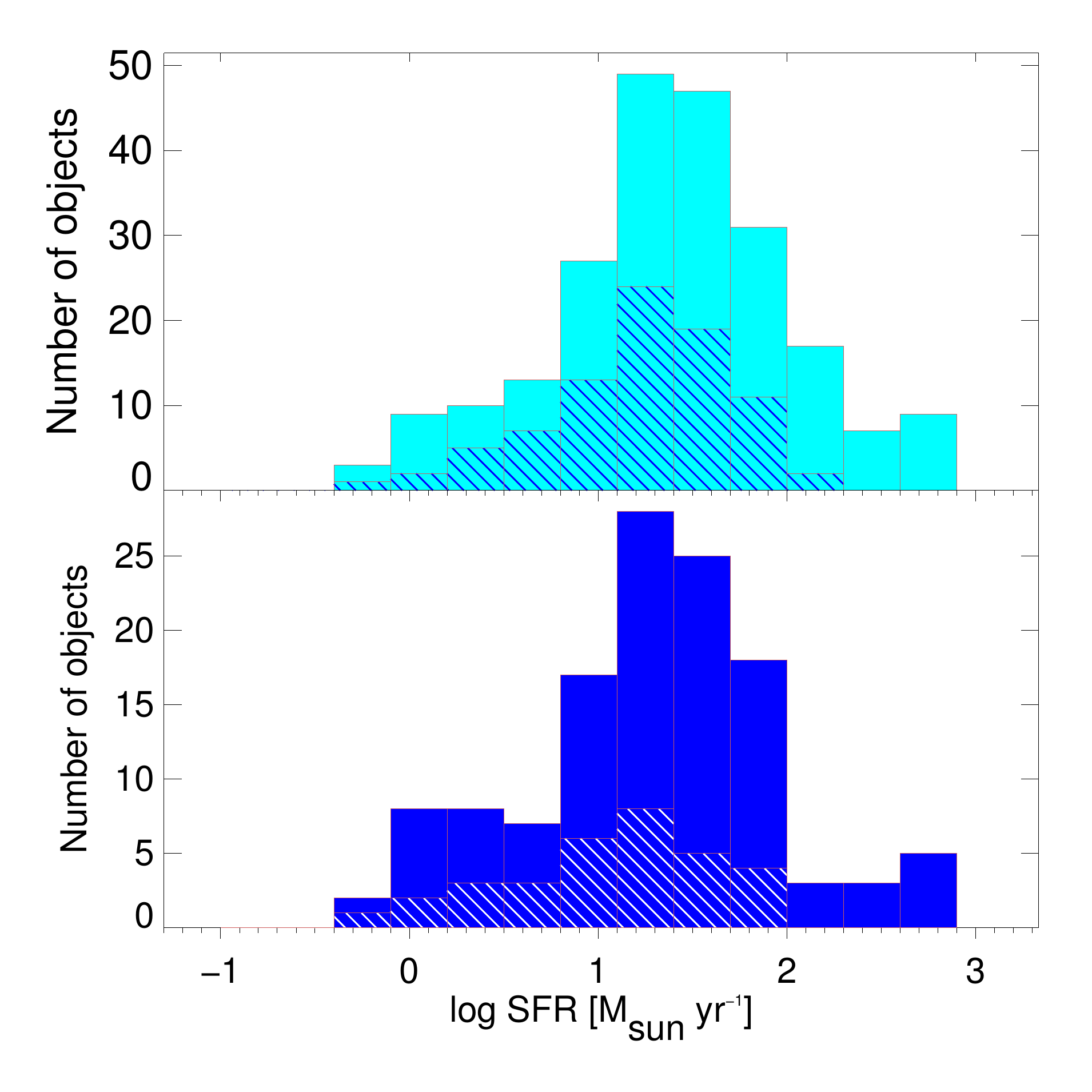}
\caption{Histogram of the star formation rate distribution for all the quasars (top panel) and only for quasars with
reliable \oiii measurements. The hatched area represent quasars with upper limit values on SFRs.}
\label{histo_sfr}}
\end{figure}

\begin{figure*}
\centering{
\includegraphics[width=7.5cm,angle=0]{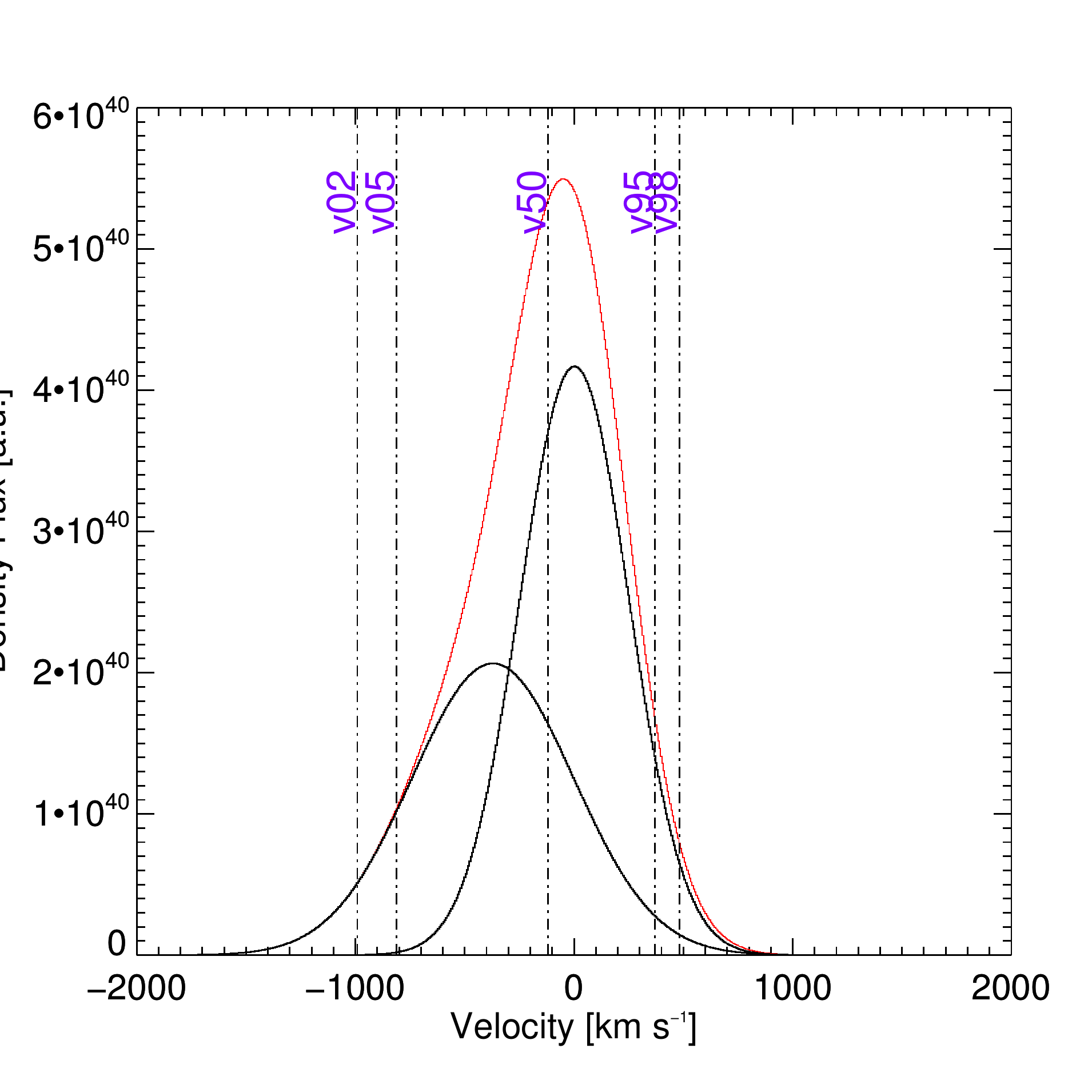}
\includegraphics[width=7.5cm,angle=0]{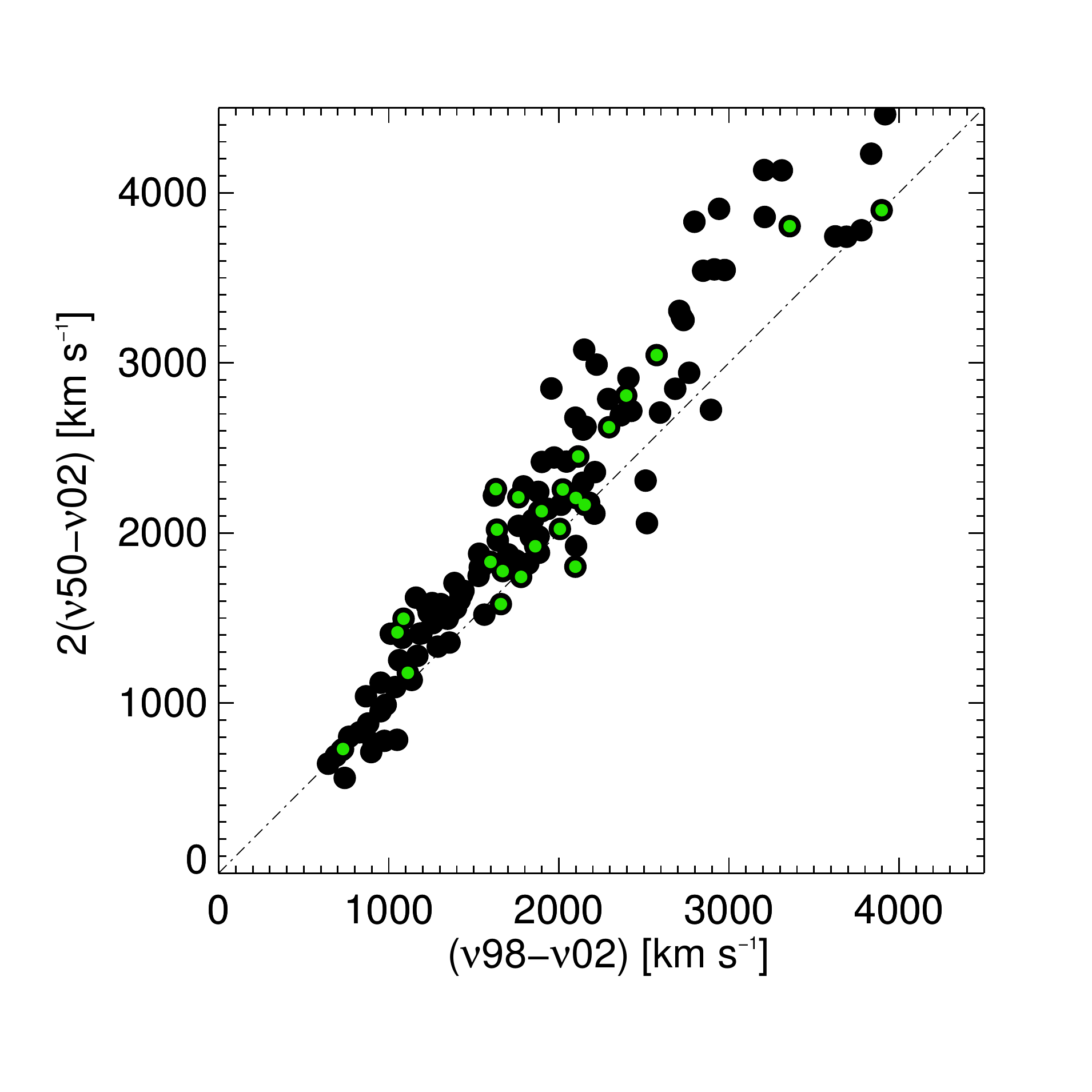}
\caption{Left panel: Example of [OIII]$\lambda$5007 line decomposition
  in a narrow and broad component and of different velocity definition
  used to characterize the properties of the outflow.  In particular
  we show velocities at different percentiles of the flux contained in the overall
  emission line profile (from left to right: 2nd; 5th, 50th, 95th, and
  98th). Right panel:  two
  times the blue tail, defined as ($v$50-$v$02) versus maximum velocity of the \oiii line ($v$98-$v$02). The dashed line is the
  bisectrix of the plane and represents the condition for symmetry.
 Almost all of the objects show a broad blue
  tail.   Green points are for radio loud quasars.}
\label{blue}}
\end{figure*}

\begin{figure*}
\centering{
\includegraphics[width=7.5cm,angle=0]{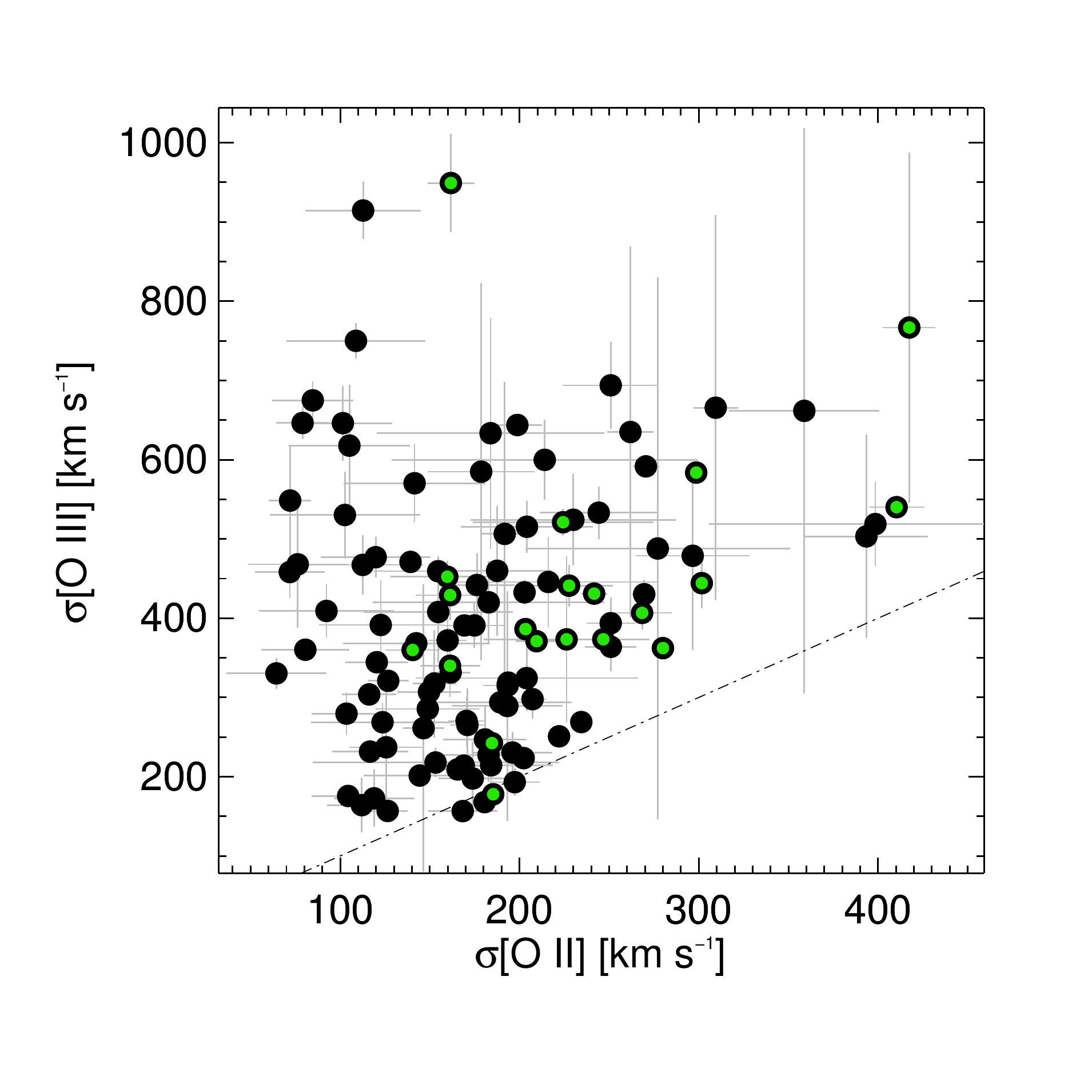}
\includegraphics[width=7.5cm,angle=0]{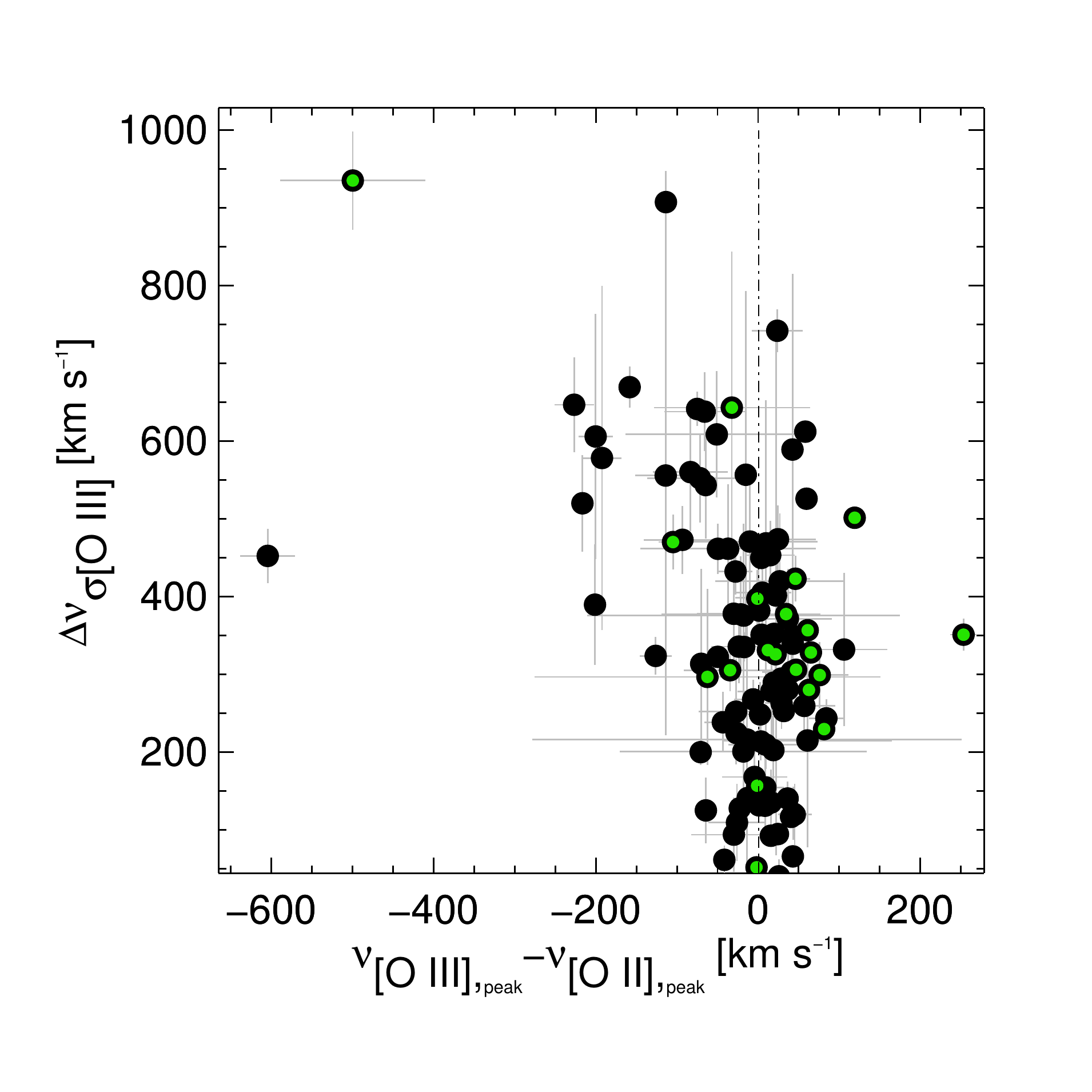}
\caption{Comparison of the [OIII] and [OII] principal emission line
  parameters.  Left panel:  Sigma of the \oiii line versus the sigma of the [O~II]. The [OIII]
  lines show broader widths. Right panel: $\sigma$ excess (defined using the square root of
  the quadratic difference of the \oiii and \oii sigma, see text) versus the relative shift of the  \oiii peak respect to \oii peak
 (both referring to their rest frame position).}
\label{shift}}
\end{figure*}

In this section we describe the analysis of the spectra used to
reproduce the profile of the [O~III]$\lambda$4959,5007 and of the
[O~II]$\lambda$3727,3729 emission line doublet in order to characterize the
outflow properties. The reduced spectra data used in this work are
available through the SDSS Data Archive Server 10 (DAS). The
wavelength coverage is $\Delta\lambda\sim$3800-9200 \AA\, with a spectral
resolution R$\sim$1800 - 2200 ($\Delta\lambda\sim$3600-10500 \AA\, and R$\sim$1400-2600 for BOSS spectra).

We employed a $\chi^2$ minimization procedure using the IDL routine
 mpfit, \citet{markwardt09}.  Since the optical emission lines show complex profiles
 diverging from a simple Gaussian profile, we fit the principal emission
 lines (i.e. the [O~III] doublet, the narrow \Hb and eventually the
 He~II $\lambda$4686 with two Gaussians, one for a narrow component and one for the
 wings.  The intensity ratio is fixed at [OIII] $\lambda$4959/[O~III]$\lambda$5007 = 0.33 (e.g.  \citealt{dimitrijevic07}) and the
 width of the [O~III]$\lambda$4959 line was fixed to be the same as
 the [O~III]$\lambda$5007 line and of the narrow \Hb line. The
 velocity offset of the narrow \Hb line are tied to those of the core
 [O~III]$\lambda$4959,5007 components. The \oii emission line doublet in general is
 unresolved in our spectra. However, we fit the profile with two emission lines with the same $\sigma$ (each one made of two Gaussian components)
 with a flux ratio of the two components that vary between 1.3 and 1.5.
 
The continuum is fitted with a power-law and the
broad \Hb emission line with a broken power-law profile
\citep{nagao06}.  A major difficulty in the spectral analysis is that the [OIII] doublet overlaps
with the broad \Hb emission and optical Fe~II emission lines (see
Fig.\ref{seds}, bottom panel).  The Fe~II emission was modelled with a linear
combination of templates, which was then convolved with a broadening
Gaussian function. The free parameters characterizing the Fe~II
emission are the weights of the template, the central velocity, and
the velocity dispersion of the Gaussian function.  The adopted Fe~II
templates were the observed IZw1 spectrum (\citealt{veron04}).

 In many cases the  \oiii and \oii lines are too weak and  confused with the continuum and the other fitting components to extract reliable kinematic information. 
Therefore, we adopt thresholds in flux and  in S/N to discard useless spectra. 
We only consider  spectra with

\begin{itemize}
\item  S/N$>$5 for the  emission line flux;
\item log(F$_{\rm [O~III]}$) $>$ -11.5 \ergscm.
\end{itemize}

For the \oii line we lower the flux threshold to -12.5 erg cm$^{-2}$ s$^{-1}$, because this line is weaker and not blended with
other strong features. We  exclude an additional 15 sources with high S/N and high F$_{\rm [O~III]}$, but with the \oiii line blended with
 Fe lines or \Hb\ and with a flux that is negligible with respect to the other components.
 The presence of  optical Fe emission lines in SDSS quasar spectra has 
 been systematically investigated  by \citet{hu08}.
 They found that  Fe~II emission lines are typically redshifted with respect to the [O~III] peak, and they suggest
  that the optical Fe emission lines trace gas  in the broad line region dominated by infall.
In Fig. \ref{selection} we represent our selection criteria on \oiii plotting the flux
versus the S/N of the line. Red points represent the rejected quasars  and 
green points quasars with bona fide clear lines and  reliable fits to the profile (confirming by eye the quality of our
selection criteria). 
Adopting this approach, 
we define a high quality subsample 1, composed of 125 QSOs with trustworthy fitted profiles in  [O~III] (out of 224 QSOs), and 
a high quality subsample 2, made of 108  QSOs with good quality fits both in \oiii and [O~II].
These quasars are used in the plots involving \oii in addition to \oiii
(we have respectively 19\% and 18\% of radio loud quasars in the two subsamples). In the following analysis we extract the outflow properties (on a case-by-case basis using the \oiii, the \oii, or sometimes both the 
line profiles) considering only the quasars that satisfy our quality selection criterium. In Fig.\ref{histo_sfr} we plot the histogram of the
star formation rate in the complete sample and in our high quality spectra sample, highlighting the objects with star formation rate  upper limits.
The fraction and the distribution of SFR undetected objects is of great  interest, since in these objects the SF-quenching might be
ongoing. In the complete sample 38\% of the quasars have a star formation rate upper limit measurement. This fraction decreases to 26\% in the subsample of quasars with reliable [O~IIII] measurements. 
We note that although the sensibility of our observations 
is almost heterogeneous, the distribution of the IR luminosity for detected and undetected quasars is similar and the IR luminosity upper limits are not located in the tail
of the distribution of detected quasars.

Owing to the uncertainties in the decomposition of the line into a
narrow and a broad component, we follow the approach of other authors
(e.g. \citealt{harrison14}, \citealt{perna15}) and we adopt a non-parametric definition
for the velocity characterizing the outflow.  We reconstructed a
synthetic line profile adding two  Gaussian
lines and we define different percentiles of the flux contained in the overall
emission profile. In particular we measure the velocity at the
2nd, 5th, 50th, 95th, and 98th percentile, containing respectively 2,
5, 50, 95, and 98\% of the overall line flux ($v$02, $v$05, $v$50, etc.). The standard deviation
($\sigma$) of the line profile is the square root of the second moment
about the mean.  In Fig. \ref{blue}, left panel, we illustrate these
definitions using an example of a [O III]$\lambda$5007 line
profile.

In the following  analysis, we select
the
following three parameters to describe the
outflow properties and to explore the connection with the SFR:

\begin{itemize}

\item  $\nu_{\rm blue}$, the maximum velocity of the blue wing, defined as |$\nu_{\rm 10}-\nu_{\rm 50}$|, linked
in some complex way to the energetic of the outflow.  In principle,
one would infer the mass outflow rate, but this measurement is highly
uncertain and requires the knowledge of the location, mass density,
geometry, and velocity of the gas outflowing.  Assuming simple
geometric arguments, the mass outflow rate is
proportional to the maximum velocity of the outflowing wind ($\sim$ 3vM/R, e.g. \citealt{maiolino12}).

\item $\Delta\nu_{\rm offset}$, the velocity offset of the
broad wings defined as $\Delta\nu_{\rm offset}=|\nu_{\rm 05}+\nu_{\rm 95}|/2-\nu_{\rm 50}$. 
For two well-separated Gaussian components, $\Delta\nu_{\rm offset}$ 
is the velocity offset of the broad component with respect to $\nu_{\rm 50}$ (\citealt{harrison14}). As usual, in the
literature we interpret the broad, blueshifted emission line with respect
to the velocity of the  \oiii narrow component as the result of an outflowing
gas moving at larger velocity with respect to the  quieter large-scale
gas.

\item \dsig =(\sigoiii$^2$-\sigoii$^2$)$^{0.5}$, the sigma excess, defined as the square root
  of the quadratic difference between the standard deviation of the
  \oiii and \oii\,lines. This
  parameter estimates the different level of the perturbation of the
  gas in the inner and outer portion of the NLR since, as we discuss in Section \ref{out},
  the bulk of the \oiii emission should be produced in a  more inner region than \oii.

\end{itemize}

\section{Results}

\subsection{Comparison of the \oiii and \oii\, line profile}
\label{out}

In the literature a blueshifted line seen in absorption is considered a clear signature of outflow. Instead,
the interpretation of blueshifted wings seen in emission is ambiguous
because different scenarios can account for the asymmetry of the line: outflows, inflows, or gas in a rotating disk partially
obscured by an asymmetric distribution of dust. Moreover, if the extinction is within the emitting clouds,
 then outflowing gas would produce a line profile with a red wing \citep{whittle85, derobertis90, veilleux91}.
However, in recent papers a blueshifted wing is considered  a  clear signature
of ionized gas moving towards our line of sight (e.g. \citealt{bian05}, \citealt{komossa08}, \citealt{brusa15}, \citealt{villar11}). 
Here we follow the latter interpretation, aware  that all our results depends on this assumption. 

In Fig.\ref{blue},
right panel, we emphasize the asymmetry of the \oiii emission lines plotting  $\nu_{\rm max}$=$\nu_{\rm 98}$--$\nu_{\rm 02}$ versus two times the blue tail (defined as
 $\nu_{\rm 50}$--$\nu_{\rm 02}$). The deviation from the bisectrix of the plane represents the
amount of asymmetry with respect to $\nu_{\rm 50}$.
 About 80\% of the quasars show asymmetric \oiii emission lines with
significant blue wings. The median deviation from symmetry is 211 \kms
but it increases up to 1033 \kms at higher $\nu_{\rm max}$. 

In various models of
line formation in the NLR, the larger widths of the ionized lines is
explained by their formation in clouds that have higher density or
higher orbital or radial velocity, likely because they are closest to
the central ionization region (\citealt{derobertis84})\footnote{ The critical density and the
ionization potential for [OIII] is 7\,10$^5$ cm$^{-3}$ and $\chi_{ev}$
=35.1, higher than the critical density for the [OII] doublet
(1.4-3.3\,10$^3$ cm$^{-3}$ eV) and the ionization potential for \oii
($\chi_{ev}$ =13.6 eV), \citet{nagao01}.}. The larger width of the \oiii with respect to \oii is consistent
with the fact that lines  of higher ionization potential and/or
higher critical densities are produced in inner regions closer to the
nucleus. If the \oii\, emitting gas likely traces the unperturbed ISM,
 the excess of the standard deviation of the \oiii line
with respect to \oii\, is an estimate of the amount of perturbation of the
gas in inner regions.

In Fig.\ref{shift}, left panel, we plot the sigma values derived from the analysis of  the  \oiii and \oii emission line profile.
The \sigoiii values span a 
range in velocity between 165 and 950 \kms, with a mean value of 405 $\pm$ 165\kms.
For comparison, \citet{mullaney13} in a sample of type 1 quasars found a value of
the order of 195 \kms (we derived this value combining their average two-Gaussian component with FWHM of 335 and 851 \kms). The value of 
\sigoiii, except for a few cases, is always larger than \sigoii. For the \oii\, emission line
we find \sigoii=185 $\pm$ 70 \kms, a value comparable with the stellar velocity dispersion, which  in quasars is of the order of 200-300 \kms
(depending on the black hole masses).

In order to characterize the gas outflow velocity with respect to the
unperturbed gas at galactic scales it is essential to determine an
accurate velocity reference. In our quasar sample the stellar absorption-line
features cannot be detected, because the continuum is dominated by the AGN.  The velocity of the
core of the \oiii line has been used to estimate the systemic
redshift of the galaxy (\citealt{villar-Martin11}) and the width of the
line has been considered a reasonable substitute for the stellar
velocity dispersion of the galaxy bulge (after properly removing the
asymmetric blue wings of the line, \citealt{greene05}).
Assuming  that the outer portion of the NLR gas follows
pure gravitational motions, we test whether the velocity of the \oii
line can be used as a reference velocity for large scale gas.
 
In Fig.\ref{shift}, right panel,  on the x-axis we show the shift in velocity
 of the \oiii peak flux density with respect to the \oii peak both referring to the rest frame wavelength . We measure a mean shift of the two peaks of  $<V_{shift}>$=-20$\pm$100\kms.
 Unfortunately,  the [OII] doublet is unresolved in the SDSS spectra. The
 ratio of the two lines depends on the electron density of the gas and it
 is between 0.35 and 1.5 \citep{pradhan05}. This introduces an
 uncertainty in our [OII] velocity reference of 2\AA, corresponding 
 to $\sim$ 80 \kms.  For most of the quasars we find relative shifts of the two peaks in this range. 
 However, a few quasars ($\sim$ 20 \% of the total) show velocity shifts larger 
 than 80 \kms. For comparison, on the y-axis we show  \dsig. The quasars that show larger \oiii shift with respect to \oii
 also have  larger \dsig\ values. We believe that this approach is promising; unfortunately,
 the \oii doublet is unresolved in our spectra and so  we prefer not to use the peak of this line as a reference frame and we choose other parameters
 to characterize the outflow properties. 
 
  \subsection{SB and AGN dominated quasars}

\begin{figure}
\centering{ 
\includegraphics[width=7.5cm,angle=0]{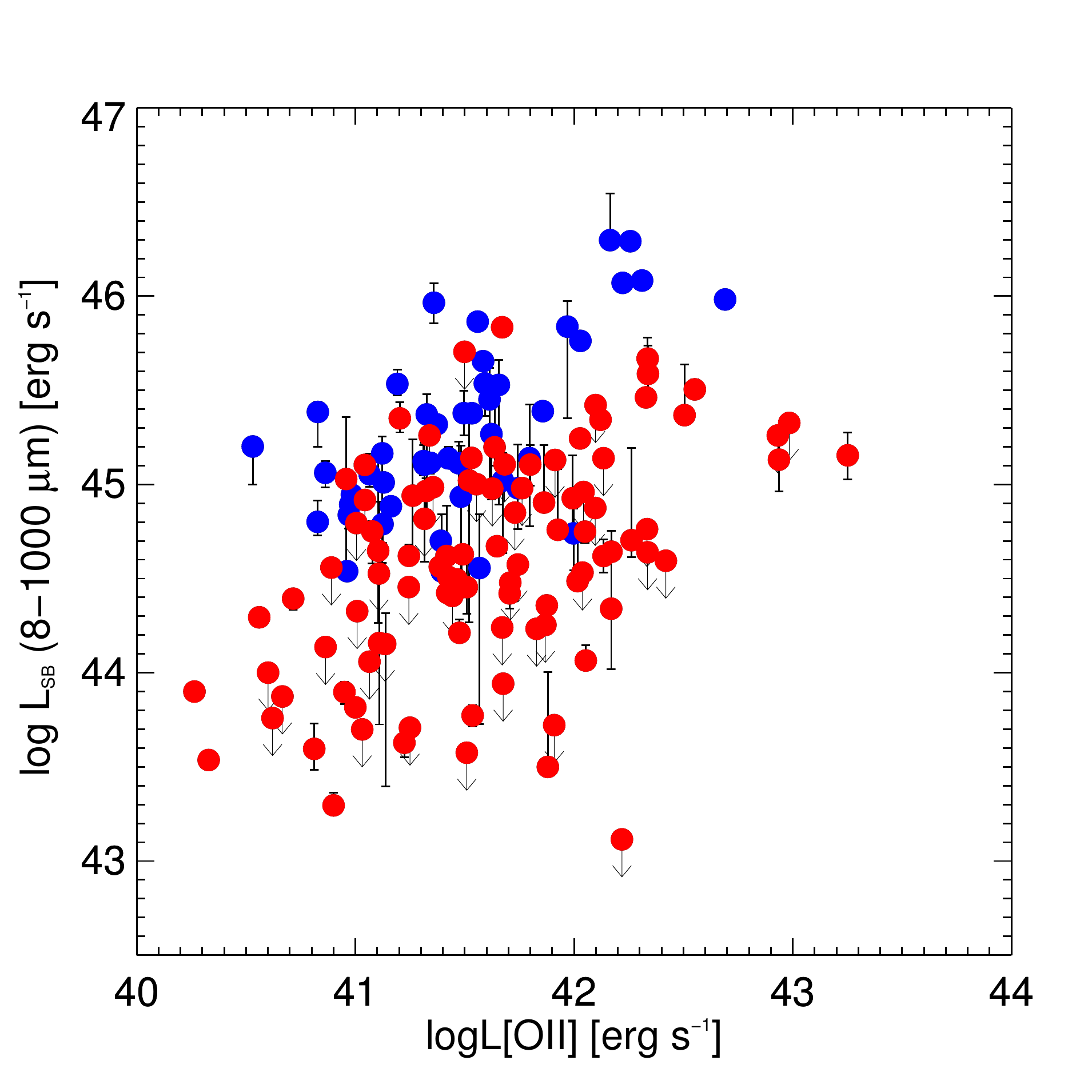}
  \includegraphics[width=7.5cm,angle=0]{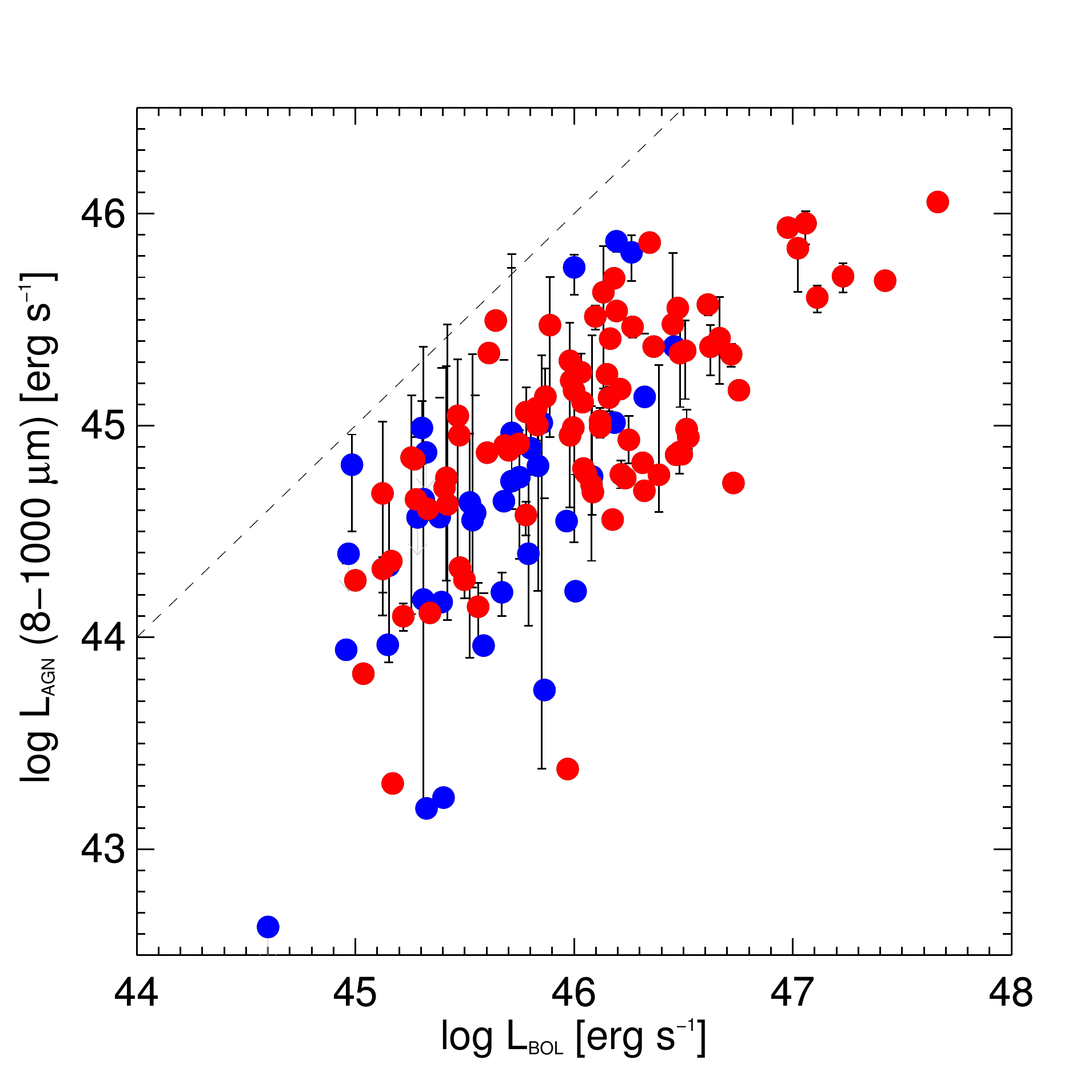}
\caption{ Upper panel: Logarithm of the infrared luminosity in the 8-1000 \mc\,
  range for the starburst component derived from the SED fitting
  versus the logarithm of the [OII] luminosity, another 
  estimator for the SFR.  Bottom panel: Logarithm of the
  infrared luminosity in the 8-1000 \mc range for the hot dust
  component derived from the SED model (likely produced by a dusty
  torus warm up by the AGN) versus the logarithm of the AGN bolometric luminosity.  Blue and red
  points represent quasars  dominated in the infrared by
  star formation or by the AGN emission, respectively.}
\label{sb_estimators}
}  
\end{figure}

\begin{figure}
\centering{
\includegraphics[width=7.5cm,angle=0]{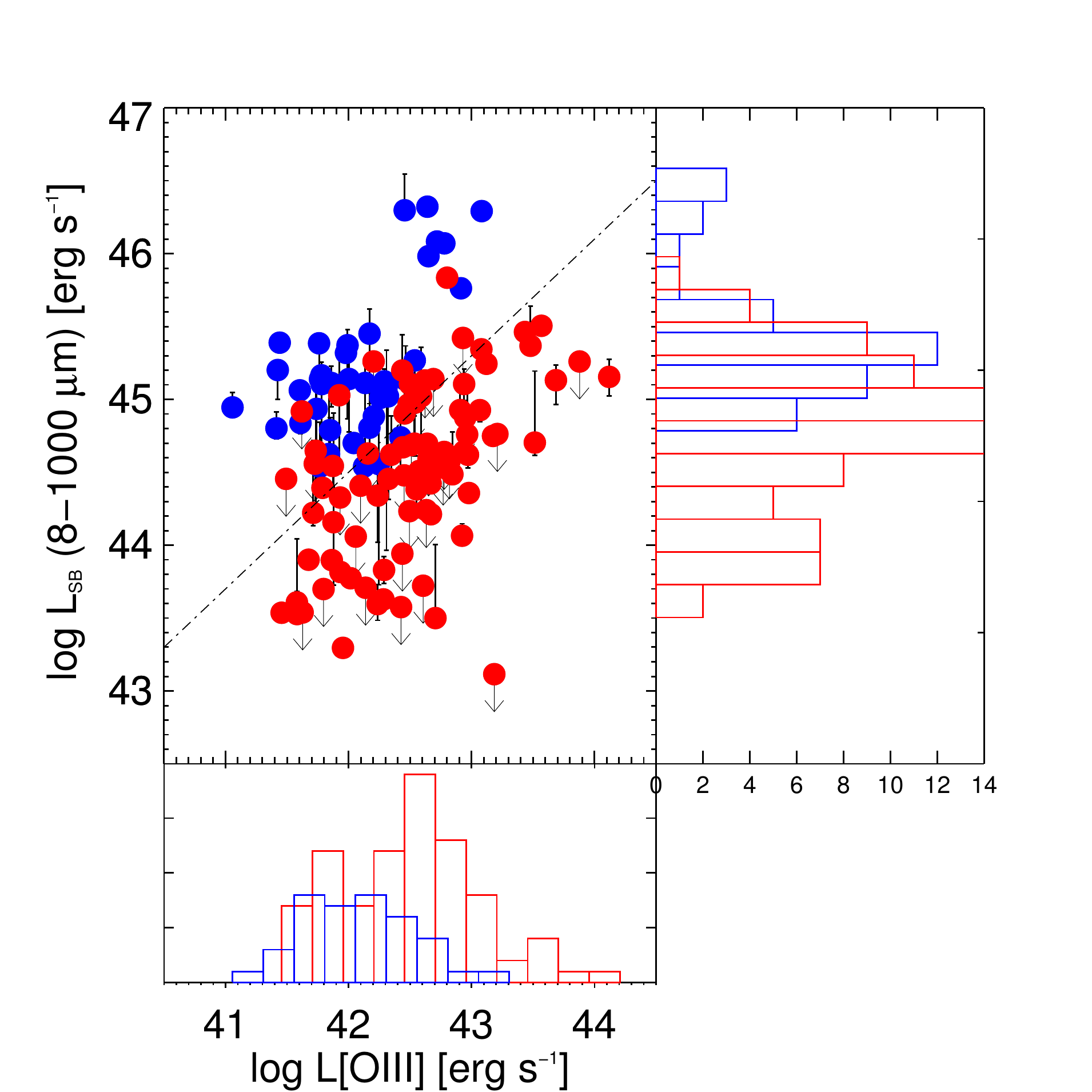}
\caption{Logarithm of the star formation rate versus the logarithm of
  the [OIII] luminosity. The  dashed line shows the correlation by
  \citet{netzer09} of slope 0.8. The normalization is from
  \citet{shao10} dividing the AGN bolometric factor by a factor of
  3500.  The  histograms at the right and  bottom represent the
  distribution of [OIII] luminosity and of star formation rates.}
\label{netzer}
}
\end{figure}

The total infrared luminosity between 8-1000 ${\rm \mu}$m is the sum of different contributions: the torus and dust heated by newly formed stars.
We divide our sample into two groups according to the dominant processes that produce the infrared-luminosity.
The first group is composed of starburst dominated objects
for which the starburst-related luminosity dominates the total
infrared (8-1000 ${\rm \mu}$m) luminosity ($L_{\rm SB}/L_{\rm tot}>$0.5) and the
second group is composed of AGN component dominated objects
($L_{\rm SB}/L_{\rm tot}<$0.5) (or quasars with upper limits in SFR). 
In this section we explore the properties of these two groups (hereafter SFd-QSO and AGNd-QSO).

We measured the star formation rate from SED fitting in the
infrared band; however,  the [O~II] emission line, corrected for
dust extinction, is another widely used estimator of the SFR. Deriving the SFR from the \oii 
luminosity is quite a challenging task because 
in type 1 quasars the \oii luminosity is contaminated by the emission 
produced in the NLR. Moreover, the conversion from luminosity to SFR
has a 
complex dependence on metallicity (\citealt{kewley04}).  In
Fig.\ref{sb_estimators}, upper panel, we compare the \oii versus the infrared luminosity in the 8-1000 \mc
derived from SED fitting.  The \oii luminosity is not corrected for dust absorption because only about half of our spectra (z
$\lessapprox$ 0.4) cover the Balmer decrement (\Ha/\Hb) typically used
to make the correction.
As expected, the dispersion is larger if we
consider all the quasars, but it is significantly reduced if we
consider only the SFd-quasars.  The red points (AGNd-quasars, 102 objects)
may have \loii\, significantly contaminated by the AGN, and in fact
they show an excess of \oii luminosity with respect to the blue points (SFd-dominated quasars, 46 objects).

In the bottom panel of Fig.\ref{sb_estimators} we compare the
8-1000 \mc\, luminosity for the AGN component versus the AGN bolometric luminosity (derived from \loiii\, as $L_{\rm BOL}=3500\times$\loiii, \citealt{heckman04}).
The ratio between these two quantities provides an estimate of 
the torus covering factor. We find a mean logarithmic value of $<$-1$\pm$ 0.5, consistent with the estimate of -0.4$\pm$ 0.2 of \citet{roseboom13} measured
in a sample of type 1 quasars (our \oiii luminosity measurements are not corrected for dust absorption and therefore we consider them as lower limit values).
In the AGNd-quasars sample (86 objects), the \loiii\, correlates with the 
infrared emission derived from torus models
(we find a linear correlation coefficient of $r$=0.6). Instead we do not find a relation in the SFd-quasars sample (39 objects, $r$=0.2).

 In  Fig.\ref{netzer} we investigate the relation between the AGN \oiii luminosity and the star
 formation rate of the host galaxy of the two groups. 
 There is a well-known close correlation between the AGN
 bolometric luminosity and the star formation rate of galaxies  (e.g. \cite{netzer09}). This
 relation is usually interpreted as a probe of the coupling between
 AGN growth and the star formation by an evolutionary mechanism,
 e.g. the merging.  Our quasars recover the correlation found by
 \citet{netzer09} $L_{AGN}\propto L_{SFR}^{0.8}$ that links the AGN
 luminosity to the SFR. 


\subsection{SB and outflow properties}

We now consider  possible relations between the SFR and the kinematic parameter characterizing the outflowing gas and 
we  discuss these results in a general context in Section \ref{discussion}.
We focus first on the SFd-QSO sample (Fig.\ref{all}, top panel).  We find an upward trend between the logarithm of the SFR
and  the three kinematic parameters considered, but no  significant correlation.
For the AGNd-QSO,  
we do not find a trend between the outflow properties and the SFR. In Fig.\ref{all}, bottom panel, 
we consider the SFR versus the three same kinematic parameters considered in the top panel.  
Even
separating the radio loud and radio quiet quasars we did not find a relation between the SFR and kinematics of the ionized gas,
as expected if the dominant source powering the outflow are supernovae winds or the AGN.
We then searched  for a connection between the kinematics of the ionized gas and the properties of the AGN (e.g. \loiii, $L_{\rm tot\, (8-1000\mu m)}$, $L_{\rm Edd}$), but again we do not find any linear relations.
 
\begin{figure*}
\centering{
\includegraphics[scale=0.3]{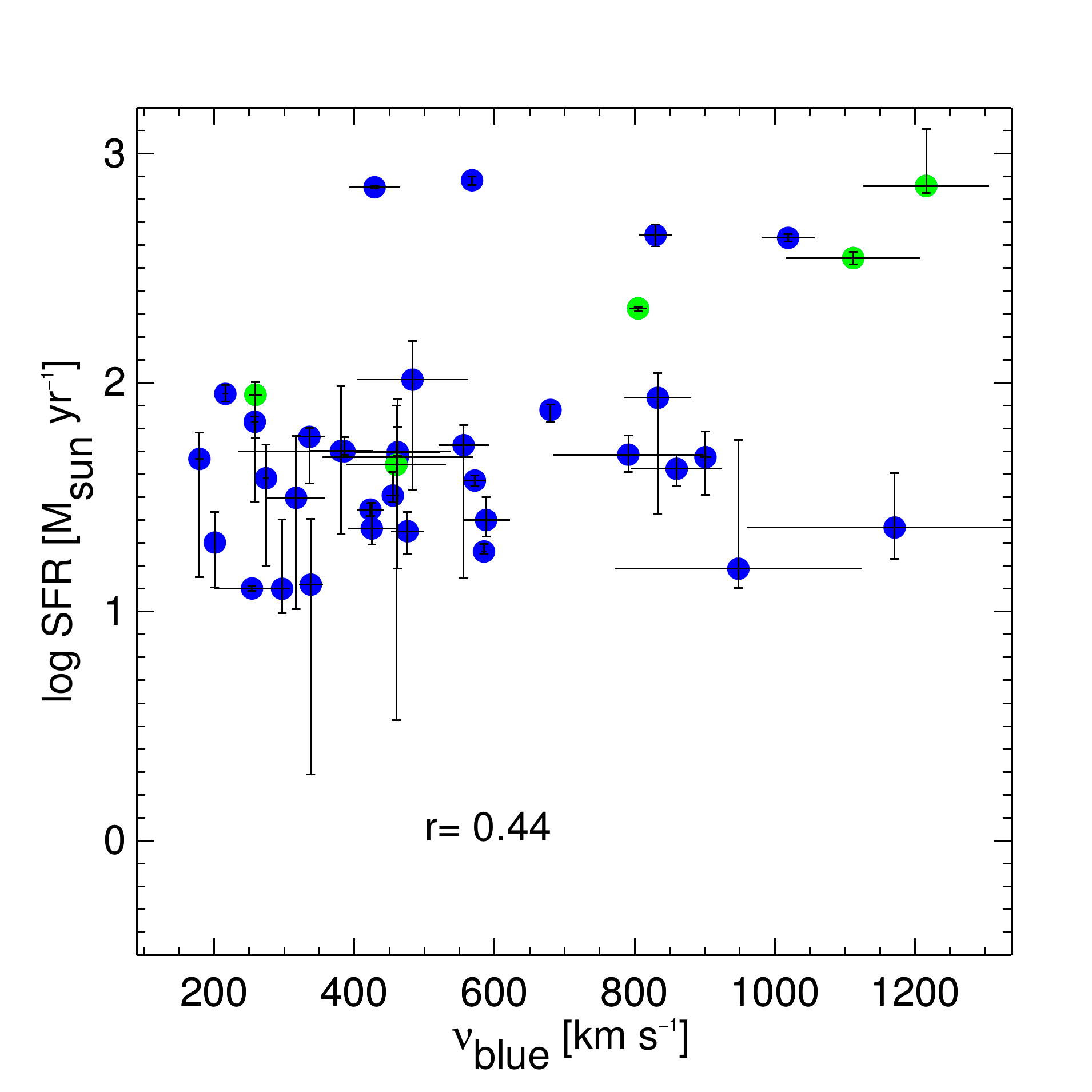}
\includegraphics[scale=0.3]{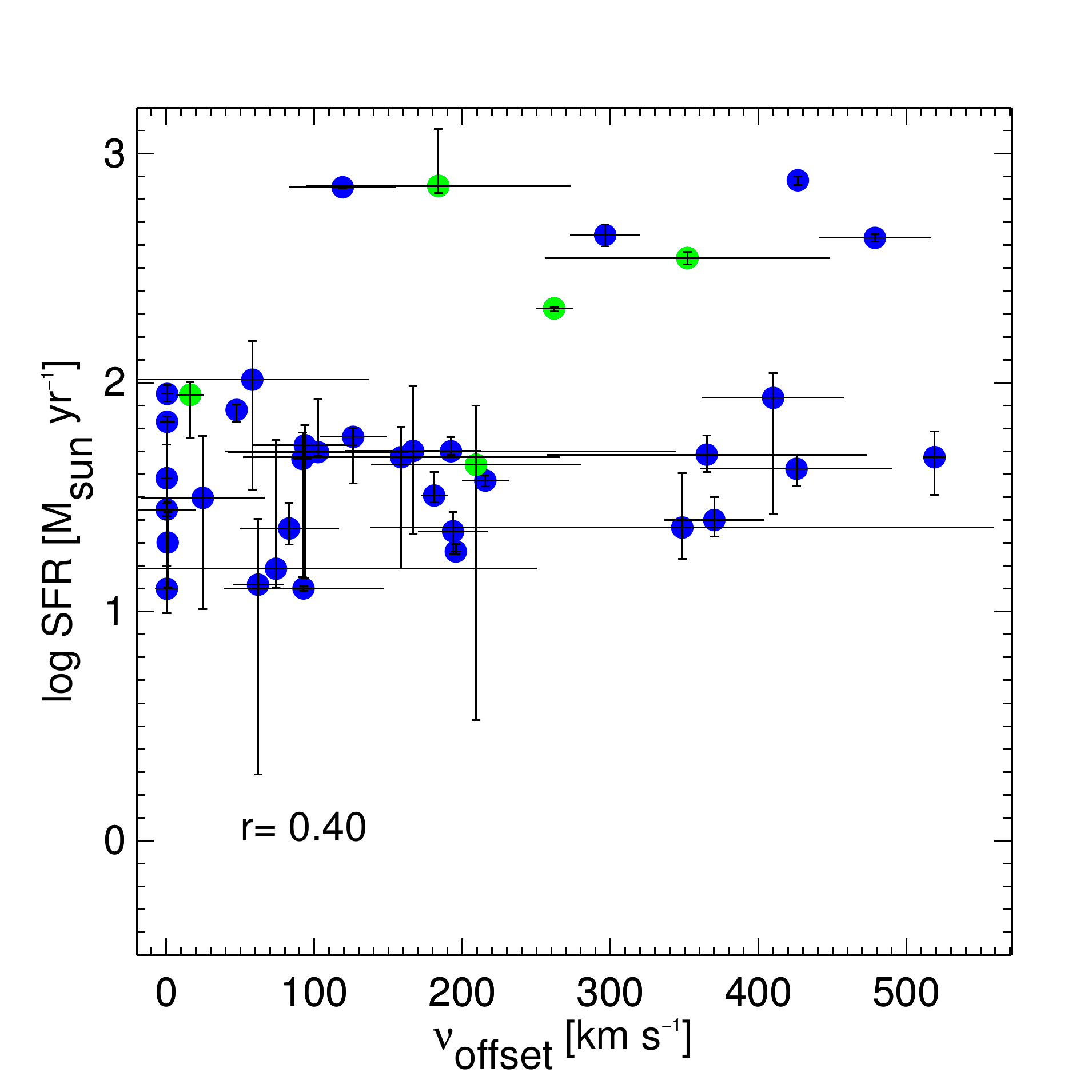}
\includegraphics[scale=0.3]{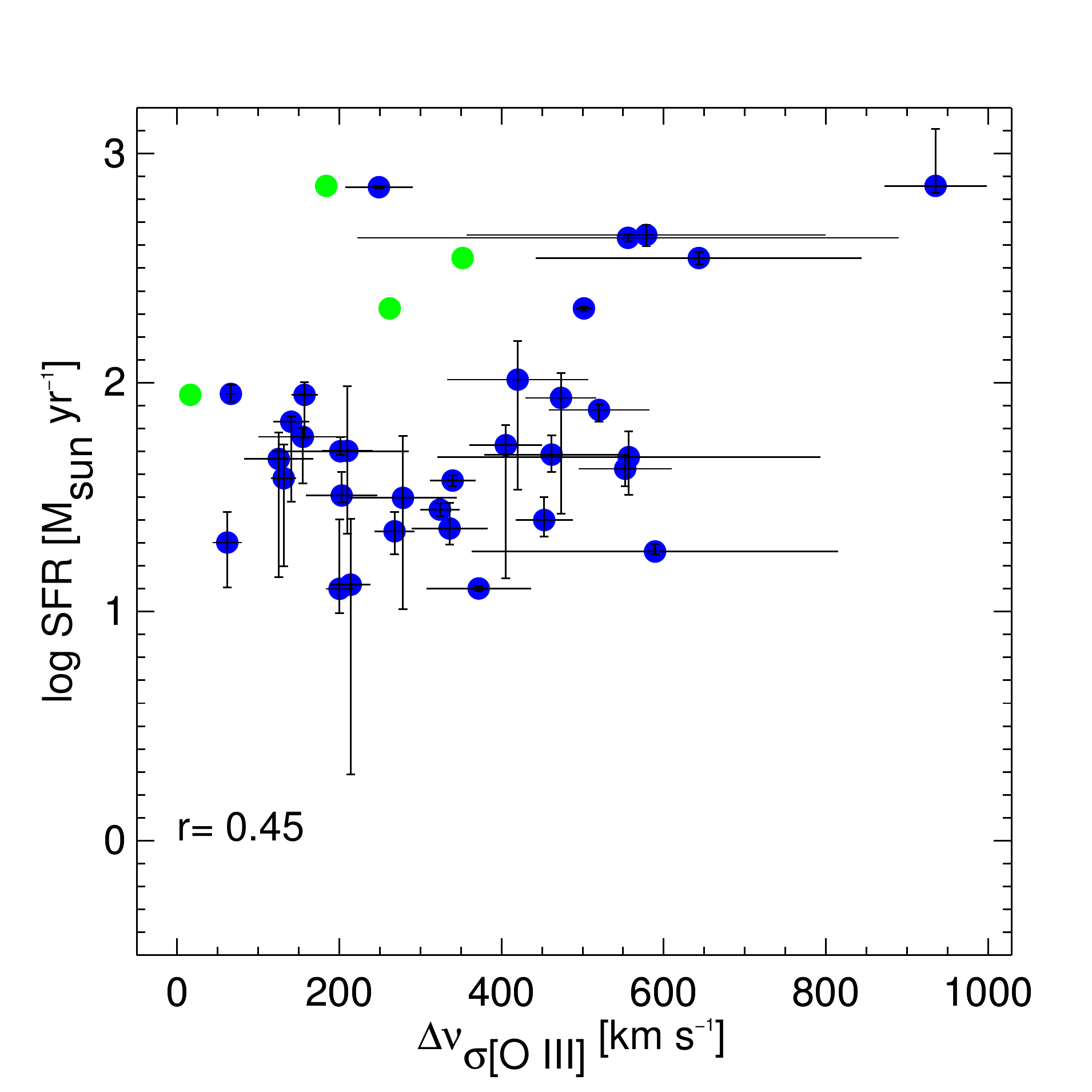}
\includegraphics[scale=0.3]{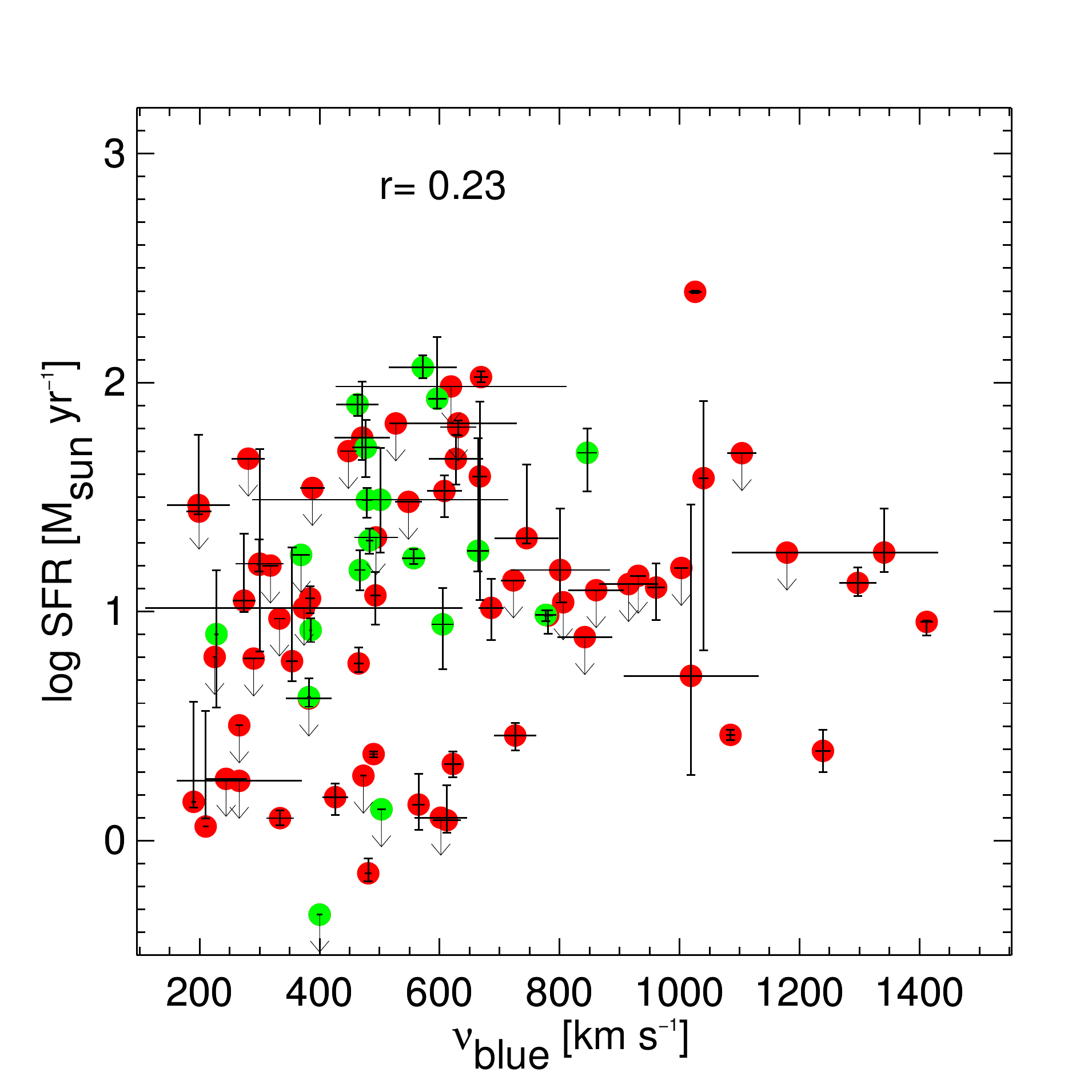}
\includegraphics[scale=0.3]{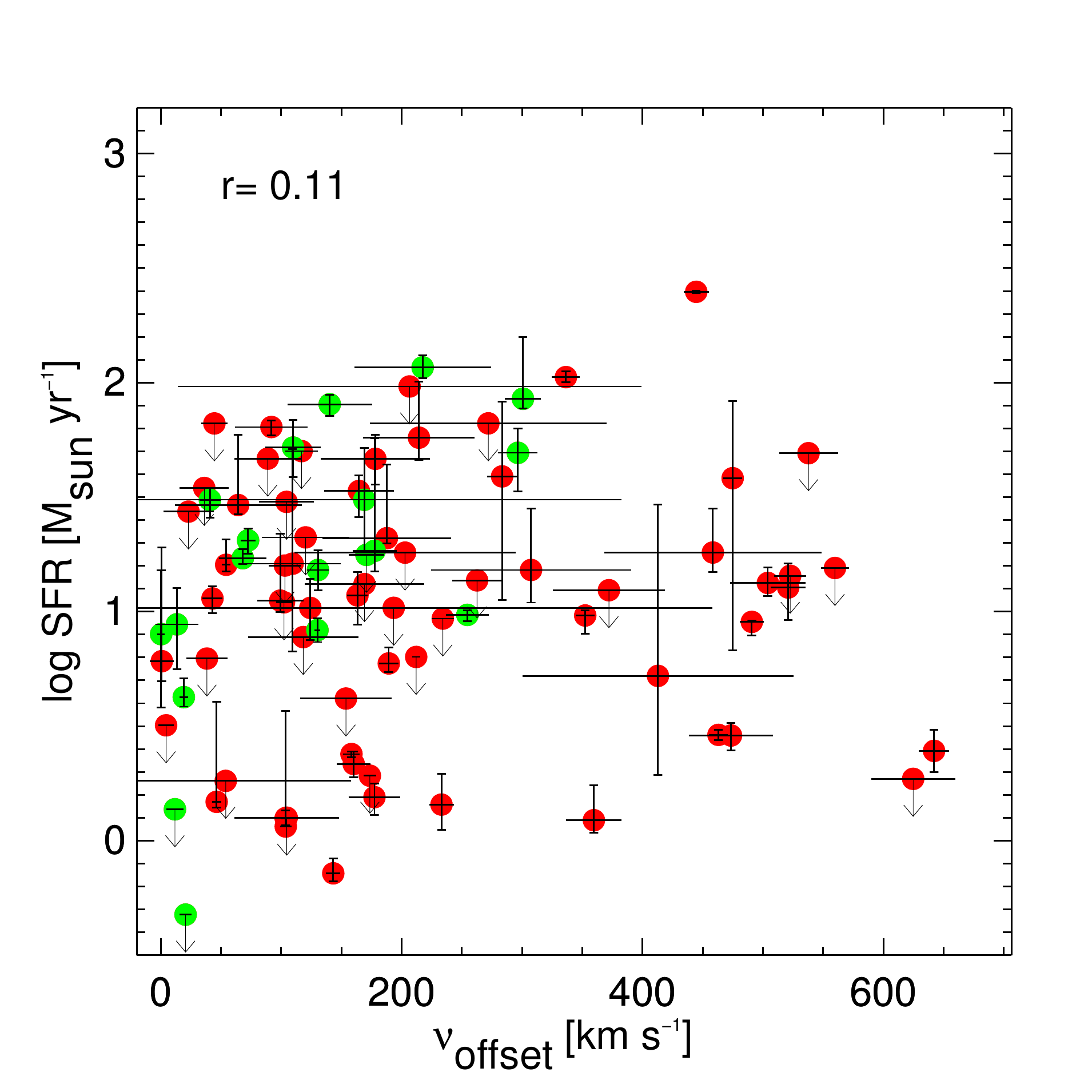}
\includegraphics[scale=0.3]{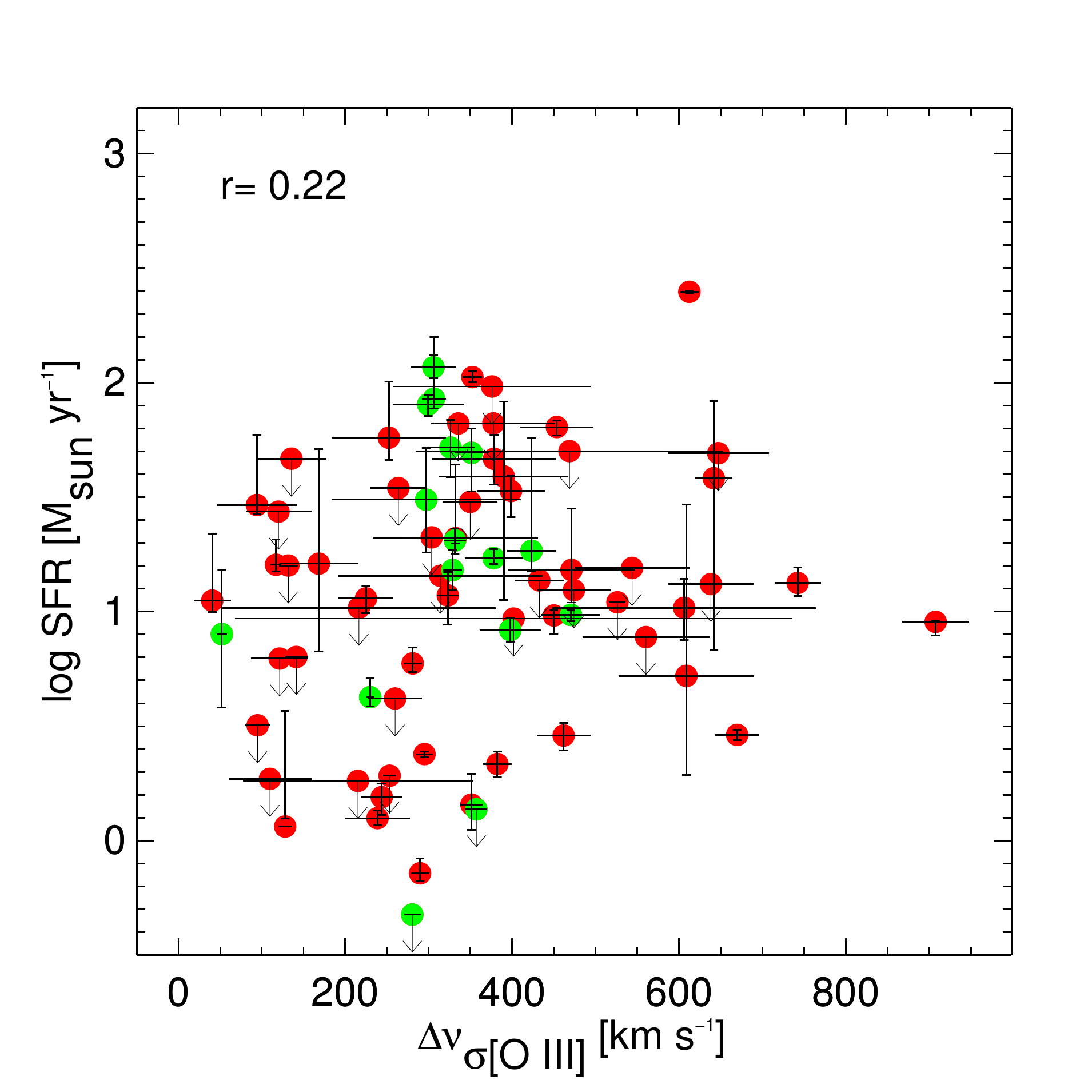}
\caption{Relation between the SFR and the three parameters
  characterizing the outflow ($\nu_{\rm
    blue}$, $\Delta\nu_{\rm offset}$, \dsig) for the starburst dominated
  quasars (blue points, top panel) and the AGN dominated quasars (red
  points, bottom panel). There is no  clear relation between outflow and SF
  considering the AGN and SB  dominated quasars. If we consider
  only the SFd-dominated objects (blue points) a positive trend
  emerges, although  with a weak significance and a large scatter.}
\label{all}}
\end{figure*}


\subsection{Testing the negative AGN feedback model}

\begin{figure*}
\includegraphics[scale=0.33,angle=0]{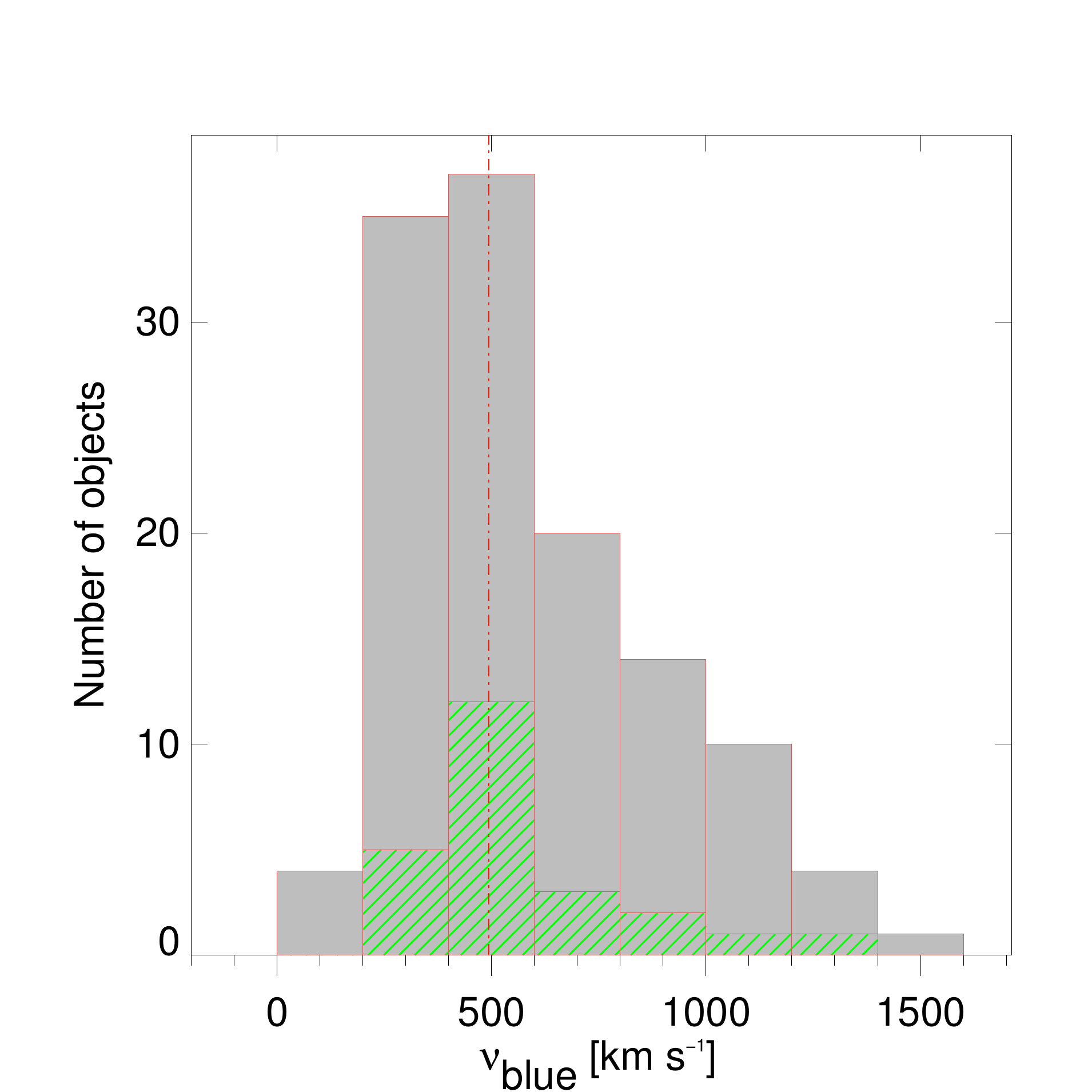}
\includegraphics[scale=0.33,angle=0]{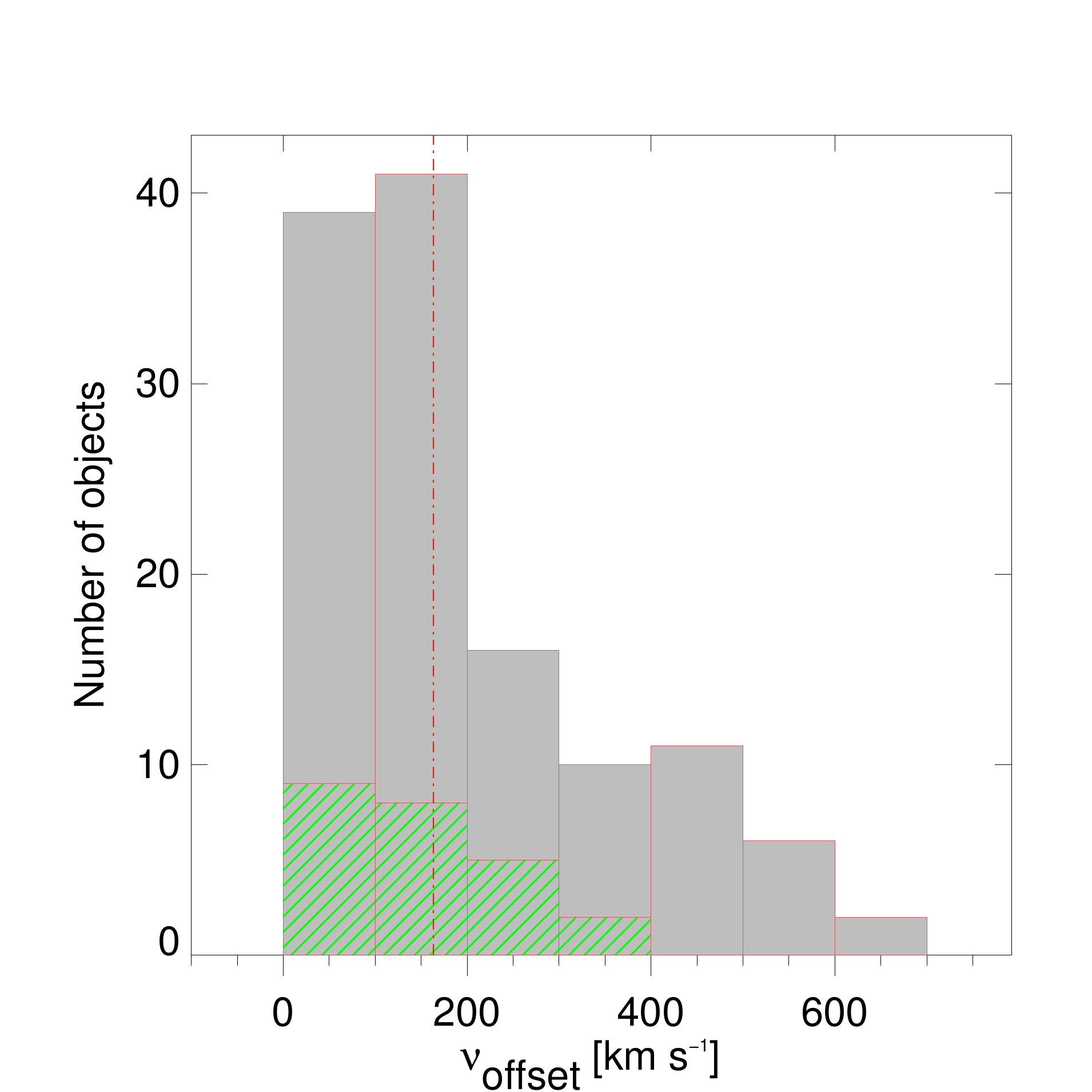}
\includegraphics[scale=0.33,angle=0]{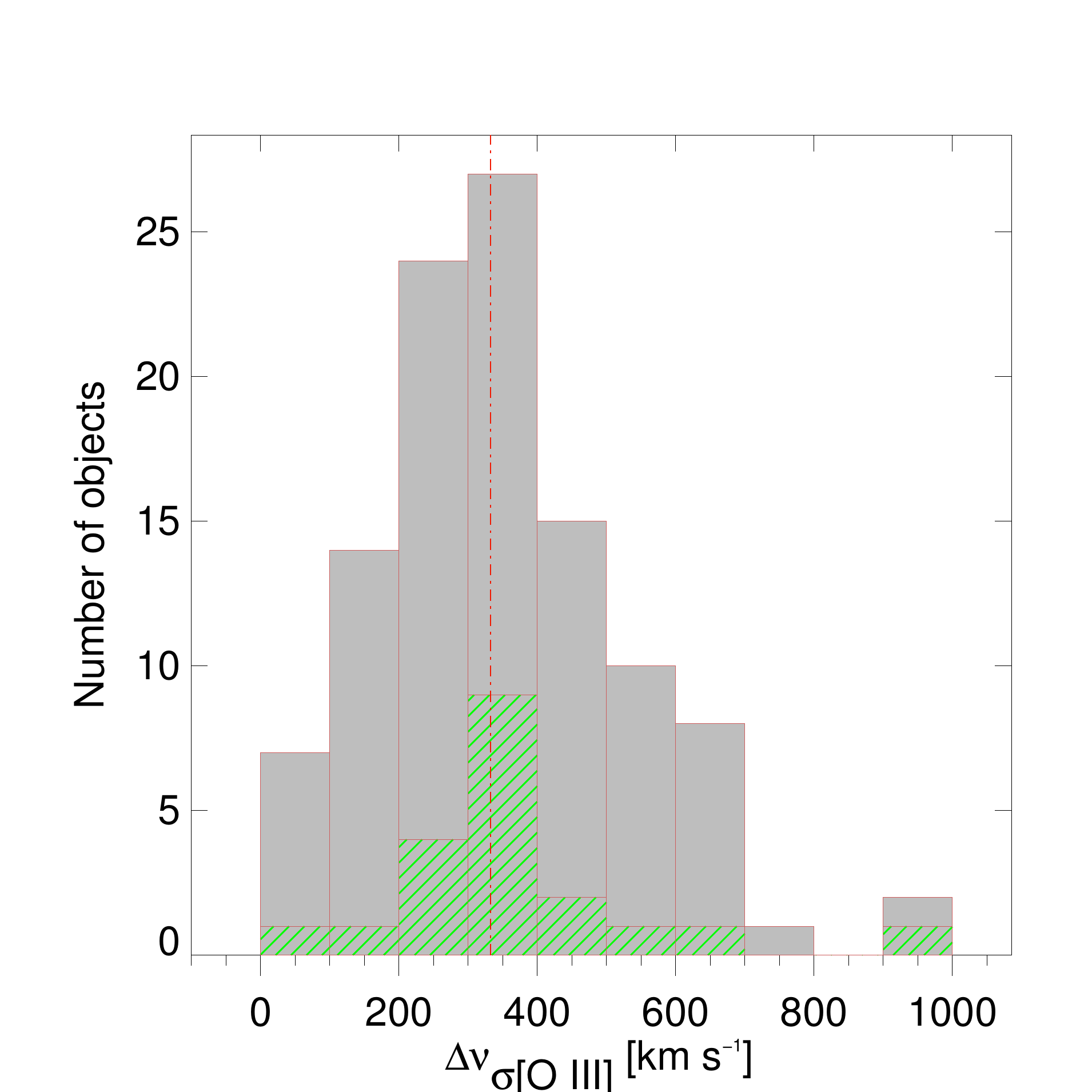}
\caption{Histogram of the distribution of the three parameters characterizing the outflow. The red dashed line marks the median of the distribution.
From left to right: \vblue, \voff, \dsig. The green hatched area represents the radio loud quasars.}
\label{velhist}
\end{figure*}

\begin{figure*}
\centering{
\includegraphics[scale=0.35,angle=0]{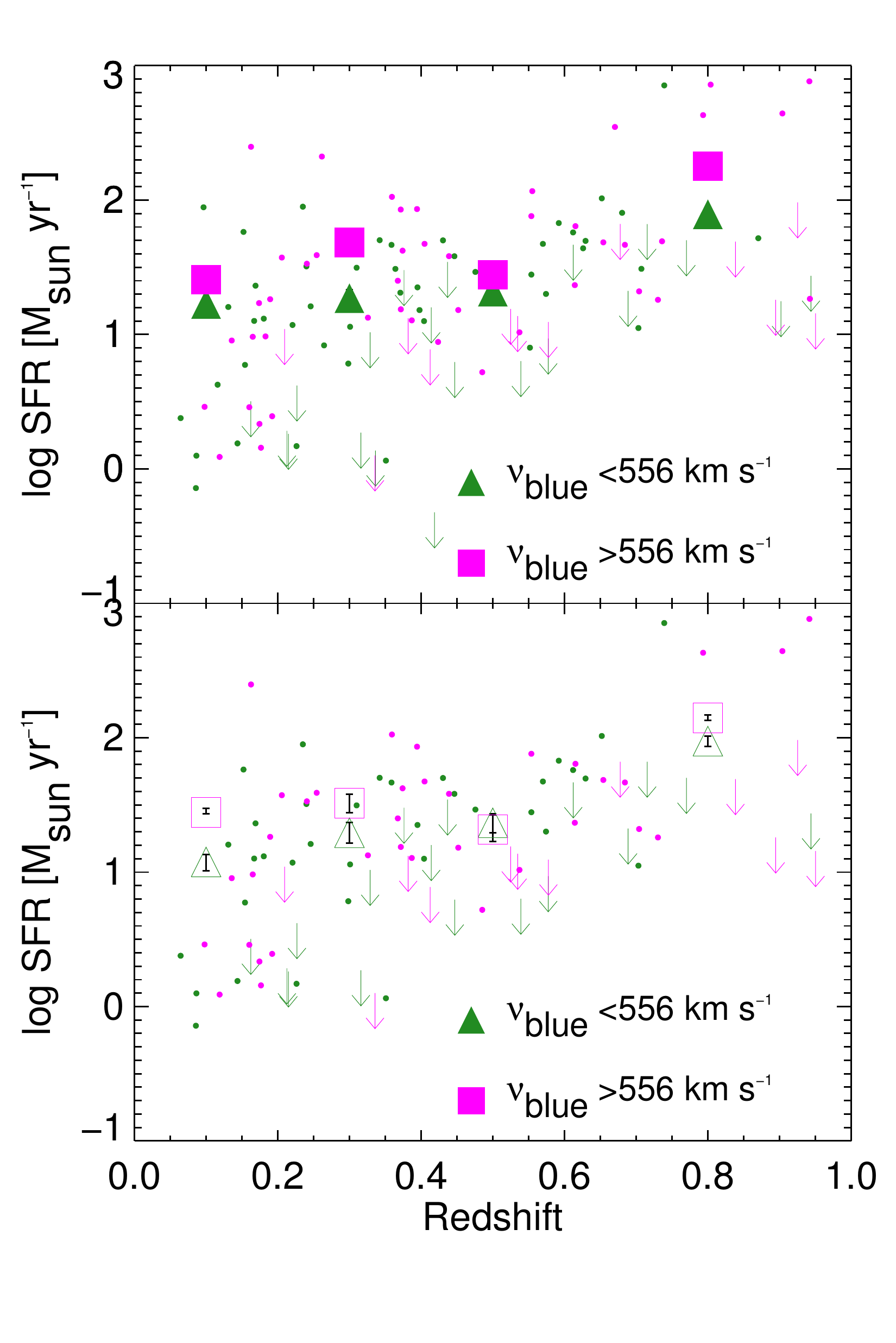}
\includegraphics[scale=0.35,angle=0]{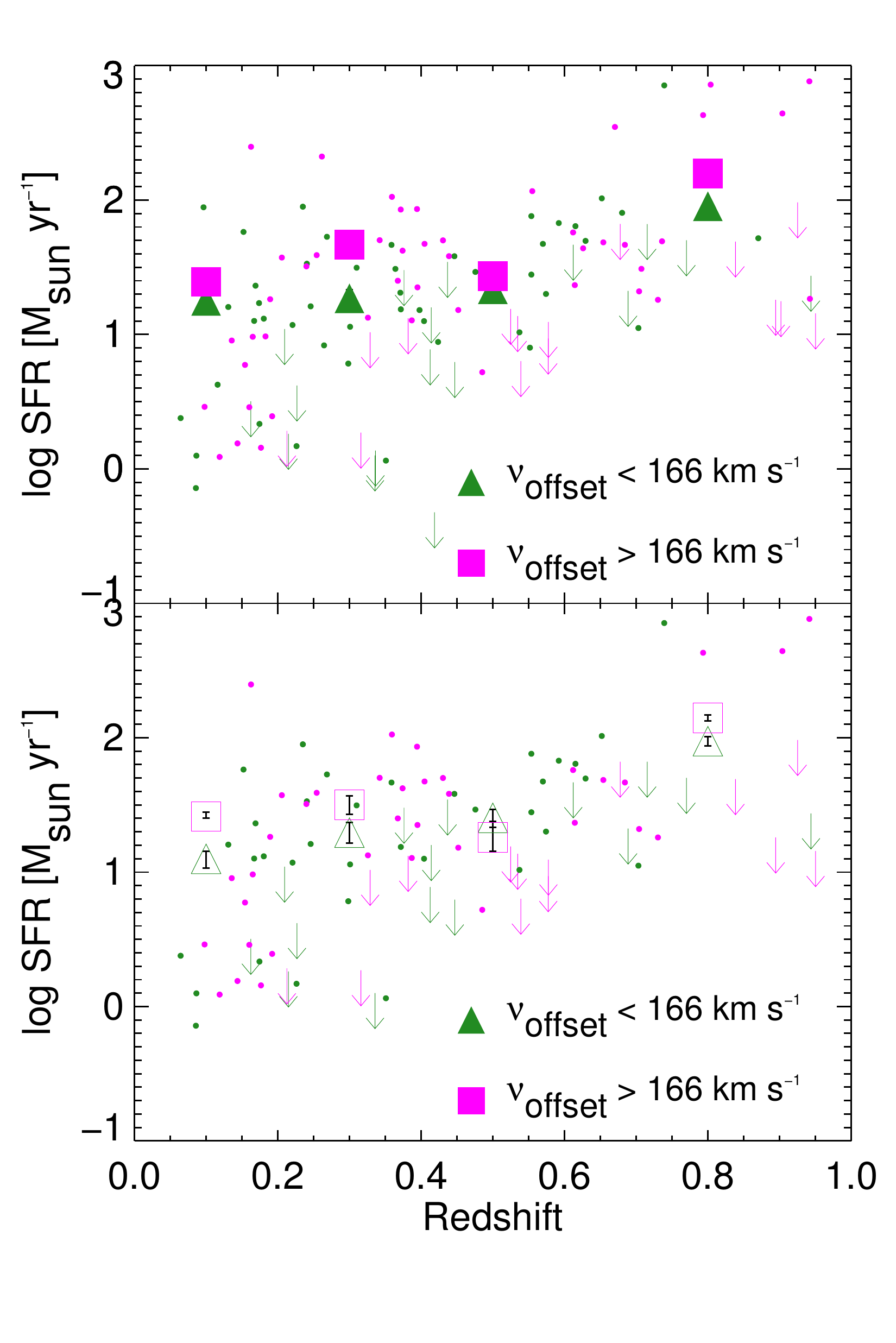}
\includegraphics[scale=0.35,angle=0]{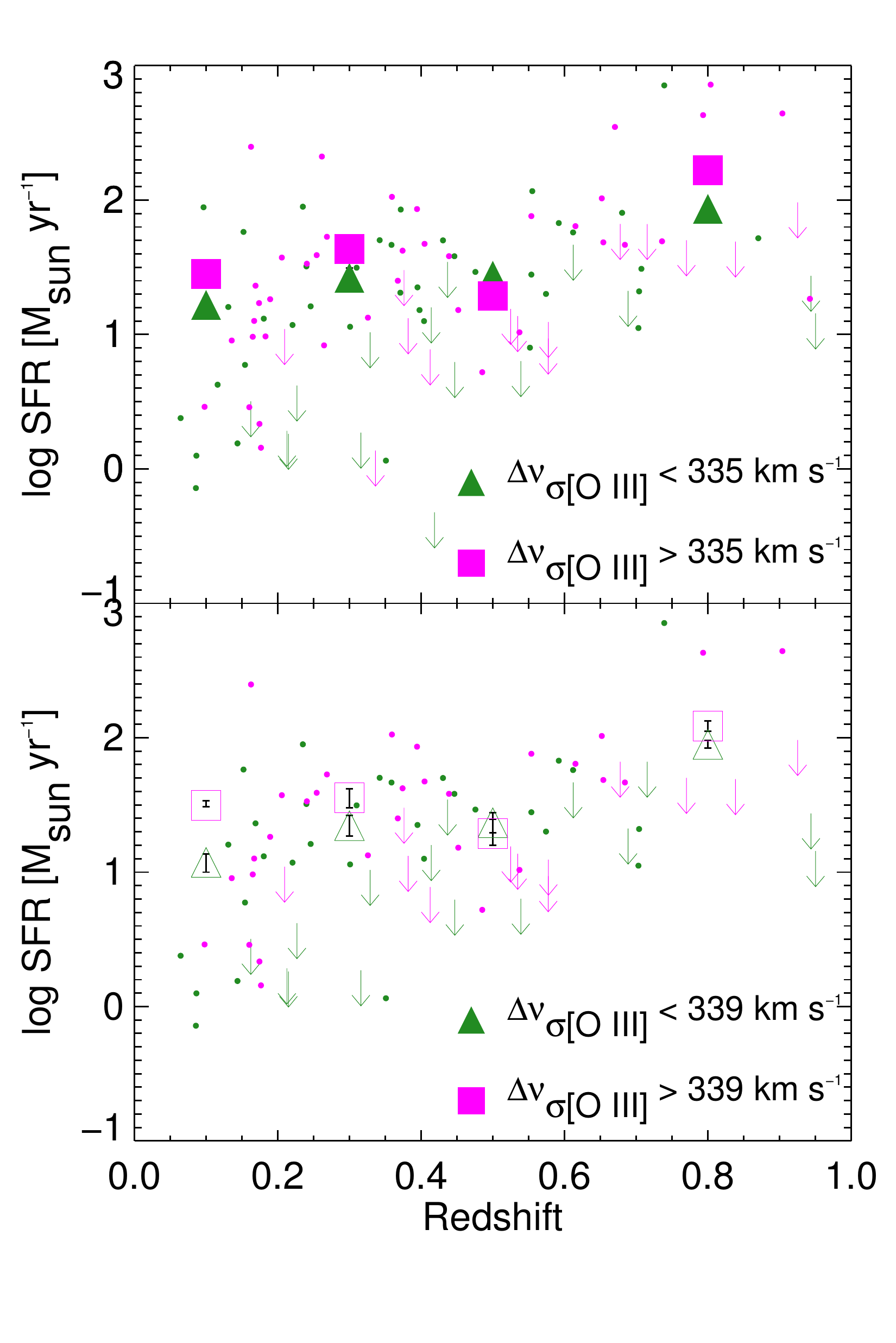}
\caption{Upper panel: Star formation rate plotted in four bins of redshift for the complete sample of quasars.  We adopt different criteria for separating
  strong-outflow (small magenta  points)  and weak-outflow galaxies (small green  points),
   respectively higher and lower than the median value of
  the distribution of the  three parameters  (\vblue (first panel), \voff (second panel), and \dsig\,).
  Bottom panel: same as top, but excluding radio loud quasars.}
  \label{stat}}
\end{figure*}

\begin{table}
\begin{tabular}{l|l|l|l|l}
\hline
  & \multicolumn{2}{c}{Weak-Out. QSO} &  \multicolumn{2}{|c}{Strong-Out. QSO} \\
\hline
$z$ bin & Det/ul & $<$SFR$>$ & Det/ul & $<$SFR$>$ \\ 
\hline
\multicolumn{5}{l}{\vblue}\\
\hline
0-0.2    & 12/1  &  1.23(1.07)  & 12/0  &   1.41 (1.44)\\
0.2-0.4  & 16/7  &  1.27(1.29)  & 13/3  &   1.69 (1.51)\\ 
0.4-0.6  & 9/6  &  1.32(1.37)  & 8/4   &   1.44 (1.31)\\
0.6-1    & 9/7   &  1.89(2.01)  & 13/5  &   2.25 (2.15) \\
\hline
\multicolumn{5}{l}{\voff}\\
\hline
0-0.2    & 12/1   & 1.25(1.10)  & 12/0    & 1.39(1.41) \\
0.2-0.4  & 16/6   & 1.27(1.29)  & 13/4   & 1.67(1.50) \\ 
0.4-0.6  & 11/5   & 1.34(1.40)  & 6/5    & 1.43(1.26) \\
0.6-1    & 7/6    & 1.95(1.97)  & 15/6   & 2.20(2.15)  \\
\hline
\multicolumn{5}{l}{\dsig}\\
\hline
0-0.2    & 10/1   & 1.22(1.07)  & 12/0  & 1.45(1.50)  \\
0.2-0.4  & 13/5  & 1.42(1.34)  & 11/4  & 1.63(1.55)  \\ 
0.4-0.6  & 9/5   & 1.44(1.36)  & 6/5   & 1.29(1.28)  \\
0.6-1    & 7/5   & 1.93(1.95)  & 10/5   & 2.18(2.09)  \\
\hline
\end{tabular}
\caption{Statistic results for the SFR plotted in bin of redshift (see Fig. \ref{stat}). For the three kinematic parameters adopted, we report in each bin of redshift (col. 1)
the number of the detection and upper limits and the mean SFR in logarithmic scale (in parenthesis the 
value obtained excluding radio loud quasars) for unperturbed quasars (col. 2 and 3) and outflow
dominated quasars (col. 4 and 5).}
\label{tab2}
\end{table}

The SFR of star-forming galaxies correlates with the stellar mass in
the local Universe,
as well as in the distant Universe out to z $\sim$ 2 (e.g. \citealt{noeske07};
 \citealt{daddi07}; \citealt{elbaz07}). This close
correlation (dispersion of $\sim 0.3$ dex) is called the main sequence  and it can be used to define the
outliers, i.e. quenched and starburst galaxies.  Later on, the
specific star formation rate (sSFR), defined as the SFR per unit stellar mass,
 is used to
investigate the dependence of the SFR with many parameters
characterizing the galaxies or the environment. For example,  a dependence of the sSFR on stellar mass and redshift
(\citealt{bauer05})  was
discovered.  Galaxies at redshift equal to one tend to have star
formation rates that are a factor of ten higher than the local ones. Also,
high redshift galaxies are typically more massive than low-redshift
star-forming galaxies (\citealt{seymour08}).

In this work, we compute   the SFR in redshift bins  independently for strong and weak 
outflow galaxies (assuming similar stellar masses in each redshift bin for
the two groups). 
In order to divide the quasars into the two groups, we consider 
the distribution of  \vblue, \voff and \dsig\, and as the threshold we choose 
the median of the respective distributions  (\vblue $>$556, \voff$>$166 and \dsig$>$335 \kms), see Fig.\ref{velhist}.
In Fig.\ref{stat} we show the logarithm of the SFR versus redshift plotting
 strong and weak outflow quasars according to these three criteria.
We have separated the quasars into four redshift bins (z$<$0.2, 0.2$<$z$<$0.4, 0.4$<$z$<$0.6, and z$>0.6$, see Table \ref{tab2})
 and in each redshift bin  we compare the mean star formation rate
 for the two classes separately,
taking error bars and
upper limits in the SFR measurements into account.

It is not an easy task to calculate a weighted mean with upper limits in a dataset.
We adopt the following strategy. For each quasar we consider the histogram distribution of SB luminosity
(i.e. the distribution of the luminosity measurements obtained through a Monte Carlo analysis). 
Some of the QSO are considered upper limits and their distribution peaks at zero luminosity,
others are detections and their distribution resemble a Gaussian distribution. 
We randomly extract 10$^4$ SB luminosities from the distributions of all quasars separating them in four classes of redshift
and in  strong and weak outflow quasars. In these four redshift bins, we obtain  two histograms of the total luminosities
that take into account detected and undetected sources. We evaluate the mean and the error at the 10\% and  90\%
percentiles of the distribution and we overplot the median points in the figure.
Since  we may have overestimated the star formation rate for radio loud quasars, and the type of the AGN feedback on the host could be different,
we repeat the same analysis excluding radio loud quasars (Fig.\ref{stat}, bottom panel).

 We note an increase in the mean
 star formation rate with redshift with a similar slope in the two
 groups. This is  expected because it is well known that the
 cosmic star formation rate peaks around z$\sim$ 2-3, after which it
 decreases by an order of magnitude to the present Universe
 (e.g. \citealt{hopkins06}). More importantly, we find that the star formation rate is similar in the two groups of quasars
 and, typically, strong-outflow quasars show values only slightly larger than weak-outflow quasars, even excluding the radio loud quasars.
In each redshift bin the discrepancies between the median values in the two groups are slightly reduced if radio loud galaxies are
excluded from the analysis. We have found no evidence that the SF in the host is suppressed in the presence of strong outflows comparing the SFR 
in bin of redshift in quasars characterized by strong or weak outflow signatures. 

\begin{table}
\begin{tabular}{l|l|l|l|}
\hline
Variables & Disp. & r(N) & Sign.\\ 
\hline
\hline
\multicolumn{4}{l}{\bf{Agn dominated quasars}} \\
\hline
log SFR vs  Max velocity                       & 0.2  & 0.44(39)  & 95\% \\  
SFR vs Wings vel. offset                      &  0.2 & 0.40(39)  & 95\% \\ 
log SFR vs ($\sigma[OIII]-\sigma[OII]$)        &  0.2 & 0.45(33)   & 95\% \\ 
\hline
\multicolumn{4}{l}{\bf{SB dominated quasars}} \\
\hline
log SFR vs Max velocity                         &  0.5  & 0.23(86)  &NO \\ 
log SFR vs Wings offset                         &  0.5  &  0.11(86) & NO \\ 
log SFR vs $\Delta$ VSIG[OIII]              &  0.5  & 0.22(75)  &NO\\  
\hline
\hline
\end{tabular}
\caption{ Col. 2: The variance of the intrinsic scatter.  Col3:the
  linear correlation coefficient between the dependent and independent
  variables (Pearson product-moment correlation coefficient). The
  maximum positive (negative) correlation is indicated by a
  correlation coefficient of 1 (-1).  An absolute value between 0.4
  and 0.7 could indicate moderate association.  Col.4: Significance of the
  linear correlation. A value of P less than 0.01 (0.05) imply a
  significant measurement at 1\%(5\%), id.e. there is 1\% (5\%)
  percentage chance that the finding is non-significant. N is the size of the sample.}
  \label{tab1}
\end{table}

\begin{figure*}
\centering{
\includegraphics[width=8cm,angle=0]{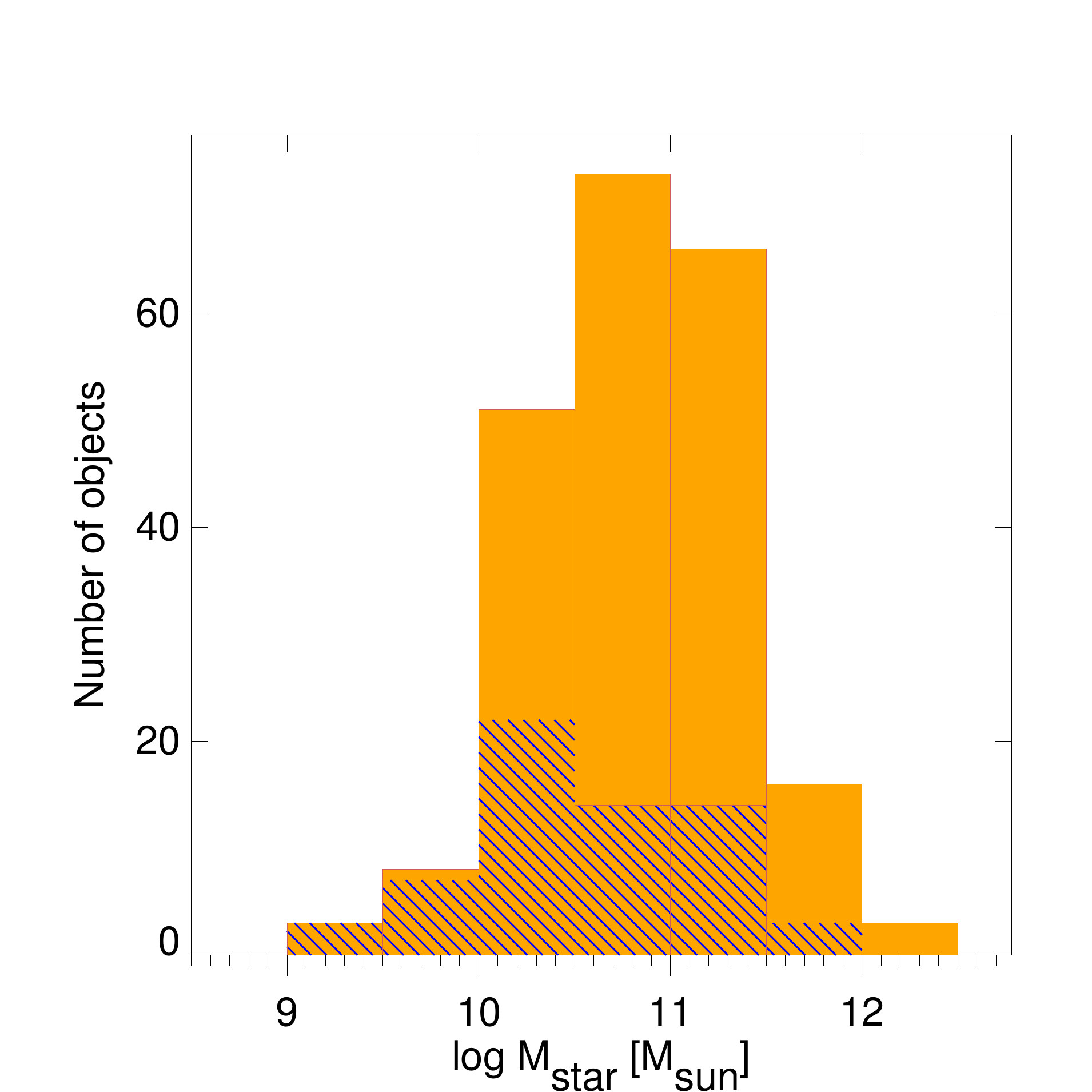}
\includegraphics[width=8cm,angle=0]{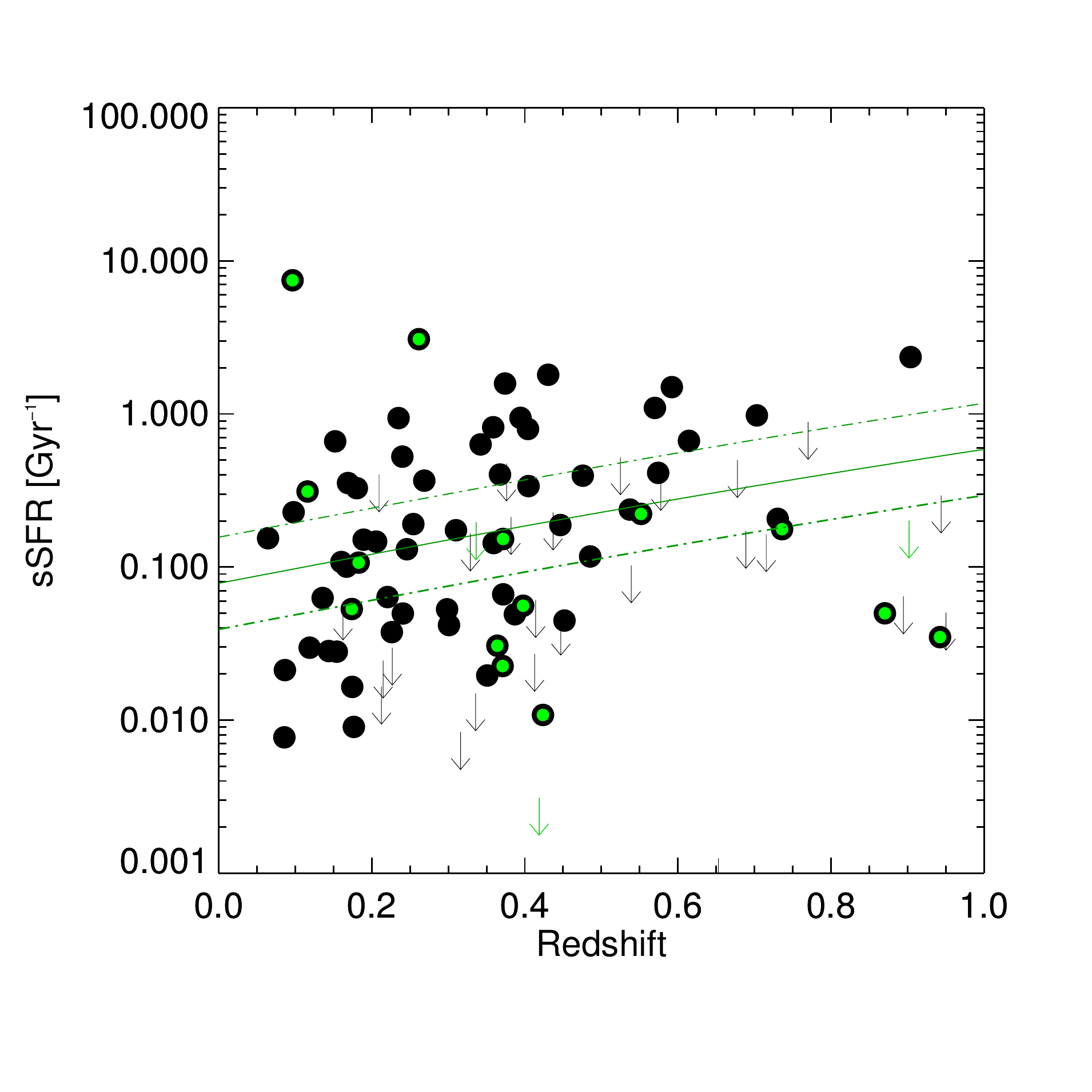}
\caption{Left panel: Histogram of the logarithm of the stellar masses in M$_\odot$ derived for our sample of type 1 quasars. The hatched area represents uncertain values. Right panel: Evolution of the main sequence SFR with redshift. The solid line represents the relation between sSFR and redshift and
the two dashed lines are a factor of
two above and below this line.}
 \label{mstarhisto}}
\end{figure*}

\begin{figure*}
\centering{
\includegraphics[scale=0.3,angle=0]{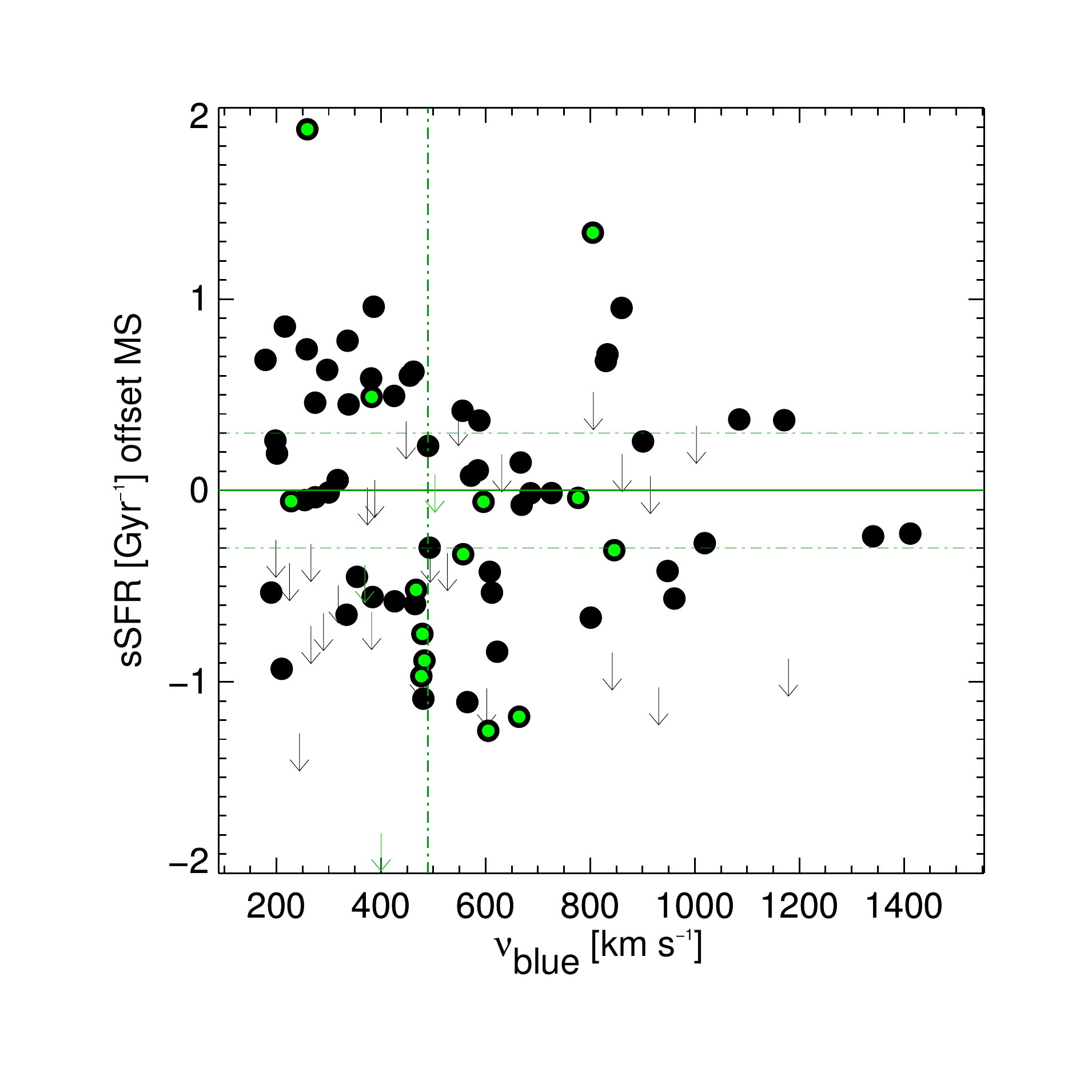}
\includegraphics[scale=0.3,angle=0]{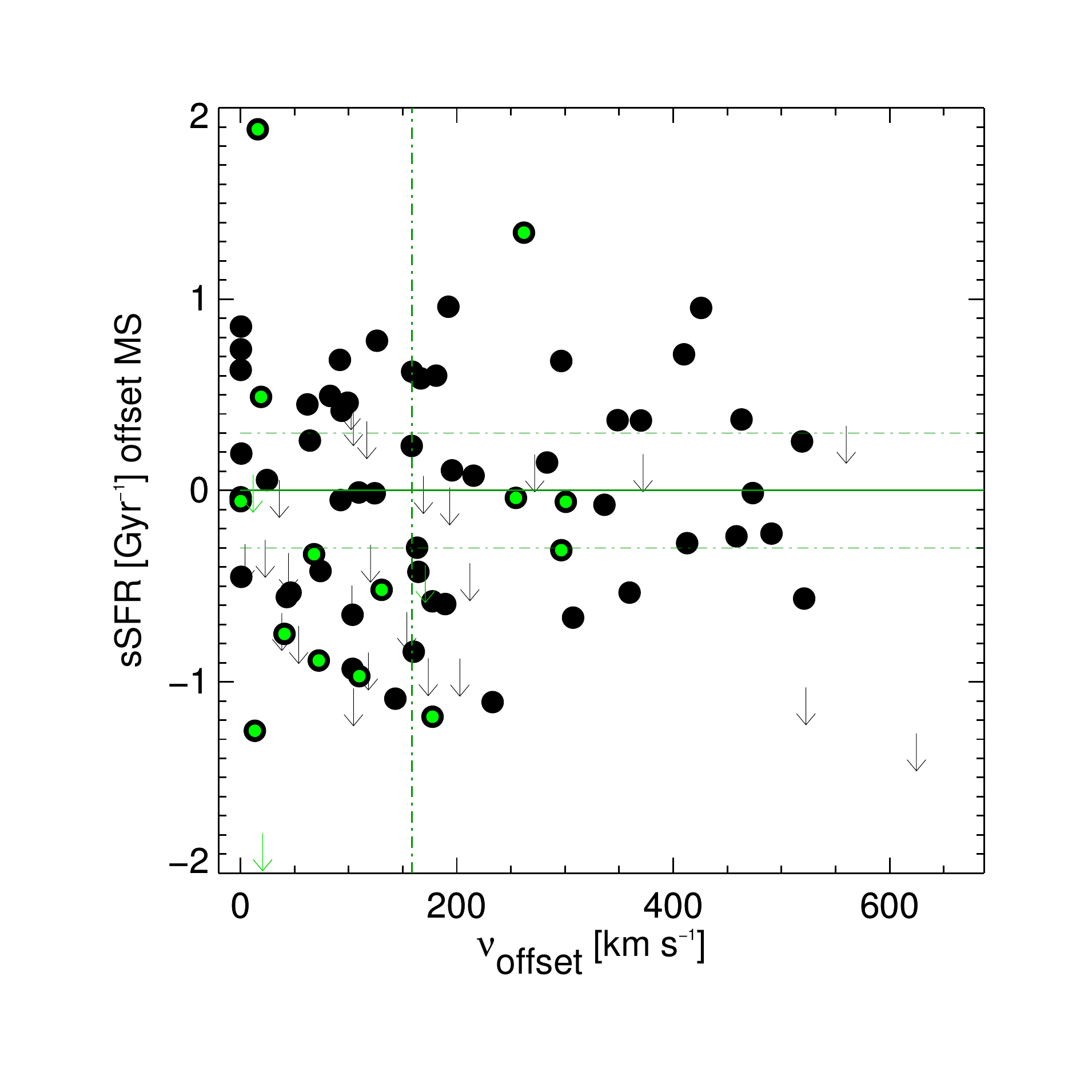}
\includegraphics[scale=0.3,angle=0]{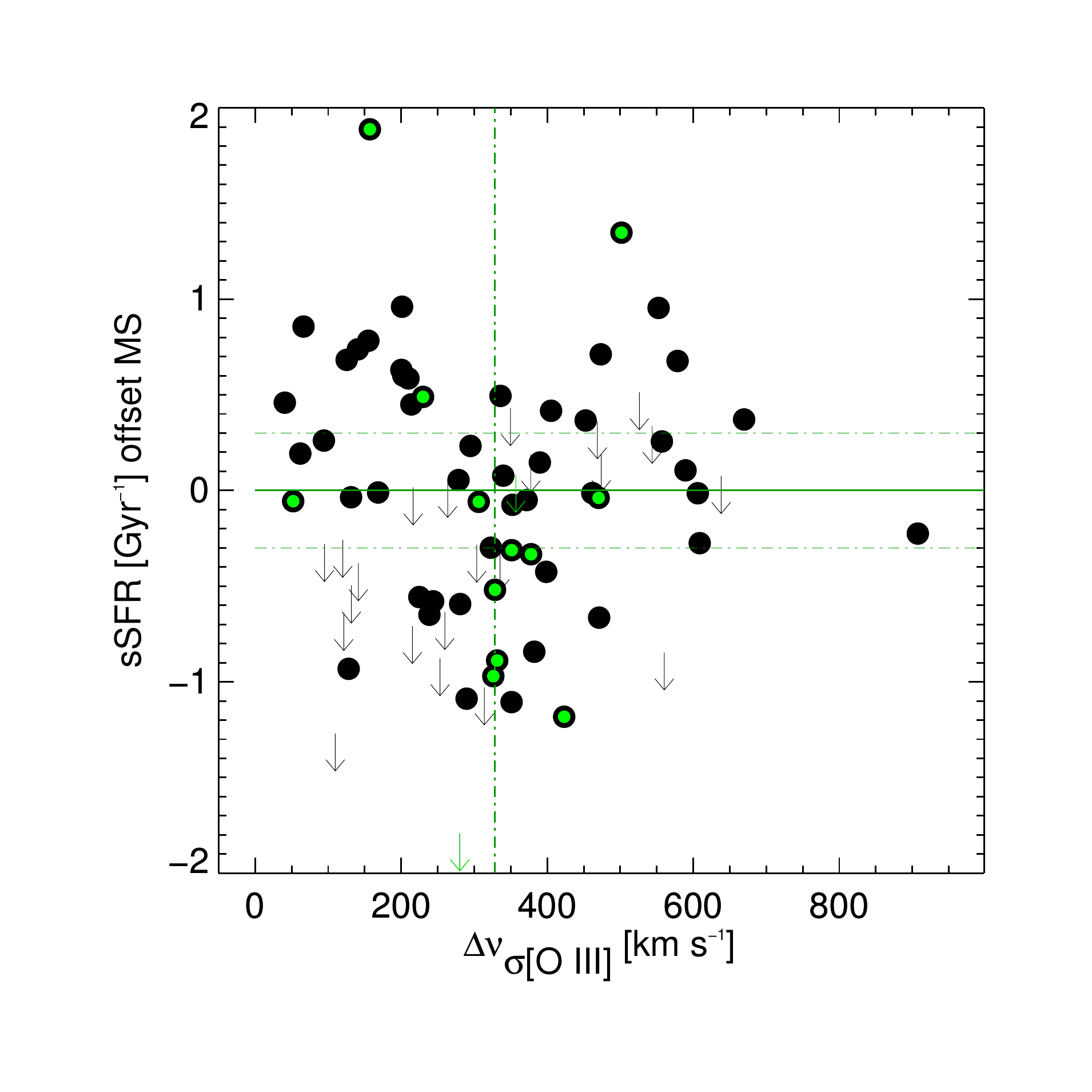}
\caption{sSFR offset from the main sequence versus the three emission line parameters characterizing the outflow: \vblue, \voff, \dsig. The green points represent the radio loud quasars.}
 \label{ms2}}
\end{figure*}

\section{ Main sequence of quasar host galaxies.}

One of the best ways to take into account the effects due to different galaxy masses is to normalize the SFR to the stellar mass obtaining the so-called 
specific star formation rate (sSFR). However, given the difficulties in measuring stellar masses in quasar host galaxies from SED fitting in the  IR band
caused by the strong AGN 
contamination, we adopt this approach as a consistency check.
The offset from the SF main sequence provides evidence 
that the SF in the galaxy is enhanced or quenched with respect to most of the  galaxies. 
Therefore, we derive the stellar mass from the normalization of the Bruzal \& Charlot stellar template.

As mentioned above, measuring stellar masses in type 1 AGNs and in quasars in particular is extremely challenging. However, in many cases the near IR SEDs of our sources around 
1-3 \mc\ are dominated by the emission of old stars,  
allowing a reliable estimate of the stellar mass.
We consider convincing the stellar mass derived from SED fitting in which the luminosity of the old star components
contribute  at least 25\% to the total luminosity in the 1-3\mc range. 
We show in Fig.\ref{mstarhisto}, right panel, the distribution of the stellar mass values obtained for  the whole sample of quasars (the dashed area represents
uncertain measurements, about 30\% of the total measurements). 
By considering only the reliable values we obtain a mean value of 10$^{10.8}$ M$_{\odot}$, consistent with the range found by \citet{mainieri11}
between 8$\times$10$^9$ and 10$^{11}$ M$_{\odot}$ for a sample of type 2 quasars from the COSMOS survey. 

In Fig. \ref{mstarhisto}, right panel,  we plot the sSFR versus redshift
and we compare these measurements with the main sequence (adopting the \citet{elbaz11} formulation\footnote{ 
$sSFR_{\rm MS}=26\times t_{\rm cosmic}^{-2.2}$ in [Gyr$^{-1}$]
with $t_{\rm cosmic}\sim\frac{28}{1+(1+z)^2}$ Gyr}). 
The sSFR of the quasars are in agreement with the $sSFR_{\rm MS}$
relation found for normal star-forming galaxies, although with a very large scatter that is   due in part to the uncertainties on the 
stellar mass measurement. 
We investigate whether there is a relation between the position in this plane and the kinematic properties
of the outflow. In Fig. \ref{ms2}, we compare the offset from the main sequence of SF (sSFR$_{\rm offset}$, a measure of the starburstiness of quasars) to the three 
kinematic parameters. Again, analogously to the SFR, we do not find a clear relation between these quantities and sSFR$_{\rm offset}$.

\section{Discussion}
\label{discussion}
\subsection{Evidence for a SB-outflow connection?}

 The origin of the fast ionized gas outflows
is controversial since in general it is not possible to establish
unambiguously whether they are produced by the intense star formation, by
the AGNs,  or  by a combination of the two processes. Galactic
winds produced by outflows driven by thermal energy from supernova
explosions or by stellar winds typically exhibit outflow velocities in
the 100-500 \kms\ range (\citealt{rupke05}) and outflows with
significantly higher velocities (|v| $>$ 1000 \kms) are generally
interpreted as driven by an AGN (e.g. \citealt{tremonti07}, \citealt{cicone14}).  
However, according to \citet{diamond12} ram pressure from
supernovae and stellar winds is sufficient to produce outflows with
velocity higher than 1000 \kms\ without  needing to invoke feedback
from an active galactic nucleus. 

Blue star-forming galaxies show a positive correlation between SFR and
outflows. For example, \citet{weiner09} found that the outflow velocity scales linearly with SFR as V$_{wind}
\sim$ SFR$^{0.3}$ in blue star-forming galaxies at z $\sim$ 1.4,
similar to the scaling found by \citet{martin05} in low-redshift ULIRGs.
For quasars  no
obvious relation between the wind velocity and infrared luminosity has been discovered.
For example, \citet{veilleux13} found that the velocity of the
OH absorption line does not correlate with the star formation rate in a
sample of ULIRGs and QSOs. A similar result is found by \citet{rupke13},
 considering both ionized and neutral gas outflow in six ULIRGs.  These authors did not find any 
 relation between the OH  outflow velocities and the star formation. 
However,  they find a non-linear relation between the velocity of the molecular outflow and the AGN luminosity, 
e.g. more blueshifted OH components are found in more AGN luminous sources.
 
We have looked for a relationship between the SFR and the strength of
the outflow, considering many different outflow parameters (e.g. \vblue, \voff, \dsig). 
We consider separately the two classes of  SFd and AGNd quasars (Fig. \ref{all}, bottom panel), in which  the driver mechanism of the outflow is probably different.
No clear trend emerges
between the star formation rate and the velocity of the outflow.
However, we do not find  any relation between the outflow and the properties of the AGNd quasars, as is expected if the
outflow is ultimately driven by the radiative output of the quasar and as was found instead for molecular outflows in quasars by  \citet{veilleux13}.
We cannot rule out that both AGN and stellar processes contribute
to driving the outflow and weaken any relations, especially in quasars with a high star-forming activity. 
This is in agreement with what has been found by other
 previously cited authors,  i.e. we do not find any
obvious relation between the wind velocity and infrared luminosity in the quasar sample.

If we separate the quasars into  two classes of  quasars with weak and strong outflow, an interesting result emerges.
We do not find that the mean star
formation rate is lower in host galaxies that exhibit the strongest outflows, as predicted by the basic AGN negative feedback model; instead, it is consistent or even higher than in weak outflow quasars. 
However, it is worth noting some important shortcomings and uncertainties of this approach.

\begin{itemize}

\item There could be a significant delay between the starburst phase and the peak of nuclear optical AGN activity
(e.g. \citealt{matsuoka11}, \citealt{davies7}; \citealt{bennert08}; \citealt{schawinski09}; \citealt{wild10}, \citealt{yesuf14}) estimated in the range of 100--400 Myr.
The quasar phase is expected to last for  a typical lifetime of 16--25 Myr (\citealt{goncalves08})
consistent with the estimate of $>10^7$ yr by \citealt{jakobsen03}. The timescale for star formation, i.e.  over which star formation remains constant,
can vary by a few 100 Myr (e.g. \citealt{grijs01}).
Therefore,
a substantial time lag between the quasars phase and the quenching of the star formation rate in the host galaxy
could wash out the most significant signatures of negative feedback. However, we note that if powerful winds are able to sweep away
the interstellar dust, we would expect an underluminous infrared galaxy host, even if the star formation proceeds.

\item Another possible interpretation of this result is that the AGN is stimulating the SFR as
predicted by a positive feedback scenario. In this interpretation
the AGN
outflow triggers star formation in the gas-rich galaxies by overcompressing cold dense gas
and thus providing positive feedback.  AGNs with pronounced radio jets
exhibit a much higher star formation rate than the purely X-ray
selected ones \citep{zinn13}.  In another case (\citealt{cresci15}), 
both types of feedback are observed in the same galaxy:
the outflow removes the gas from the host galaxy (negative feedback), but also triggers star formation  inducing pressure at the edges of the outflow (positive feedback).
We expect to see the positive feedback effect especially in
radio loud quasars. Interestingly, excluding radio loud quasars in our analysis,  the difference between the mean SFR
in the two groups decreases. 

\item  The blueshifted wings of [O~III] are typically  seen only in the
inner kpc of NLRs in local AGNs. So their detections imply outflows
over regions that are much smaller than the star-forming portion of
typical host galaxies and therefore these outflows may not be able to quench the star formation rate. Integral field spectroscopic data are crucial to answering to this question.

\item Since the Herschel spatial resolution does not allow us to
measure the SFR in the nuclear region (on a subkilo parsec scale), we cannot rule out the possibility that 
  the impact of negative feedback even in the most luminous systems only has an affect  in the very central regions (below 100 pc), as
  suggested recently in the model of \citet{roos15}. 

\item  In this work we have considered the SFR, since the stellar mass in type 1 quasars is a difficult parameter to determine.
It is possible that the sSFR correlates with the mass loading factor
 or the kinetic power of the outflow, not with the velocity. However, we cannot measure it with this data so we use 
 the velocity as a proxy of the strength of the outflow.  
 
\end{itemize}

Keeping in mind the discussed caveats, our results seem  to disfavour 
 the AGN negative feedback scenario, according to which AGN outflows on the QSO phase
sweep out the gas from the host galaxy suppressing the star formation.

Our results are more in agreement with
works that seems to demonstrate that AGNs do not regulate 
starbursts in the overall galaxy. For example \citet{debuhr10}, found that SMBH
feedback has little effect on the number of stars formed. In the
theoretical model of \citet{cen12} starburst and AGN growth are not
coeval in this model and AGN activities are expected to outlive the
starburst,  in agreement with observations (e.g. \citealt{georgakakis08}).
 In a recent simulation \citet{gabor14} found that AGN feedback
drives bursty, high velocity, hot outflows with mass rates peaking
briefly near the SFR but that have little impact on the star-forming
gas disk. The authors conclude that the star formation is not quenched
by AGN feedback, and they suggested that the key physical element
required to produce a galaxy red sequence is not  expelling gas from
galaxies, but rather  preventing gas from accreting.
Other processes than AGN negative feedback are able to transform a star-forming galaxy into a quiescent system,  for example
the so-called strangulation mechanism (e.g \citealt{larson80}, \citealt{ balogh00}). 
In this case star formation in the host continues until the  available gas is completely used up and the star formation
is not suddenly quenched as in the negative feedback model. In a recent analysis \citealt{peng15}
found that strangulation is the primary mechanism acting in local galaxies.

The increased probability of an AGN being hosted by a
star-forming galaxy (e.g. \citealt{santini12}) may be a consequence of the
relationship between gas content and AGN activity. If this is the
case,  the relationship between AGN bolometric
luminosity and the SFR found by \citet{netzer09} naturally arises. This relationship might  ultimately be caused by secular mechanisms
of gas inflow or merger mechanisms between galaxies. 
What emerges clearly is that the complex interplay between AGN outflows and star formation rate is far from being understood.

\section{Summary and conclusions}

The aim of this paper is to test the negative AGN feedback
scenario in its simplest version according to which QSO driven outflows are able to suddenly quench  the
SFR in the host, clearing out or heating the interstellar medium of the
galaxy.  We have considered 132 quasars from the SDSS quasars 
 observed in photometric mode by the infrared satellite
{\rm Herschel} at redshifts of  less than 1 (to cover the \oiii line in the SDSS
spectra) with high quality fits and spectra.

To investigate the presence of an outflow, we focus on the \oiii  and \oii emission line, in particular on  three
parameters (the maximum blue velocity, \vblue=|$\nu$10-$v$50|; the
offset of the broad emission, \voff=|$v$95+$v$5|/2-$v$50; and the $\sigma$ excess with respect to the \oii line, \dsig=$(\sigma$(\oiii)$^2$ -
$\sigma$(\oii)$^2)^{0.5}$) assuming that a broad, blueshifted wing is a signature of an outflow.

Since the SFR evolves with redshift and it is dependent on the stellar
mass of the galaxy (galaxies at higher redshift are typically more
massive and more star forming), we consider the mean SFR in bins of
redshift for strong and weak outflow galaxies. We do not
see that strong outflow galaxies have a SFR lower than weak-outflow 
galaxies as was predicted by the negative AGN feedback scenario.  Instead,
we found that the SFRs in the two groups are comparable or even higher
for the outflow dominated galaxies. We found an analogous result considering the offset
from the main sequence of the sSFR. The stellar masses derived from SED fitting in type 1 quasars is
uncertain because of the AGN contamination in IR. However, for the quasars with trustable measurements
 in which the stellar component dominates in the range 1-3 \mc, we find that
the redshift evolution of the specific SFR for
the hosts in our sample of type 1 QSOs is in agreement with
that observed for star-forming galaxies. 

To investigate the main driver of the outflow, we divided our sample into two groups: the AGN-dominated
quasars (defined as L$_{\rm SB}$/L$_{\rm tot}<$0.5) and the SF-dominated
quasars (L$_{\rm SB}$/L$_{\rm tot}>$0.5). We find that by focusing  only
on SF-dominated quasars we did not find a correlation between the SFR and the velocity, as found for starburst galaxies; rather, a slight positive trend was found.
  Instead in the AGN-dominated
quasars do not find any clear relationship between the outflow
velocities and the SFR or the AGN properties  (Eddington ratio, \oiii
luminosity, or total infrared luminosity).

We conclude that the negative
feedback scenario is disfavoured by our results and that other possible mechanisms
could be responsible for quenching star formation. In our large sample of 
quasars,  a high level of SFR coexists with powerful outflow and, moreover, outflow dominated host galaxies
have star formation rates consistent with  or slightly larger than galaxies showing weak outflows. Different possibilities
can explain the lack of observational evidence of negative feedback, including the possibility  that AGN driven outflows are not the main mechanism
by which the star formation in blue star-forming galaxies is suppressed. 

\begin{acknowledgements} 
We thank the referee for his/her accurate revision of the paper.\\

BB acknowledges support from grant PRIN-INAF 2011 "Black hole growth and AGN feedback through the cosmic time".
MB acknowledges support from the FP7 Career Integration Grant ``eEASy'': "supermasssive black holes through cosmic time: from current surveys to eROSITA-Euclid Synergies" (CIG 321913).\\

Funding for SDSS-III has been provided by the Alfred P. Sloan Foundation,
the Participating Institutions, the National Science Foundation, and the
U.S. Department of Energy Office of Science. The SDSS-III web site is
http://www.sdss3.org/.
SDSS-III is managed by the Astrophysical Research Consortium for the
Participating Institutions of the SDSS-III Collaboration including the University
of Arizona, the Brazilian Participation Group, Brookhaven National Laboratory,
Carnegie Mellon University, University of Florida, the French Participation
Group, the German Participation Group, Harvard University, the Instituto de
Astrofisica de Canarias, the Michigan State/Notre Dame/JINA Participation
Group, Johns Hopkins University, Lawrence Berkeley National Laboratory,
Max Planck Institute for Astrophysics, Max Planck Institute for Extraterrestrial
Physics, New Mexico State University, New York University, Ohio State
University, Pennsylvania State University, University of Portsmouth, Princeton
University, the Spanish Participation Group, University of Tokyo, University of
Utah, Vanderbilt University, University of Virginia, University of Washington,
and Yale University.
\end{acknowledgements}

\bibliographystyle{aa} 

\begin{thebibliography}{113}
\expandafter\ifx\csname natexlab\endcsname\relax\def\natexlab#1{#1}\fi

\bibitem[{{Alexander} {et~al.}(2010){Alexander}, {Swinbank}, {Smail},
  {McDermid}, \& {Nesvadba}}]{alexander10}
{Alexander}, D.~M., {Swinbank}, A.~M., {Smail}, I., {McDermid}, R., \&
  {Nesvadba}, N.~P.~H. 2010, \mnras, 402, 2211

\bibitem[{{Balogh} {et~al.}(2000){Balogh}, {Navarro}, \& {Morris}}]{balogh00}
{Balogh}, M.~L., {Navarro}, J.~F., \& {Morris}, S.~L. 2000, \apj, 540, 113

\bibitem[{{Bauer} {et~al.}(2005){Bauer}, {Drory}, {Hill}, \&
  {Feulner}}]{bauer05}
{Bauer}, A.~E., {Drory}, N., {Hill}, G.~J., \& {Feulner}, G. 2005, \apjl, 621,
  L89

\bibitem[{{Bennert} {et~al.}(2008){Bennert}, {Canalizo}, {Jungwiert},
  {Stockton}, {Schweizer}, {Peng}, \& {Lacy}}]{bennert08}
{Bennert}, N., {Canalizo}, G., {Jungwiert}, B., {et~al.} 2008, \apj, 677, 846

\bibitem[{{Benson} {et~al.}(2003){Benson}, {Frenk}, {Baugh}, {Cole}, \&
  {Lacey}}]{benson03}
{Benson}, A.~J., {Frenk}, C.~S., {Baugh}, C.~M., {Cole}, S., \& {Lacey}, C.~G.
  2003, \mnras, 343, 679

\bibitem[{{Bian} {et~al.}(2005){Bian}, {Yuan}, \& {Zhao}}]{bian05}
{Bian}, W., {Yuan}, Q., \& {Zhao}, Y. 2005, \mnras, 364, 187

\bibitem[{{Bicknell} {et~al.}(2000){Bicknell}, {Sutherland}, {van Breugel},
  {Dopita}, {Dey}, \& {Miley}}]{bicknell00}
{Bicknell}, G.~V., {Sutherland}, R.~S., {van Breugel}, W.~J.~M., {et~al.} 2000,
  \apj, 540, 678

\bibitem[{{Bolzonella} {et~al.}(2010){Bolzonella}, {Kova{\v c}}, {Pozzetti},
  {Zucca}, {Cucciati}, {Lilly}, {Peng}, {Iovino}, {Zamorani}, {Vergani},
  {Tasca}, {Lamareille}, {Oesch}, {Caputi}, {Kampczyk}, {Bardelli}, {Maier},
  {Abbas}, {Knobel}, {Scodeggio}, {Carollo}, {Contini}, {Kneib}, {Le
  F{\`e}vre}, {Mainieri}, {Renzini}, {Bongiorno}, {Coppa}, {de la Torre}, {de
  Ravel}, {Franzetti}, {Garilli}, {Le Borgne}, {Le Brun}, {Mignoli},
  {Pell{\'o}}, {Perez-Montero}, {Ricciardelli}, {Silverman}, {Tanaka},
  {Tresse}, {Bottini}, {Cappi}, {Cassata}, {Cimatti}, {Guzzo}, {Koekemoer},
  {Leauthaud}, {Maccagni}, {Marinoni}, {McCracken}, {Memeo}, {Meneux},
  {Porciani}, {Scaramella}, {Aussel}, {Capak}, {Halliday}, {Ilbert},
  {Kartaltepe}, {Salvato}, {Sanders}, {Scarlata}, {Scoville}, {Taniguchi}, \&
  {Thompson}}]{bolzonella10}
{Bolzonella}, M., {Kova{\v c}}, K., {Pozzetti}, L., {et~al.} 2010, \aap, 524,
  A76

\bibitem[{{Brusa} {et~al.}(2015){Brusa}, {Bongiorno}, {Cresci}, {Perna},
  {Marconi}, {Mainieri}, {Maiolino}, {Salvato}, {Lusso}, {Santini}, {Comastri},
  {Fiore}, {Gilli}, {La Franca}, {Lanzuisi}, {Lutz}, {Merloni}, {Mignoli},
  {Onori}, {Piconcelli}, {Rosario}, {Vignali}, \& {Zamorani}}]{brusa15}
{Brusa}, M., {Bongiorno}, A., {Cresci}, G., {et~al.} 2015, \mnras, 446, 2394

\bibitem[{{Bruzual} \& {Charlot}(2003)}]{bruzual03}
{Bruzual}, G. \& {Charlot}, S. 2003, \mnras, 344, 1000

\bibitem[{{Cano-D{\'{\i}}az} {et~al.}(2012){Cano-D{\'{\i}}az}, {Maiolino},
  {Marconi}, {Netzer}, {Shemmer}, \& {Cresci}}]{cano12}
{Cano-D{\'{\i}}az}, M., {Maiolino}, R., {Marconi}, A., {et~al.} 2012, \aap,
  537, L8

\bibitem[{{Cen}(2012)}]{cen12}
{Cen}, R. 2012, \apj, 755, 28

\bibitem[{{Chabrier}(2003)}]{chabrier03}
{Chabrier}, G. 2003, \pasp, 115, 763

\bibitem[{{Chary} \& {Elbaz}(2001)}]{chary01}
{Chary}, R. \& {Elbaz}, D. 2001, \apj, 556, 562

\bibitem[{{Cicone} {et~al.}(2013){Cicone}, {Maiolino}, {Sturm},
  {Graci{\'a}-Carpio}, {Feruglio}, {Neri}, {Aalto}, {Davies}, {Fiore},
  {Fischer}, {Garc{\'{\i}}a-Burillo}, {Gonz{\'a}lez-Alfonso},
  {Hailey-Dunsheath}, {Piconcelli}, \& {Veilleux}}]{cicone13}
{Cicone}, C., {Maiolino}, R., {Sturm}, E., {et~al.} 2013, ArXiv e-prints

\bibitem[{{Cicone} {et~al.}(2014){Cicone}, {Maiolino}, {Sturm},
  {Graci{\'a}-Carpio}, {Feruglio}, {Neri}, {Aalto}, {Davies}, {Fiore},
  {Fischer}, {Garc{\'{\i}}a-Burillo}, {Gonz{\'a}lez-Alfonso},
  {Hailey-Dunsheath}, {Piconcelli}, \& {Veilleux}}]{cicone14}
---. 2014, \aap, 562, A21

\bibitem[{{Ciotti} \& {Ostriker}(2007)}]{ciotti07}
{Ciotti}, L. \& {Ostriker}, J.~P. 2007, \apj, 665, 1038

\bibitem[{{Cresci} {et~al.}(2015){Cresci}, {Mainieri}, {Brusa}, {Marconi},
  {Perna}, {Mannucci}, {Piconcelli}, {Maiolino}, {Feruglio}, {Fiore},
  {Bongiorno}, {Lanzuisi}, {Merloni}, {Schramm}, {Silverman}, \&
  {Civano}}]{cresci15}
{Cresci}, G., {Mainieri}, V., {Brusa}, M., {et~al.} 2015, \apj, 799, 82

\bibitem[{{Croton}(2006)}]{croton06}
{Croton}, D.~J. 2006, \mnras, 369, 1808

\bibitem[{{Daddi} {et~al.}(2007){Daddi}, {Dickinson}, {Morrison}, {Chary},
  {Cimatti}, {Elbaz}, {Frayer}, {Renzini}, {Pope}, {Alexander}, {Bauer},
  {Giavalisco}, {Huynh}, {Kurk}, \& {Mignoli}}]{daddi07}
{Daddi}, E., {Dickinson}, M., {Morrison}, G., {et~al.} 2007, \apj, 670, 156

\bibitem[{{Dale} {et~al.}(2014){Dale}, {Helou}, {Magdis}, {Armus},
  {D{\'{\i}}az-Santos}, \& {Shi}}]{dale14}
{Dale}, D.~A., {Helou}, G., {Magdis}, G.~E., {et~al.} 2014, \apj, 784, 83

\bibitem[{{Davies} {et~al.}(2007){Davies}, {M{\"u}ller S{\'a}nchez}, {Genzel},
  {Tacconi}, {Hicks}, {Friedrich}, \& {Sternberg}}]{davies7}
{Davies}, R.~I., {M{\"u}ller S{\'a}nchez}, F., {Genzel}, R., {et~al.} 2007,
  \apj, 671, 1388

\bibitem[{{de Grijs}(2001)}]{grijs01}
{de Grijs}, R. 2001, Astronomy and Geophysics, 42, 12

\bibitem[{{De Robertis} \& {Osterbrock}(1984)}]{derobertis84}
{De Robertis}, M.~M. \& {Osterbrock}, D.~E. 1984, \apj, 286, 171

\bibitem[{{De Robertis} \& {Shaw}(1990)}]{derobertis90}
{De Robertis}, M.~M. \& {Shaw}, R.~A. 1990, \apj, 348, 421

\bibitem[{{Debuhr} {et~al.}(2010){Debuhr}, {Quataert}, {Ma}, \&
  {Hopkins}}]{debuhr10}
{Debuhr}, J., {Quataert}, E., {Ma}, C.-P., \& {Hopkins}, P. 2010, \mnras, 406,
  L55

\bibitem[{{Di Matteo} {et~al.}(2005){Di Matteo}, {Springel}, \&
  {Hernquist}}]{dimatteo05}
{Di Matteo}, T., {Springel}, V., \& {Hernquist}, L. 2005, \nat, 433, 604

\bibitem[{{Diamond-Stanic} {et~al.}(2012){Diamond-Stanic}, {Moustakas},
  {Tremonti}, {Coil}, {Hickox}, {Robaina}, {Rudnick}, \& {Sell}}]{diamond12}
{Diamond-Stanic}, A.~M., {Moustakas}, J., {Tremonti}, C.~A., {et~al.} 2012,
  \apjl, 755, L26

\bibitem[{{Dimitrijevi{\'c}} {et~al.}(2007){Dimitrijevi{\'c}}, {Popovi{\'c}},
  {Kova{\v c}evi{\'c}}, {Da{\v c}i{\'c}}, \& {Ili{\'c}}}]{dimitrijevic07}
{Dimitrijevi{\'c}}, M.~S., {Popovi{\'c}}, L.~{\v C}., {Kova{\v c}evi{\'c}}, J.,
  {Da{\v c}i{\'c}}, M., \& {Ili{\'c}}, D. 2007, \mnras, 374, 1181

\bibitem[{{Draine} \& {Li}(2007)}]{draine07}
{Draine}, B.~T. \& {Li}, A. 2007, \apj, 657, 810

\bibitem[{{Elbaz} {et~al.}(2007){Elbaz}, {Daddi}, {Le Borgne}, {Dickinson},
  {Alexander}, {Chary}, {Starck}, {Brandt}, {Kitzbichler}, {MacDonald},
  {Nonino}, {Popesso}, {Stern}, \& {Vanzella}}]{elbaz07}
{Elbaz}, D., {Daddi}, E., {Le Borgne}, D., {et~al.} 2007, \aap, 468, 33

\bibitem[{{Elbaz} {et~al.}(2011){Elbaz}, {Dickinson}, {Hwang},
  {D{\'{\i}}az-Santos}, {Magdis}, {Magnelli}, {Le Borgne}, {Galliano},
  {Pannella}, {Chanial}, {Armus}, {Charmandaris}, {Daddi}, {Aussel}, {Popesso},
  {Kartaltepe}, {Altieri}, {Valtchanov}, {Coia}, {Dannerbauer}, {Dasyra},
  {Leiton}, {Mazzarella}, {Alexander}, {Buat}, {Burgarella}, {Chary}, {Gilli},
  {Ivison}, {Juneau}, {Le Floc'h}, {Lutz}, {Morrison}, {Mullaney}, {Murphy},
  {Pope}, {Scott}, {Brodwin}, {Calzetti}, {Cesarsky}, {Charlot}, {Dole},
  {Eisenhardt}, {Ferguson}, {F{\"o}rster Schreiber}, {Frayer}, {Giavalisco},
  {Huynh}, {Koekemoer}, {Papovich}, {Reddy}, {Surace}, {Teplitz}, {Yun}, \&
  {Wilson}}]{elbaz11}
{Elbaz}, D., {Dickinson}, M., {Hwang}, H.~S., {et~al.} 2011, \aap, 533, A119

\bibitem[{{Fabian}(2012)}]{fabian12}
{Fabian}, A.~C. 2012, \araa, 50, 455

\bibitem[{{Ferrarese} \& {Merritt}(2000)}]{ferrarese00}
{Ferrarese}, L. \& {Merritt}, D. 2000, \apjl, 539, L9

\bibitem[{{Feruglio} {et~al.}(2015){Feruglio}, {Fiore}, {Carniani},
  {Piconcelli}, {Zappacosta}, {Bongiorno}, {Cicone}, {Maiolino}, {Marconi},
  {Menci}, {Puccetti}, \& {Veilleux}}]{feruglio15}
{Feruglio}, C., {Fiore}, F., {Carniani}, S., {et~al.} 2015, ArXiv e-prints

\bibitem[{{Gabor} \& {Dav{\'e}}(2014)}]{gabor14}
{Gabor}, J.~M. \& {Dav{\'e}}, R. 2014, ArXiv e-prints

\bibitem[{{Gebhardt} {et~al.}(2000){Gebhardt}, {Bender}, {Bower}, {Dressler},
  {Faber}, {Filippenko}, {Green}, {Grillmair}, {Ho}, {Kormendy}, {Lauer},
  {Magorrian}, {Pinkney}, {Richstone}, \& {Tremaine}}]{gebhardt00}
{Gebhardt}, K., {Bender}, R., {Bower}, G., {et~al.} 2000, \apjl, 539, L13

\bibitem[{{Georgakakis} {et~al.}(2008){Georgakakis}, {Nandra}, {Yan},
  {Willner}, {Lotz}, {Pierce}, {Cooper}, {Laird}, {Koo}, {Barmby}, {Newman},
  {Primack}, \& {Coil}}]{georgakakis08}
{Georgakakis}, A., {Nandra}, K., {Yan}, R., {et~al.} 2008, \mnras, 385, 2049

\bibitem[{{Gon{\c c}alves} {et~al.}(2008){Gon{\c c}alves}, {Steidel}, \&
  {Pettini}}]{goncalves08}
{Gon{\c c}alves}, T.~S., {Steidel}, C.~C., \& {Pettini}, M. 2008, \apj, 676,
  816

\bibitem[{{Granato} {et~al.}(2004){Granato}, {De Zotti}, {Silva}, {Bressan}, \&
  {Danese}}]{granato04}
{Granato}, G.~L., {De Zotti}, G., {Silva}, L., {Bressan}, A., \& {Danese}, L.
  2004, \apj, 600, 580

\bibitem[{{Greene} \& {Ho}(2005)}]{greene05}
{Greene}, J.~E. \& {Ho}, L.~C. 2005, \apj, 627, 721

\bibitem[{{Griffin} {et~al.}(2010){Griffin}, {Abergel}, {Abreu}, {Ade},
  {Andr{\'e}}, {Augueres}, {Babbedge}, {Bae}, {Baillie}, {Baluteau}, {Barlow},
  {Bendo}, {Benielli}, {Bock}, {Bonhomme}, {Brisbin}, {Brockley-Blatt},
  {Caldwell}, {Cara}, {Castro-Rodriguez}, {Cerulli}, {Chanial}, {Chen},
  {Clark}, {Clements}, {Clerc}, {Coker}, {Communal}, {Conversi}, {Cox},
  {Crumb}, {Cunningham}, {Daly}, {Davis}, {de Antoni}, {Delderfield}, {Devin},
  {di Giorgio}, {Didschuns}, {Dohlen}, {Donati}, {Dowell}, {Dowell}, {Duband},
  {Dumaye}, {Emery}, {Ferlet}, {Ferrand}, {Fontignie}, {Fox}, {Franceschini},
  {Frerking}, {Fulton}, {Garcia}, {Gastaud}, {Gear}, {Glenn}, {Goizel},
  {Griffin}, {Grundy}, {Guest}, {Guillemet}, {Hargrave}, {Harwit}, {Hastings},
  {Hatziminaoglou}, {Herman}, {Hinde}, {Hristov}, {Huang}, {Imhof}, {Isaak},
  {Israelsson}, {Ivison}, {Jennings}, {Kiernan}, {King}, {Lange}, {Latter},
  {Laurent}, {Laurent}, {Leeks}, {Lellouch}, {Levenson}, {Li}, {Li},
  {Lilienthal}, {Lim}, {Liu}, {Lu}, {Madden}, {Mainetti}, {Marliani}, {McKay},
  {Mercier}, {Molinari}, {Morris}, {Moseley}, {Mulder}, {Mur}, {Naylor},
  {Nguyen}, {O'Halloran}, {Oliver}, {Olofsson}, {Olofsson}, {Orfei}, {Page},
  {Pain}, {Panuzzo}, {Papageorgiou}, {Parks}, {Parr-Burman}, {Pearce},
  {Pearson}, {P{\'e}rez-Fournon}, {Pinsard}, {Pisano}, {Podosek}, {Pohlen},
  {Polehampton}, {Pouliquen}, {Rigopoulou}, {Rizzo}, {Roseboom}, {Roussel},
  {Rowan-Robinson}, {Rownd}, {Saraceno}, {Sauvage}, {Savage}, {Savini},
  {Sawyer}, {Scharmberg}, {Schmitt}, {Schneider}, {Schulz}, {Schwartz},
  {Shafer}, {Shupe}, {Sibthorpe}, {Sidher}, {Smith}, {Smith}, {Smith},
  {Spencer}, {Stobie}, {Sudiwala}, {Sukhatme}, {Surace}, {Stevens}, {Swinyard},
  {Trichas}, {Tourette}, {Triou}, {Tseng}, {Tucker}, {Turner}, {Vaccari},
  {Valtchanov}, {Vigroux}, {Virique}, {Voellmer}, {Walker}, {Ward}, {Waskett},
  {Weilert}, {Wesson}, {White}, {Whitehouse}, {Wilson}, {Winter}, {Woodcraft},
  {Wright}, {Xu}, {Zavagno}, {Zemcov}, {Zhang}, \& {Zonca}}]{griffin10}
{Griffin}, M.~J., {Abergel}, A., {Abreu}, A., {et~al.} 2010, \aap, 518, L3

\bibitem[{{Hainline} {et~al.}(2011){Hainline}, {Blain}, {Smail}, {Alexander},
  {Armus}, {Chapman}, \& {Ivison}}]{hainline11}
{Hainline}, L.~J., {Blain}, A.~W., {Smail}, I., {et~al.} 2011, \apj, 740, 96

\bibitem[{{Harrison} {et~al.}(2014){Harrison}, {Alexander}, {Mullaney}, \&
  {Swinbank}}]{harrison14}
{Harrison}, C.~M., {Alexander}, D.~M., {Mullaney}, J.~R., \& {Swinbank}, A.~M.
  2014, \mnras, 441, 3306

\bibitem[{{Harrison} {et~al.}(2012){Harrison}, {Alexander}, {Swinbank},
  {Smail}, {Alaghband-Zadeh}, {Bauer}, {Chapman}, {Del Moro}, {Hickox},
  {Ivison}, {Men{\'e}ndez-Delmestre}, {Mullaney}, \& {Nesvadba}}]{harrison12}
{Harrison}, C.~M., {Alexander}, D.~M., {Swinbank}, A.~M., {et~al.} 2012,
  \mnras, 426, 1073

\bibitem[{{Heckman} {et~al.}(2004){Heckman}, {Kauffmann}, {Brinchmann},
  {Charlot}, {Tremonti}, \& {White}}]{heckman04}
{Heckman}, T.~M., {Kauffmann}, G., {Brinchmann}, J., {et~al.} 2004, \apj, 613,
  109

\bibitem[{{Hopkins} {et~al.}(2006){Hopkins}, {Somerville}, {Hernquist}, {Cox},
  {Robertson}, \& {Li}}]{hopkins06}
{Hopkins}, P.~F., {Somerville}, R.~S., {Hernquist}, L., {et~al.} 2006, \apj,
  652, 864

\bibitem[{{Hu} {et~al.}(2008){Hu}, {Wang}, {Ho}, {Chen}, {Zhang}, {Bian}, \&
  {Xue}}]{hu08}
{Hu}, C., {Wang}, J.-M., {Ho}, L.~C., {et~al.} 2008, \apj, 687, 78

\bibitem[{{Ishibashi} \& {Fabian}(2012)}]{ishibashi12}
{Ishibashi}, W. \& {Fabian}, A.~C. 2012, \mnras, 427, 2998

\bibitem[{{Jahnke} \& {Macci{\`o}}(2011)}]{jahnke11}
{Jahnke}, K. \& {Macci{\`o}}, A.~V. 2011, \apj, 734, 92

\bibitem[{{Jakobsen} {et~al.}(2003){Jakobsen}, {Jansen}, {Wagner}, \&
  {Reimers}}]{jakobsen03}
{Jakobsen}, P., {Jansen}, R.~A., {Wagner}, S., \& {Reimers}, D. 2003, \aap,
  397, 891

\bibitem[{{Jiang} {et~al.}(2007){Jiang}, {Fan}, {Ivezi{\'c}}, {Richards},
  {Schneider}, {Strauss}, \& {Kelly}}]{jiang07}
{Jiang}, L., {Fan}, X., {Ivezi{\'c}}, {\v Z}., {et~al.} 2007, \apj, 656, 680

\bibitem[{{Kalfountzou} {et~al.}(2012){Kalfountzou}, {Jarvis}, {Bonfield}, \&
  {Hardcastle}}]{kalfountzou12}
{Kalfountzou}, E., {Jarvis}, M.~J., {Bonfield}, D.~G., \& {Hardcastle}, M.~J.
  2012, \mnras, 427, 2401

\bibitem[{{Kewley} {et~al.}(2004){Kewley}, {Geller}, \& {Jansen}}]{kewley04}
{Kewley}, L.~J., {Geller}, M.~J., \& {Jansen}, R.~A. 2004, \aj, 127, 2002

\bibitem[{{Komossa} {et~al.}(2008){Komossa}, {Xu}, {Zhou}, {Storchi-Bergmann},
  \& {Binette}}]{komossa08}
{Komossa}, S., {Xu}, D., {Zhou}, H., {Storchi-Bergmann}, T., \& {Binette}, L.
  2008, \apj, 680, 926

\bibitem[{{Kroupa}(2001)}]{kroupa01}
{Kroupa}, P. 2001, \mnras, 322, 231

\bibitem[{{Larson} {et~al.}(1980){Larson}, {Tinsley}, \& {Caldwell}}]{larson80}
{Larson}, R.~B., {Tinsley}, B.~M., \& {Caldwell}, C.~N. 1980, \apj, 237, 692

\bibitem[{{Leipski} {et~al.}(2014){Leipski}, {Meisenheimer}, {Walter}, {Klaas},
  {Dannerbauer}, {De Rosa}, {Fan}, {Haas}, {Krause}, \& {Rix}}]{leipski14}
{Leipski}, C., {Meisenheimer}, K., {Walter}, F., {et~al.} 2014, \apj, 785, 154

\bibitem[{{Mainieri} {et~al.}(2011){Mainieri}, {Bongiorno}, {Merloni}, {Aller},
  {Carollo}, {Iwasawa}, {Koekemoer}, {Mignoli}, {Silverman}, {Bolzonella},
  {Brusa}, {Comastri}, {Gilli}, {Halliday}, {Ilbert}, {Lusso}, {Salvato},
  {Vignali}, {Zamorani}, {Contini}, {Kneib}, {Le F{\`e}vre}, {Lilly},
  {Renzini}, {Scodeggio}, {Balestra}, {Bardelli}, {Caputi}, {Coppa},
  {Cucciati}, {de la Torre}, {de Ravel}, {Franzetti}, {Garilli}, {Iovino},
  {Kampczyk}, {Knobel}, {Kova{\v c}}, {Lamareille}, {Le Borgne}, {Le Brun},
  {Maier}, {Nair}, {Pello}, {Peng}, {Perez Montero}, {Pozzetti},
  {Ricciardelli}, {Tanaka}, {Tasca}, {Tresse}, {Vergani}, {Zucca}, {Aussel},
  {Capak}, {Cappelluti}, {Elvis}, {Fiore}, {Hasinger}, {Impey}, {Le Floc'h},
  {Scoville}, {Taniguchi}, \& {Trump}}]{mainieri11}
{Mainieri}, V., {Bongiorno}, A., {Merloni}, A., {et~al.} 2011, \aap, 535, A80

\bibitem[{{Maiolino} {et~al.}(2012){Maiolino}, {Gallerani}, {Neri}, {Cicone},
  {Ferrara}, {Genzel}, {Lutz}, {Sturm}, {Tacconi}, {Walter}, {Feruglio},
  {Fiore}, \& {Piconcelli}}]{maiolino12}
{Maiolino}, R., {Gallerani}, S., {Neri}, R., {et~al.} 2012, \mnras, 425, L66

\bibitem[{{Markwardt}(2009)}]{markwardt09}
{Markwardt}, C.~B. 2009, in Astronomical Society of the Pacific Conference
  Series, Vol. 411, Astronomical Data Analysis Software and Systems XVIII, ed.
  D.~A. {Bohlender}, D.~{Durand}, \& P.~{Dowler}, 251

\bibitem[{{Martig} {et~al.}(2009){Martig}, {Bournaud}, {Teyssier}, \&
  {Dekel}}]{martig09}
{Martig}, M., {Bournaud}, F., {Teyssier}, R., \& {Dekel}, A. 2009, \apj, 707,
  250

\bibitem[{{Martin}(2005)}]{martin05}
{Martin}, C.~L. 2005, \apj, 621, 227

\bibitem[{{Matsuoka} {et~al.}(2011){Matsuoka}, {Nagao}, {Marconi}, {Maiolino},
  \& {Taniguchi}}]{matsuoka11}
{Matsuoka}, K., {Nagao}, T., {Marconi}, A., {Maiolino}, R., \& {Taniguchi}, Y.
  2011, \aap, 527, A100

\bibitem[{{Meidt} {et~al.}(2012){Meidt}, {Schinnerer}, {Knapen}, {Bosma},
  {Athanassoula}, {Sheth}, {Buta}, {Zaritsky}, {Laurikainen}, {Elmegreen},
  {Elmegreen}, {Gadotti}, {Salo}, {Regan}, {Ho}, {Madore}, {Hinz}, {Skibba},
  {Gil de Paz}, {Mu{\~n}oz-Mateos}, {Men{\'e}ndez-Delmestre}, {Seibert}, {Kim},
  {Mizusawa}, {Laine}, \& {Comer{\'o}n}}]{meidt12}
{Meidt}, S.~E., {Schinnerer}, E., {Knapen}, J.~H., {et~al.} 2012, \apj, 744, 17

\bibitem[{{Mullaney} {et~al.}(2013){Mullaney}, {Alexander}, {Fine}, {Goulding},
  {Harrison}, \& {Hickox}}]{mullaney13}
{Mullaney}, J.~R., {Alexander}, D.~M., {Fine}, S., {et~al.} 2013, \mnras, 433,
  622, interessante

\bibitem[{{Murphy} {et~al.}(2011){Murphy}, {Condon}, {Schinnerer}, {Kennicutt},
  {Calzetti}, {Armus}, {Helou}, {Turner}, {Aniano}, {Beir{\~a}o}, {Bolatto},
  {Brandl}, {Croxall}, {Dale}, {Donovan Meyer}, {Draine}, {Engelbracht},
  {Hunt}, {Hao}, {Koda}, {Roussel}, {Skibba}, \& {Smith}}]{murphy11}
{Murphy}, E.~J., {Condon}, J.~J., {Schinnerer}, E., {et~al.} 2011, \apj, 737,
  67

\bibitem[{{Nagao} {et~al.}(2006){Nagao}, {Marconi}, \& {Maiolino}}]{nagao06}
{Nagao}, T., {Marconi}, A., \& {Maiolino}, R. 2006, \aap, 447, 157

\bibitem[{{Nagao} {et~al.}(2001){Nagao}, {Murayama}, \& {Taniguchi}}]{nagao01}
{Nagao}, T., {Murayama}, T., \& {Taniguchi}, Y. 2001, \apj, 549, 155

\bibitem[{{Nenkova} {et~al.}(2008){Nenkova}, {Sirocky}, {Ivezi{\'c}}, \&
  {Elitzur}}]{nenkova08}
{Nenkova}, M., {Sirocky}, M.~M., {Ivezi{\'c}}, {\v Z}., \& {Elitzur}, M. 2008,
  \apj, 685, 147

\bibitem[{{Netzer}(2009)}]{netzer09}
{Netzer}, H. 2009, \mnras, 399, 1907

\bibitem[{{Noeske} {et~al.}(2007){Noeske}, {Weiner}, {Faber}, {Papovich},
  {Koo}, {Somerville}, {Bundy}, {Conselice}, {Newman}, {Schiminovich}, {Le
  Floc'h}, {Coil}, {Rieke}, {Lotz}, {Primack}, {Barmby}, {Cooper}, {Davis},
  {Ellis}, {Fazio}, {Guhathakurta}, {Huang}, {Kassin}, {Martin}, {Phillips},
  {Rich}, {Small}, {Willmer}, \& {Wilson}}]{noeske07}
{Noeske}, K.~G., {Weiner}, B.~J., {Faber}, S.~M., {et~al.} 2007, \apjl, 660,
  L43

\bibitem[{{Osterbrock}(1989)}]{osterbrock89}
{Osterbrock}, D.~E. 1989, {Astrophysics of gaseous nebulae and active galactic
  nuclei}

\bibitem[{{P{\^a}ris} {et~al.}(2012){P{\^a}ris}, {Petitjean}, {Aubourg},
  {Bailey}, {Ross}, {Myers}, {Strauss}, {Anderson}, {Arnau}, {Bautista},
  {Bizyaev}, {Bolton}, {Bovy}, {Brandt}, {Brewington}, {Browstein}, {Busca},
  {Capellupo}, {Carithers}, {Croft}, {Dawson}, {Delubac}, {Ebelke},
  {Eisenstein}, {Engelke}, {Fan}, {Filiz Ak}, {Finley}, {Font-Ribera}, {Ge},
  {Gibson}, {Hall}, {Hamann}, {Hennawi}, {Ho}, {Hogg}, {Ivezi{\'c}}, {Jiang},
  {Kimball}, {Kirkby}, {Kirkpatrick}, {Lee}, {Le Goff}, {Lundgren}, {MacLeod},
  {Malanushenko}, {Malanushenko}, {Maraston}, {McGreer}, {McMahon},
  {Miralda-Escud{\'e}}, {Muna}, {Noterdaeme}, {Oravetz},
  {Palanque-Delabrouille}, {Pan}, {Perez-Fournon}, {Pieri}, {Richards},
  {Rollinde}, {Sheldon}, {Schlegel}, {Schneider}, {Slosar}, {Shelden}, {Shen},
  {Simmons}, {Snedden}, {Suzuki}, {Tinker}, {Viel}, {Weaver}, {Weinberg},
  {White}, {Wood-Vasey}, \& {Y{\`e}che}}]{paris12}
{P{\^a}ris}, I., {Petitjean}, P., {Aubourg}, {\'E}., {et~al.} 2012, \aap, 548,
  A66

\bibitem[{{Peng} {et~al.}(2015){Peng}, {Maiolino}, \& {Cochrane}}]{peng15}
{Peng}, Y., {Maiolino}, R., \& {Cochrane}, R. 2015, \nat, 521, 192

\bibitem[{{Perna} {et~al.}(2015){Perna}, {Brusa}, {Cresci}, {Comastri},
  {Lanzuisi}, {Lusso}, {Marconi}, {Salvato}, {Zamorani}, {Bongiorno},
  {Mainieri}, {Maiolino}, \& {Mignoli}}]{perna15}
{Perna}, M., {Brusa}, M., {Cresci}, G., {et~al.} 2015, \aap, 574, A82

\bibitem[{{Poglitsch} {et~al.}(2010){Poglitsch}, {Waelkens}, {Geis},
  {Feuchtgruber}, {Vandenbussche}, {Rodriguez}, {Krause}, {Renotte}, {van
  Hoof}, {Saraceno}, {Cepa}, {Kerschbaum}, {Agn{\`e}se}, {Ali}, {Altieri},
  {Andreani}, {Augueres}, {Balog}, {Barl}, {Bauer}, {Belbachir}, {Benedettini},
  {Billot}, {Boulade}, {Bischof}, {Blommaert}, {Callut}, {Cara}, {Cerulli},
  {Cesarsky}, {Contursi}, {Creten}, {De Meester}, {Doublier}, {Doumayrou},
  {Duband}, {Exter}, {Genzel}, {Gillis}, {Gr{\"o}zinger}, {Henning},
  {Herreros}, {Huygen}, {Inguscio}, {Jakob}, {Jamar}, {Jean}, {de Jong},
  {Katterloher}, {Kiss}, {Klaas}, {Lemke}, {Lutz}, {Madden}, {Marquet},
  {Martignac}, {Mazy}, {Merken}, {Montfort}, {Morbidelli}, {M{\"u}ller},
  {Nielbock}, {Okumura}, {Orfei}, {Ottensamer}, {Pezzuto}, {Popesso},
  {Putzeys}, {Regibo}, {Reveret}, {Royer}, {Sauvage}, {Schreiber}, {Stegmaier},
  {Schmitt}, {Schubert}, {Sturm}, {Thiel}, {Tofani}, {Vavrek}, {Wetzstein},
  {Wieprecht}, \& {Wiezorrek}}]{poglitsch10}
{Poglitsch}, A., {Waelkens}, C., {Geis}, N., {et~al.} 2010, \aap, 518, L2

\bibitem[{{Pozzetti} {et~al.}(2010){Pozzetti}, {Bolzonella}, {Zucca},
  {Zamorani}, {Lilly}, {Renzini}, {Moresco}, {Mignoli}, {Cassata}, {Tasca},
  {Lamareille}, {Maier}, {Meneux}, {Halliday}, {Oesch}, {Vergani}, {Caputi},
  {Kova{\v c}}, {Cimatti}, {Cucciati}, {Iovino}, {Peng}, {Carollo}, {Contini},
  {Kneib}, {Le F{\'e}vre}, {Mainieri}, {Scodeggio}, {Bardelli}, {Bongiorno},
  {Coppa}, {de la Torre}, {de Ravel}, {Franzetti}, {Garilli}, {Kampczyk},
  {Knobel}, {Le Borgne}, {Le Brun}, {Pell{\`o}}, {Perez Montero},
  {Ricciardelli}, {Silverman}, {Tanaka}, {Tresse}, {Abbas}, {Bottini}, {Cappi},
  {Guzzo}, {Koekemoer}, {Leauthaud}, {Maccagni}, {Marinoni}, {McCracken},
  {Memeo}, {Porciani}, {Scaramella}, {Scarlata}, \& {Scoville}}]{pozzetti10}
{Pozzetti}, L., {Bolzonella}, M., {Zucca}, E., {et~al.} 2010, \aap, 523, A13

\bibitem[{{Pradhan} {et~al.}(2006){Pradhan}, {Montenegro}, {Nahar}, \&
  {Eissner}}]{pradhan05}
{Pradhan}, A.~K., {Montenegro}, M., {Nahar}, S.~N., \& {Eissner}, W. 2006,
  \mnras, 366, L6

\bibitem[{{Richards} {et~al.}(2002){Richards}, {Fan}, {Newberg}, {Strauss},
  {Vanden Berk}, {Schneider}, {Yanny}, {Boucher}, {Burles}, {Frieman}, {Gunn},
  {Hall}, {Ivezi{\'c}}, {Kent}, {Loveday}, {Lupton}, {Rockosi}, {Schlegel},
  {Stoughton}, {SubbaRao}, \& {York}}]{richards02}
{Richards}, G.~T., {Fan}, X., {Newberg}, H.~J., {et~al.} 2002, \aj, 123, 2945

\bibitem[{{Rieke} {et~al.}(2009){Rieke}, {Alonso-Herrero}, {Weiner},
  {P{\'e}rez-Gonz{\'a}lez}, {Blaylock}, {Donley}, \& {Marcillac}}]{rieke09}
{Rieke}, G.~H., {Alonso-Herrero}, A., {Weiner}, B.~J., {et~al.} 2009, \apj,
  692, 556

\bibitem[{{Roche} {et~al.}(1991){Roche}, {Aitken}, \& {Smith}}]{roche91}
{Roche}, P.~F., {Aitken}, D.~K., \& {Smith}, C.~H. 1991, \mnras, 252, 282

\bibitem[{{Roos} {et~al.}(2015){Roos}, {Juneau}, {Bournaud}, \&
  {Gabor}}]{roos15}
{Roos}, O., {Juneau}, S., {Bournaud}, F., \& {Gabor}, J.~M. 2015, \apj, 800, 19

\bibitem[{{Roseboom} {et~al.}(2013){Roseboom}, {Lawrence}, {Elvis}, {Petty},
  {Shen}, \& {Hao}}]{roseboom13}
{Roseboom}, I.~G., {Lawrence}, A., {Elvis}, M., {et~al.} 2013, \mnras, 429,
  1494

\bibitem[{{Rupke} {et~al.}(2005){Rupke}, {Veilleux}, \& {Sanders}}]{rupke05}
{Rupke}, D.~S., {Veilleux}, S., \& {Sanders}, D.~B. 2005, \apj, 632, 751

\bibitem[{{Rupke} \& {Veilleux}(2013)}]{rupke13}
{Rupke}, D.~S.~N. \& {Veilleux}, S. 2013, \apj, 768, 75

\bibitem[{{Santini} {et~al.}(2012){Santini}, {Rosario}, {Shao}, {Lutz},
  {Maiolino}, {Alexander}, {Altieri}, {Andreani}, {Aussel}, {Bauer}, {Berta},
  {Bongiovanni}, {Brandt}, {Brusa}, {Cepa}, {Cimatti}, {Daddi}, {Elbaz},
  {Fontana}, {F{\"o}rster Schreiber}, {Genzel}, {Grazian}, {Le Floc'h},
  {Magnelli}, {Mainieri}, {Nordon}, {P{\'e}rez Garcia}, {Poglitsch}, {Popesso},
  {Pozzi}, {Riguccini}, {Rodighiero}, {Salvato}, {Sanchez-Portal}, {Sturm},
  {Tacconi}, {Valtchanov}, \& {Wuyts}}]{santini12}
{Santini}, P., {Rosario}, D.~J., {Shao}, L., {et~al.} 2012, \aap, 540, A109

\bibitem[{{Schawinski} {et~al.}(2009){Schawinski}, {Virani}, {Simmons}, {Urry},
  {Treister}, {Kaviraj}, \& {Kushkuley}}]{schawinski09}
{Schawinski}, K., {Virani}, S., {Simmons}, B., {et~al.} 2009, \apjl, 692, L19

\bibitem[{{Schlegel} {et~al.}(2007){Schlegel}, {Blanton}, {Eisenstein},
  {Gillespie}, {Gunn}, {Harding}, {McDonald}, {Nichol}, {Padmanabhan},
  {Percival}, {Richards}, {Rockosi}, {Roe}, {Ross}, {Schneider}, {Strauss},
  {Weinberg}, \& {White}}]{schlegel07}
{Schlegel}, D.~J., {Blanton}, M., {Eisenstein}, D., {et~al.} 2007, in Bulletin
  of the American Astronomical Society, Vol.~39, American Astronomical Society
  Meeting Abstracts, 132.29

\bibitem[{{Schneider} {et~al.}(2010){Schneider}, {Richards}, {Hall}, {Strauss},
  {Anderson}, {Boroson}, {Ross}, {Shen}, {Brandt}, {Fan}, {Inada}, {Jester},
  {Knapp}, {Krawczyk}, {Thakar}, {Vanden Berk}, {Voges}, {Yanny}, {York},
  {Bahcall}, {Bizyaev}, {Blanton}, {Brewington}, {Brinkmann}, {Eisenstein},
  {Frieman}, {Fukugita}, {Gray}, {Gunn}, {Hibon}, {Ivezi{\'c}}, {Kent}, {Kron},
  {Lee}, {Lupton}, {Malanushenko}, {Malanushenko}, {Oravetz}, {Pan}, {Pier},
  {Price}, {Saxe}, {Schlegel}, {Simmons}, {Snedden}, {SubbaRao}, {Szalay}, \&
  {Weinberg}}]{schneider10}
{Schneider}, D.~P., {Richards}, G.~T., {Hall}, P.~B., {et~al.} 2010, \aj, 139,
  2360

\bibitem[{{Seymour} {et~al.}(2008){Seymour}, {Dwelly}, {Moss}, {McHardy},
  {Zoghbi}, {Rieke}, {Page}, {Hopkins}, \& {Loaring}}]{seymour08}
{Seymour}, N., {Dwelly}, T., {Moss}, D., {et~al.} 2008, \mnras, 386, 1695

\bibitem[{{Shao} {et~al.}(2010){Shao}, {Lutz}, {Nordon}, {Maiolino},
  {Alexander}, {Altieri}, {Andreani}, {Aussel}, {Bauer}, {Berta},
  {Bongiovanni}, {Brandt}, {Brusa}, {Cava}, {Cepa}, {Cimatti}, {Daddi},
  {Dominguez-Sanchez}, {Elbaz}, {F{\"o}rster Schreiber}, {Geis}, {Genzel},
  {Grazian}, {Gruppioni}, {Magdis}, {Magnelli}, {Mainieri}, {P{\'e}rez
  Garc{\'{\i}}a}, {Poglitsch}, {Popesso}, {Pozzi}, {Riguccini}, {Rodighiero},
  {Rovilos}, {Saintonge}, {Salvato}, {Sanchez Portal}, {Santini}, {Sturm},
  {Tacconi}, {Valtchanov}, {Wetzstein}, \& {Wieprecht}}]{shao10}
{Shao}, L., {Lutz}, D., {Nordon}, R., {et~al.} 2010, \aap, 518, L26

\bibitem[{{Shen} {et~al.}(2011){Shen}, {Richards}, {Strauss}, {Hall},
  {Schneider}, {Snedden}, {Bizyaev}, {Brewington}, {Malanushenko},
  {Malanushenko}, {Oravetz}, {Pan}, \& {Simmons}}]{shen11}
{Shen}, Y., {Richards}, G.~T., {Strauss}, M.~A., {et~al.} 2011, \apjs, 194, 45

\bibitem[{{Siebenmorgen} {et~al.}(2004){Siebenmorgen}, {Kr{\"u}gel}, \&
  {Spoon}}]{siebenmorgen04}
{Siebenmorgen}, R., {Kr{\"u}gel}, E., \& {Spoon}, H.~W.~W. 2004, \aap, 414, 123

\bibitem[{{Silva} {et~al.}(1998){Silva}, {Granato}, {Bressan}, \&
  {Danese}}]{silva98}
{Silva}, L., {Granato}, G.~L., {Bressan}, A., \& {Danese}, L. 1998, \apj, 509,
  103

\bibitem[{{Slone} \& {Netzer}(2012)}]{slone12}
{Slone}, O. \& {Netzer}, H. 2012, \mnras, 426, 656

\bibitem[{{Stanley} {et~al.}(2015){Stanley}, {Harrison}, {Alexander},
  {Swinbank}, {Aird}, {Del Moro}, {Hickox}, \& {Mullaney}}]{stanley15}
{Stanley}, F., {Harrison}, C.~M., {Alexander}, D.~M., {et~al.} 2015, ArXiv
  e-prints

\bibitem[{{Tombesi} {et~al.}(2015){Tombesi}, {Mel{\'e}ndez}, {Veilleux},
  {Reeves}, {Gonz{\'a}lez-Alfonso}, \& {Reynolds}}]{tombesi15}
{Tombesi}, F., {Mel{\'e}ndez}, M., {Veilleux}, S., {et~al.} 2015, \nat, 519,
  436

\bibitem[{{Tombesi} {et~al.}(2010){Tombesi}, {Sambruna}, {Reeves}, {Braito},
  {Ballo}, {Gofford}, {Cappi}, \& {Mushotzky}}]{tombesi10}
{Tombesi}, F., {Sambruna}, R.~M., {Reeves}, J.~N., {et~al.} 2010, \apj, 719,
  700

\bibitem[{{Tremonti} {et~al.}(2007){Tremonti}, {Moustakas}, \&
  {Diamond-Stanic}}]{tremonti07}
{Tremonti}, C.~A., {Moustakas}, J., \& {Diamond-Stanic}, A.~M. 2007, \apjl,
  663, L77

\bibitem[{{Turnshek} {et~al.}(1988){Turnshek}, {Grillmair}, {Foltz}, \&
  {Weymann}}]{turnshek88}
{Turnshek}, D.~A., {Grillmair}, C.~J., {Foltz}, C.~B., \& {Weymann}, R.~J.
  1988, \apj, 325, 651

\bibitem[{{Veilleux}(1991)}]{veilleux91}
{Veilleux}, S. 1991, \apj, 369, 331

\bibitem[{{Veilleux} {et~al.}(2013){Veilleux}, {Mel{\'e}ndez}, {Sturm},
  {Gracia-Carpio}, {Fischer}, {Gonz{\'a}lez-Alfonso}, {Contursi}, {Lutz},
  {Poglitsch}, {Davies}, {Genzel}, {Tacconi}, {de Jong}, {Sternberg}, {Netzer},
  {Hailey-Dunsheath}, {Verma}, {Rupke}, {Maiolino}, {Teng}, \&
  {Polisensky}}]{veilleux13}
{Veilleux}, S., {Mel{\'e}ndez}, M., {Sturm}, E., {et~al.} 2013, \apj, 776, 27

\bibitem[{{Veilleux} {et~al.}(2009){Veilleux}, {Rupke}, {Kim}, {Genzel},
  {Sturm}, {Lutz}, {Contursi}, {Schweitzer}, {Tacconi}, {Netzer}, {Sternberg},
  {Mihos}, {Baker}, {Mazzarella}, {Lord}, {Sanders}, {Stockton}, {Joseph}, \&
  {Barnes}}]{veilleux09}
{Veilleux}, S., {Rupke}, D.~S.~N., {Kim}, D.-C., {et~al.} 2009, \apjs, 182, 628

\bibitem[{{V{\'e}ron-Cetty} {et~al.}(2004){V{\'e}ron-Cetty}, {Joly}, \&
  {V{\'e}ron}}]{veron04}
{V{\'e}ron-Cetty}, M.-P., {Joly}, M., \& {V{\'e}ron}, P. 2004, \aap, 417, 515

\bibitem[{{Villar-Mart{\'{\i}}n}
  {et~al.}(2011{\natexlab{a}}){Villar-Mart{\'{\i}}n}, {Humphrey}, {Delgado},
  {Colina}, \& {Arribas}}]{villar11}
{Villar-Mart{\'{\i}}n}, M., {Humphrey}, A., {Delgado}, R.~G., {Colina}, L., \&
  {Arribas}, S. 2011{\natexlab{a}}, \mnras, 418, 2032

\bibitem[{{Villar-Mart{\'{\i}}n}
  {et~al.}(2011{\natexlab{b}}){Villar-Mart{\'{\i}}n}, {Humphrey}, {Delgado},
  {Colina}, \& {Arribas}}]{villar-Martin11}
---. 2011{\natexlab{b}}, \mnras, 418, 2032

\bibitem[{{Weiner} {et~al.}(2009){Weiner}, {Coil}, {Prochaska}, {Newman},
  {Cooper}, {Bundy}, {Conselice}, {Dutton}, {Faber}, {Koo}, {Lotz}, {Rieke}, \&
  {Rubin}}]{weiner09}
{Weiner}, B.~J., {Coil}, A.~L., {Prochaska}, J.~X., {et~al.} 2009, \apj, 692,
  187

\bibitem[{{Whittle}(1985)}]{whittle85}
{Whittle}, M. 1985, \mnras, 216, 817

\bibitem[{{Wild} {et~al.}(2010){Wild}, {Heckman}, \& {Charlot}}]{wild10}
{Wild}, V., {Heckman}, T., \& {Charlot}, S. 2010, \mnras, 405, 933

\bibitem[{{Yesuf} {et~al.}(2014){Yesuf}, {Faber}, {Trump}, {Koo}, {Fang},
  {Liu}, {Wild}, \& {Hayward}}]{yesuf14}
{Yesuf}, H.~M., {Faber}, S.~M., {Trump}, J.~R., {et~al.} 2014, \apj, 792, 84

\bibitem[{{Zakamska} \& {Greene}(2014)}]{zakamska14}
{Zakamska}, N.~L. \& {Greene}, J.~E. 2014, ArXiv e-prints

\bibitem[{{Zinn} {et~al.}(2013){Zinn}, {Middelberg}, {Norris}, \&
  {Dettmar}}]{zinn13}
{Zinn}, P.-C., {Middelberg}, E., {Norris}, R.~P., \& {Dettmar}, R.-J. 2013,
  \apj, 774, 66

\end{thebibliography}

\begin{appendix}
\section{}

We explored the accuracy of our SED fitting code, since the best model that reproduces the photometric data points can  potentially 
be subject to model degeneracy. 
To evaluate the uncertainties on  the main quantities derived from the fit, i.e. $L_{\rm AGN}$,
$L_{\rm SF}$, and $M_{\rm star}$, in relation to the photometric uncertainties and model degeneracy, we performed some tests. 
 
 First of all, we built various  theoretical SEDs, extracting photometric fluxes from our best fitting models.
 In each case the integrated luminosity of the SF and AGN component were known.
 
 Second, we reproduced real cases, adding observed, realistic random errors to the theoretical photometric points 
 (substituting randomly some photometric points with upper limits). Then we fit this SED
 obtaining L$_{\rm SF}$ L$_{\rm AGN}$ observed. We repeated this procedure for 100 times.
  
 In Fig.A1, left panel, we show an example of a result of this analysis for one theoretical SED. 
 We plotted the distribution of the values obtained in 100 iterations, and we compared the obtained values with 
 the theoretical value shown with a dashed line. In each iteration the best model is composed of different templates; 
 however, the distribution of the
 integrated luminosities are centred on the expected value. In Fig.A1, right panel, we report in percentage the distribution of the relative errors 
 for the main quantities $L_{\rm AGN}$,
$L_{\rm SF}$, and $M_{\rm star}$ in all the
 iterations compared to the errors that we  obtained in the theoretical case.
 
 As a consistency check, we compare the colour obtained by \citet{veilleux09} for a sample of
 ultraluminous infrared galaxies and Palomar
Green quasars with Spitzer. We derive the
flux ratios at different wavelengths from the SEDs best fitting model. In Fig.\ref{veilleux}, left panel,
 we compare $F_{\rm 30 \mu m}/F_{\rm 15 \mu m}$ with respect to the flux ratio between 5 and 25 \mc (mid-IR) and between 40 and 120 \mc  (far-IR).
 In the right panel, we plot
  $F_{\rm 15 \mu m}/F_{\rm 6 \mu m}$ with respect to 
  $F_{\rm 30 \mu m}/F_{\rm 6 \mu m}$. In both cases
 our quasars extend  the relation found by \citet{veilleux09} downward.
 
\begin{figure*}
\centering{
\includegraphics[scale=0.4,angle=0]{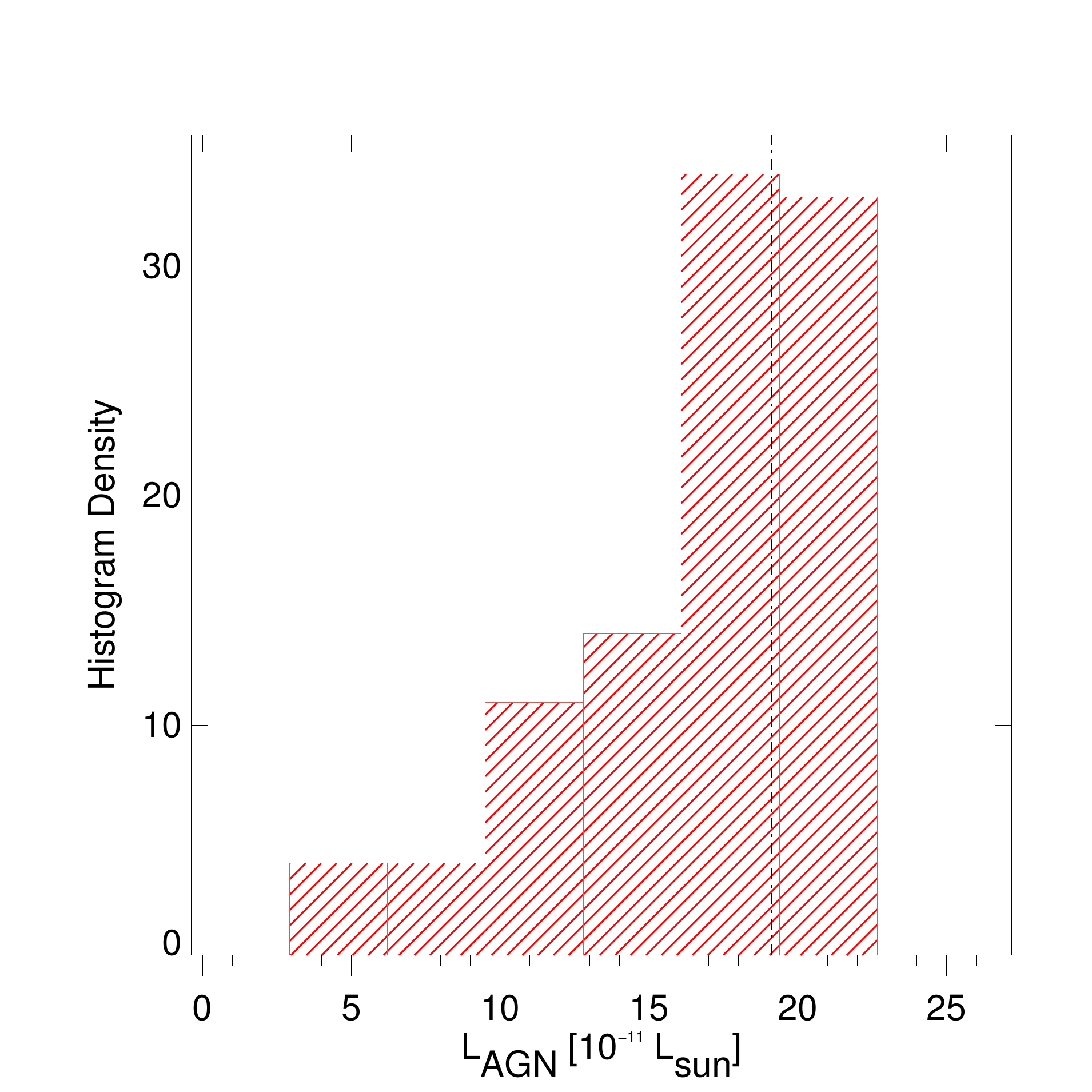}
\includegraphics[scale=0.4,angle=0]{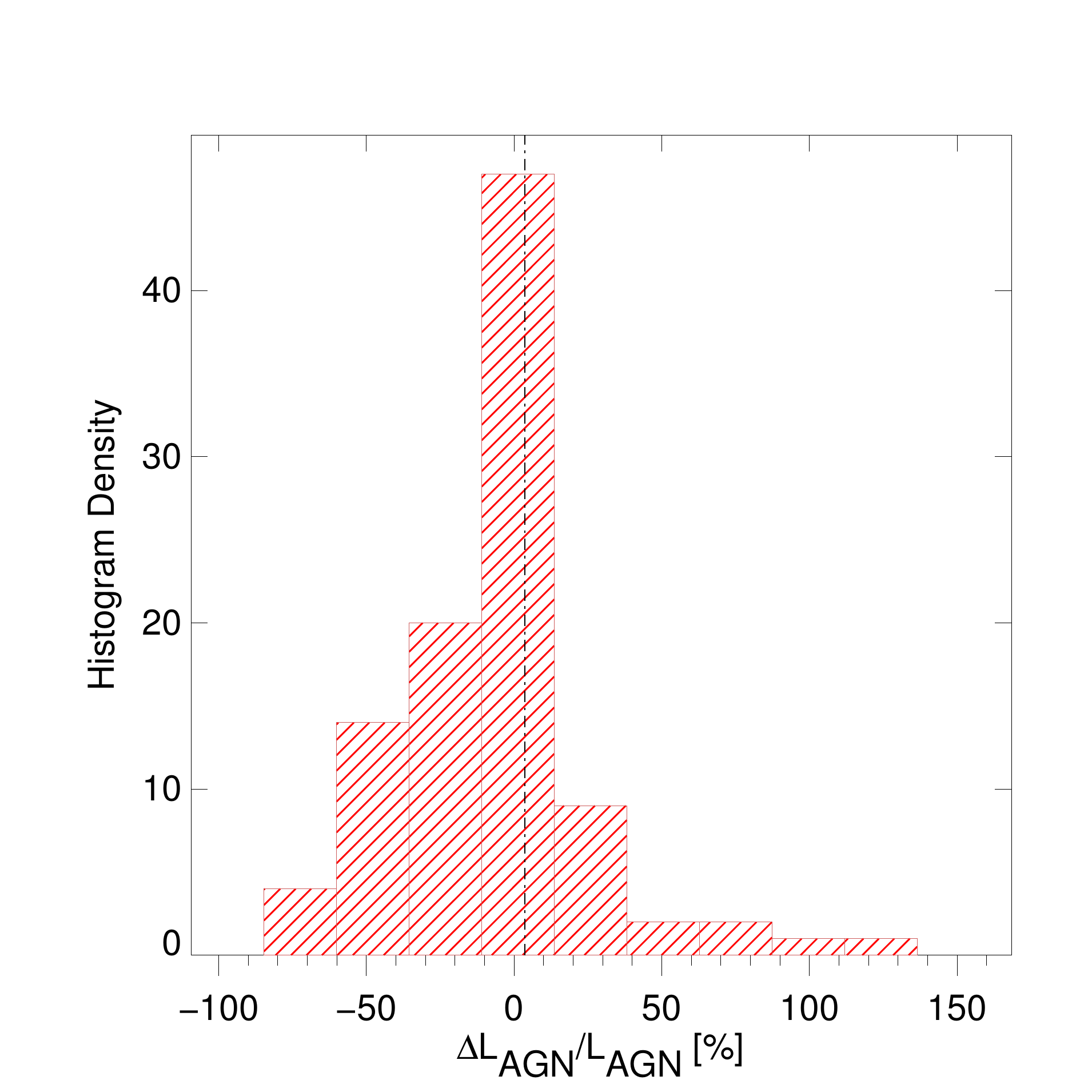}
\includegraphics[scale=0.4,angle=0]{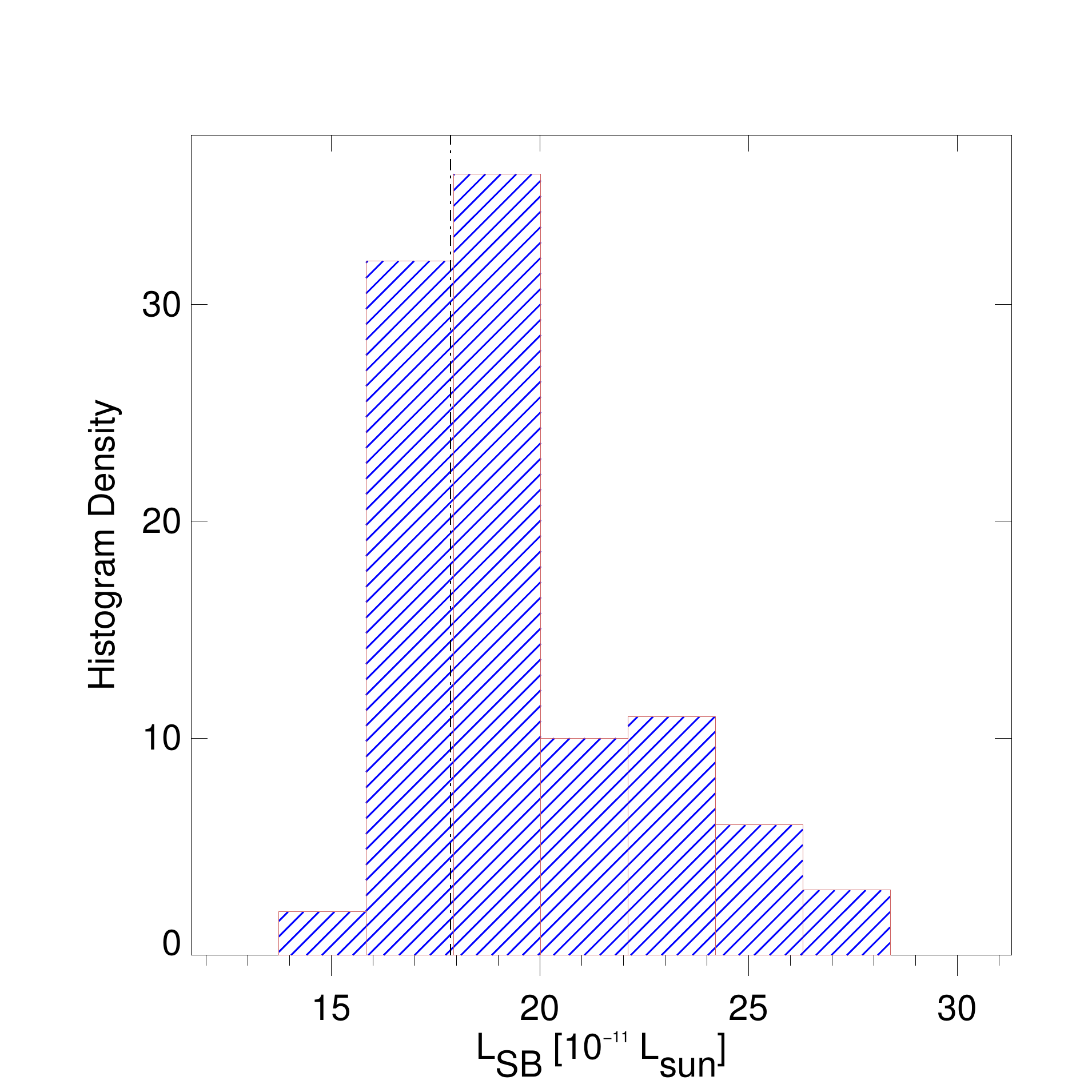}
\includegraphics[scale=0.4,angle=0]{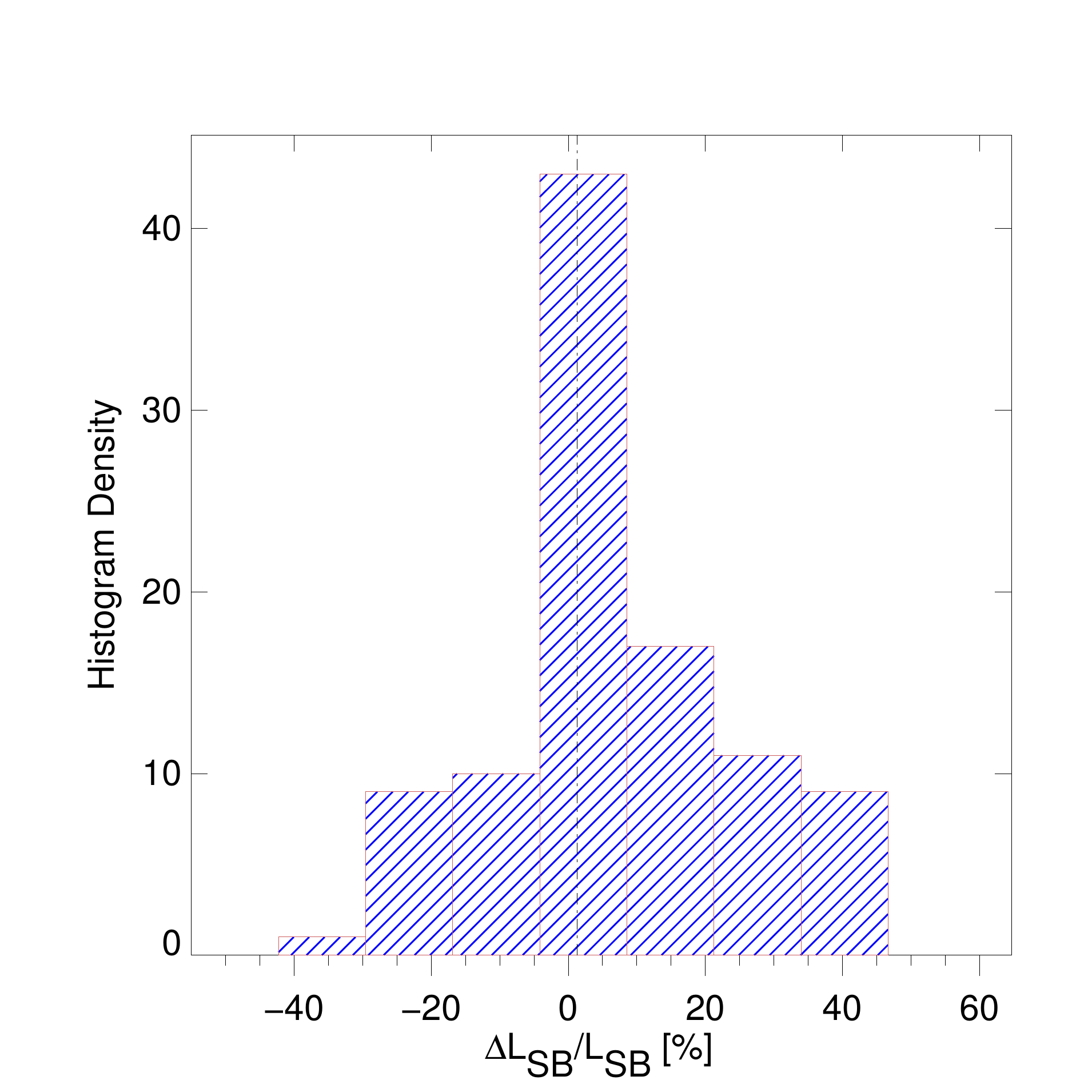}
\includegraphics[scale=0.4,angle=0]{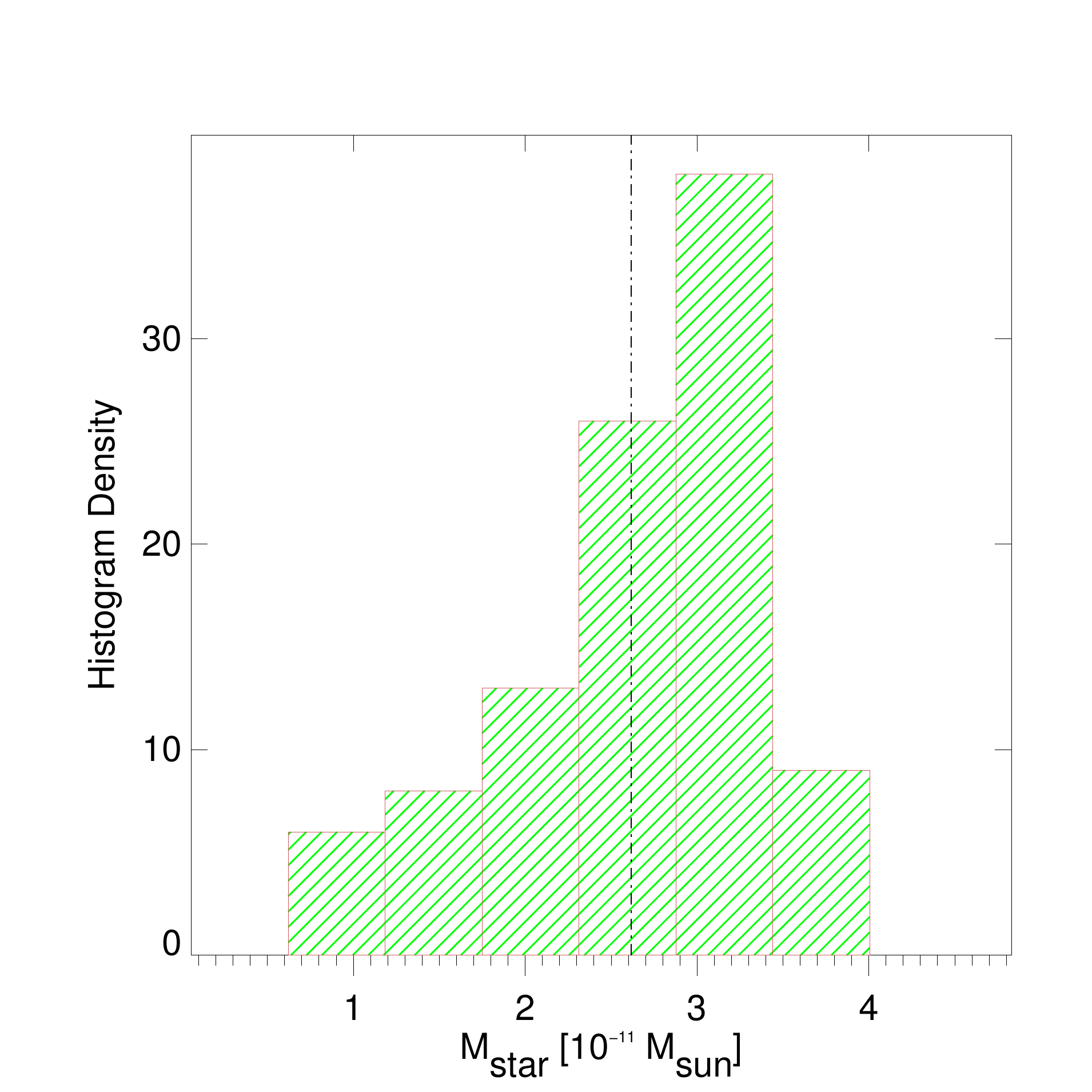}
\includegraphics[scale=0.4,angle=0]{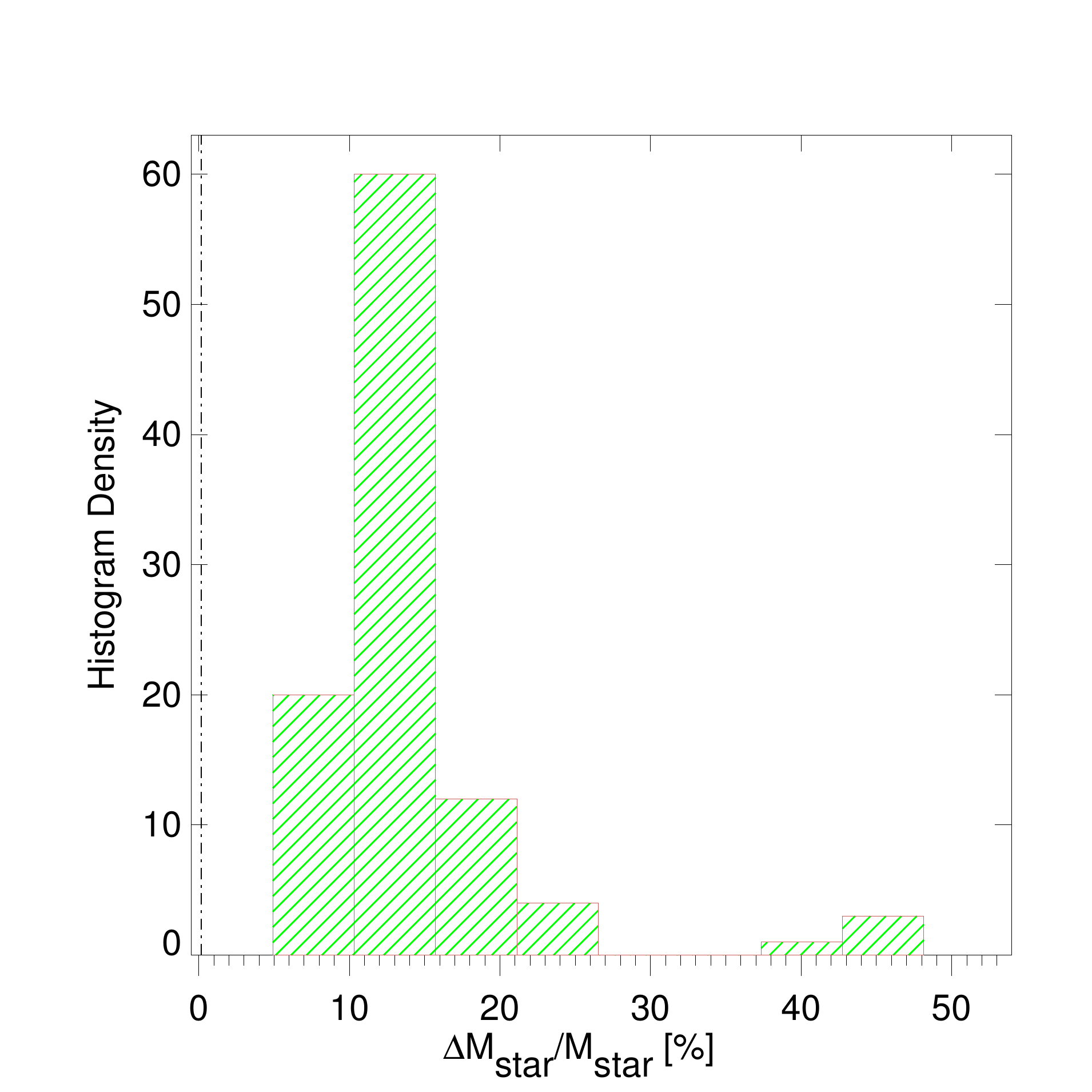}
\caption{Example  of a SED fitting test.
Left panel: Distribution of the AGN luminosity, SB luminosity, and stellar mass obtained in 100 iterations obtained
from  theoretical SEDs, and 
by varying the photometric points according to the observed uncertainties compared to the theoretical value represented by the dashed line. 
Right panel: as before, but we plot the distribution in percentage of the relative errors for the  parameters in each iteration.}
}
\end{figure*}

\begin{figure*}
\centering{
\includegraphics[scale=0.4,angle=0]{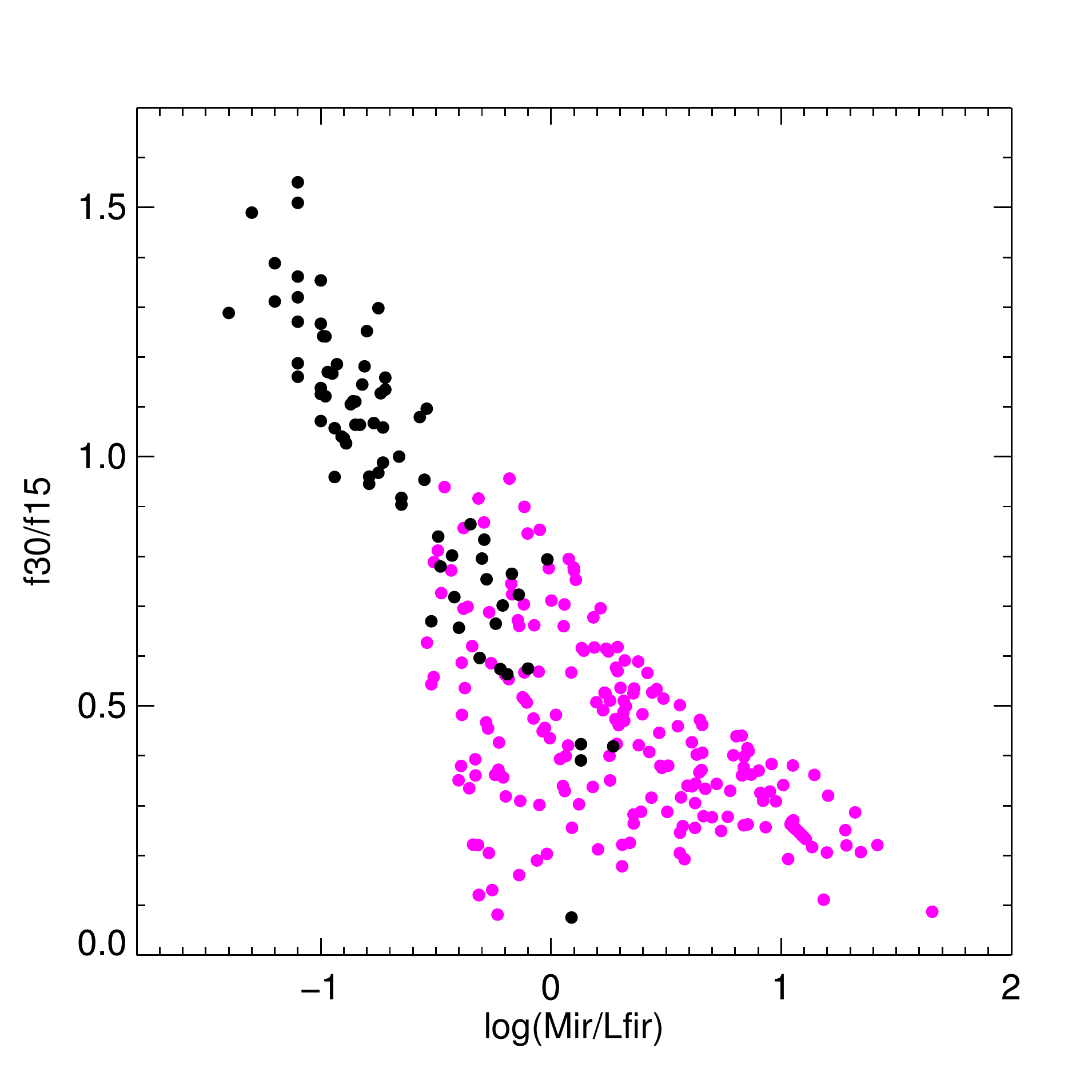}
\includegraphics[scale=0.4,angle=0]{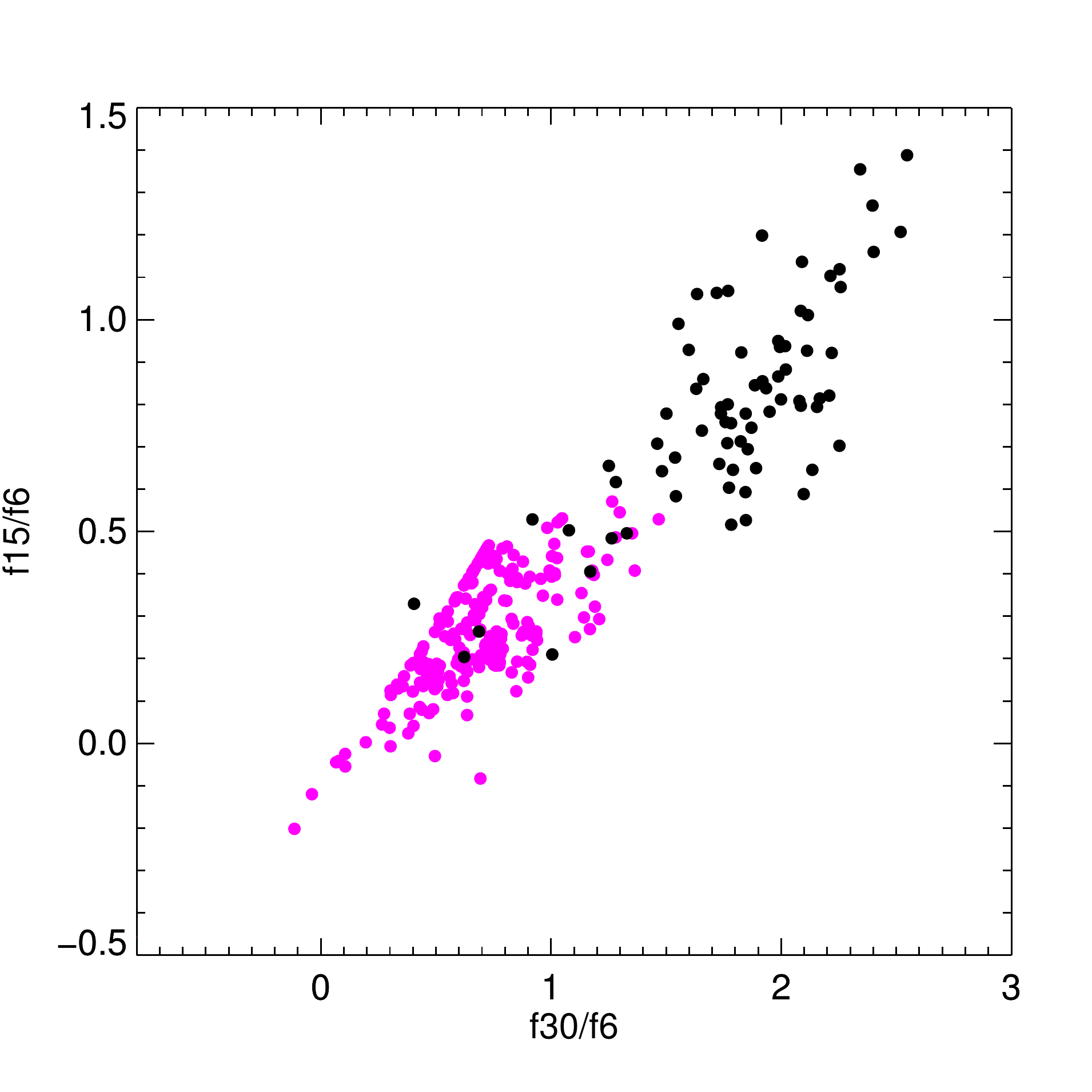}
}
\caption{Comparison of SED colours derived from the SED with respect to a sample of PG quasars with Spitzer measurements and
ULIR galaxies (black points, from Veilleux et al. 2009) and our sample of quasars (magenta points).}
\label{veilleux}
\end{figure*}
\end{appendix}

\end{document}